\documentclass[nonacm,autogobble,dvipsnames,acmsmall,screen]{acmart}
\makeatletter

\@ifclasswith{acmart}{review}{
  \newcommand{\reviewmode}{}
}{
}

\usepackage{amsmath}
\usepackage{xparse}

\usepackage{mathpartir}
\usepackage[capitalise]{cleveref}
\usepackage{booktabs}   %
\usepackage{csquotes}
\usepackage{fancyvrb}
\usepackage{stmaryrd}
\usepackage{adjustbox}
\usepackage{xspace}
\usepackage{relsize}
\usepackage{comment}

\usepackage{savesym}
\savesymbol{program}
\usepackage{amsmath}
\usepackage{semantic}
\usepackage{mdframed}

\numberwithin{desiderata}{section}
\crefname{desiderata}{Desideratum}{Desiderata}

\usepackage[geometry]{ifsym}

\DeclareMathAlphabet{\mymathbb}{U}{bbold}{m}{n}

\newcommand\doubleplus{+\kern-1.3ex+\kern0.8ex}

\NewDocumentCommand\defineMeta{mmmm}{%
    \expandafter\DeclareDocumentCommand\csname#1#2\endcsname{ E{^_}{{}{}} }{%
    #3{#4^{#3{##1}}_{#3{##2}}}%
    }%
}
\NewDocumentCommand\sgMeta{mG{#1}}{
    \defineMeta{g}{#1}{\gradualstyle}{#2}
    \defineMeta{s}{#1}{\staticstyle}{#2}
    \defineMeta{r}{#1}{\surfstyle}{#2}
}

\NewDocumentCommand\defineStaticFun{mG{#1}}{%
    \expandafter\NewDocumentCommand\csname#1\endcsname{od()}{%
    \staticstyle{\mathtt{#2}\ {\IfNoValueF{##1}{##1}}{\IfNoValueF{##2}{(##2)}} }
    }%
}

\sgMeta{x}
\sgMeta{y}
\sgMeta{z}
\sgMeta{C}
\sgMeta{D}
\sgMeta{N}
\sgMeta{M}
\sgMeta{ff}{f}
\sgMeta{pf}{p}
\sgMeta{P}
\sgMeta{el}{\overline{e}}
\sgMeta{X}
\sgMeta{Y}
\sgMeta{Z}
\sgMeta{t}
\sgMeta{T}
\sgMeta{s}
\sgMeta{S}
\sgMeta{m}
\sgMeta{n}
\sgMeta{u}
\sgMeta{U}
\sgMeta{v}
\sgMeta{V}
\sgMeta{A}
\sgMeta{B}
\sgMeta{J}{\text{J}}
\sgMeta{Nat}{\text{Nat}}
\sgMeta{Int}{\text{Int}}
\sgMeta{Bool}{\text{Bool}}
\sgMeta{Zero}{0}
\sgMeta{NatElim}{\text{NatElim}}
\sgMeta{VecElim}{\text{EqElim}}
\sgMeta{EqElim}{\text{VecElim}}
\sgMeta{Refl}{\text{Refl}}
\sgMeta{subst}{\text{subst}}

\sgMeta{G}{\Gamma}
\sgMeta{ep}{\varepsilon}
\sgMeta{times}{\times}
\sgMeta{lambda}{\lambda}
\sgMeta{Unit}{\mathbf{1}}
\sgMeta{unit}{()}

\sgMeta{spine}{\seq{e}}

\mathlig{||}{\bnfalt}
\mathlig{::=}{\bnfdef}
\mathlig{+::=}{\bnfadd}
\mathlig{-::=}{\bnfsub}
\mathlig{|>}{\vartriangleright}
\mathlig{<|}{\vartriangleleft}
\mathlig{<=}{\Leftarrow}
\mathlig{=>}{\Rightarrow}
\mathlig{>->}{\rightarrowtail}
\mathlig{|->}{\mapsto}
\mathlig{|=>}{\Mapsto}
\mathlig{==}{=}
\mathlig{**}{\times}
\mathlig{<<}{\langle}
\mathlig{>>}{\rangle}

\mathlig{|=}{\vDash}
\newcommand{\set}[1]{{\{ #1 \}}}

\usepackage{mathtools}

\usepackage{pifont}

\newcommand{\gradualcolor}{\color{RoyalBlue}}
\newcommand{\surfcolor}{\color{Emerald}}
\newcommand{\staticcolor}{\color{RedOrange}}

\newcommand{\gradualstyle}[1]{{\ensuremath{\gradualcolor\mathbf{{#1}}}}}
\newcommand{\surfstyle}[1]{{\ensuremath{\surfcolor\mathit{{#1}}}}}
\newcommand{\staticstyle}[1]{{\ensuremath{\staticcolor\mathsf{{#1}}}}}

\newcommand{\staticdesc}{\staticstyle{\textsf{red sans-serif font}}\xspace}
\newcommand{\gradualdesc}{\gradualstyle{\textbf{blue, bold serif font}}\xspace}
\newcommand{\surfdesc}{\surfstyle{\textit{green, italic serif font}}\xspace}

\newcommand{\g}[1]{\gradualstyle{#1}}
\newcommand{\s}[1]{\staticstyle{#1}}
\renewcommand{\r}[1]{\surfstyle{#1}}

\NewDocumentCommand{\sType}{m}{{\staticstyle{\mathbf{Type}_{#1}}}}
\NewDocumentCommand{\gType}{m}{{\gradualstyle{\mathbf{Type}_{#1}}}}
\NewDocumentCommand{\rType}{m}{{\surfstyle{\mathbf{Type}_{#1}}}}

\newcommand{\stepstostar}{\longrightarrow^{*}}

\newcommand{\stepsto}{\longrightarrow}

\newcommand{\bnfalt}{\mathbf{\,\,\mid\,\,}}
\newcommand{\bnfdef}{\mathbf{\ \Coloneqq\ }}
\newcommand{\defbnf}{\bnfdef} %
\newcommand{\bnfadd}{{\mathbf{\ +\!\!\Coloneqq\ }}}
\newcommand{\bnfsub}{{\bf - : : =}}

\makeatletter
\newcommand{\mathboxed}[1]{\text{\fboxsep=.2em\fbox{\m@th$\displaystyle#1$}}}
\makeatother

\newcommand{\ie}{i.e.\ }

\newcommand{\errsym}{\mho}
\newcommand\gqm{\gradualstyle{\normalfont{\textbf{?}}}}
\newcommand\rqm{\surfstyle{\normalfont{\textbf{?}}}}
\newcommand\rqmat[1]{\surfstyle{\normalfont{\textit{?}}_{@#1}}}
\newcommand\gqmat[1]{\gradualstyle{\normalfont{\textbf{?}}_{#1}}}
\newcommand\err{\gradualstyle{\errsym}}
\newcommand\errat[1]{\gradualstyle{\errsym}_{\g{#1}}}

\newcommand{\rrule}{\textsc}

\newcommand\blackoline[1]{\colorlet{temp}{.}\color{black}\overline{\color{temp}#1}\color{temp}}
\newcommand{\seq}[1]{\blackoline{#1}}

\newcommand{\bN}{\mymathbb{N}}

\newcommand{\bB}{\mymathbb{B}}

\definecolor{lightgray}{gray}{0.90}
\newcommand{\Gbox}[1]{\colorbox{lightgray}{$#1$}}

\newmdenv[
usetwoside=false,
topline=false,
bottomline=false,
rightline=false,
leftmargin=0.2in,
linewidth=0.75pt,
skipabove=\topsep,
skipbelow=\topsep,
nobreak=false
]{leftrule}

\newcommand{\case}[2]{
  \noindent $\blacktriangleright$ \textbf{Case} \text{#1} \textbf{:}
  {
    \begin{leftrule}
      #2
    \end{leftrule}
  }
  \noindent \ignorespaces
}

\newcommand{\rcase}[2]{\case{\rrule{#1}}{#2}}

\newsavebox{\saveboxedarray}
\newenvironment{inferbox}[0]
{\begin{minipage}{\textwidth}\mprset{center}\begin{mathpar}}
    {\end{mathpar}\end{minipage}}

 \newenvironment{boxedarray}[1]
 {\begin{lrbox}{\saveboxedarray}\begin{math}\begin{array}{#1}}
                                              {\end{array}\end{math}\end{lrbox}\fbox{\usebox{\saveboxedarray}}}
                                        \reservestyle{\command}{\textsf}

\usepackage{wrapfig}

\newcommand{\ifapx}[1]{}
\newcommand{\ifnotapx}[1]{#1}
\newcommand{\mname}[1]{\ifapx{#1}}

\usepackage{tikz-cd}
\usepackage{letltxmacro}

\LetLtxMacro{\oldfigure}{\figure}
\LetLtxMacro{\oldendfigure}{\endfigure}

\LetLtxMacro{\oldcaption}{\caption}

\renewenvironment{figure}
{\oldfigure}
{\vspace{-2ex}\oldendfigure}

\renewcommand{\caption}[1]{\vspace{-0.25ex}\vspace{-\baselineskip}\oldcaption{#1}}

\newcommand{\ifapxCaption}[1]{\caption{#1}}

\usepackage[createShortEnv]{proof-at-the-end}
\pgfkeys{/prAtEnd/apxproof/.style={
restate,no link to proof,proof at the end,no link to theorem
  }
}

\pgfkeys{/prAtEnd/apxlem/.style={
all end
  }
}

\pgfkeys{/prAtEnd/global custom defaults/.style={
}
}

\usepackage{iftex}
\ifPDFTeX
\usepackage[utf8]{inputenc}
\usepackage[inline]{enumitem}
\DeclareUnicodeCharacter{2029}{}
\fi

\usepackage{underscore}

\usepackage[nomain,
            order=word,
            hyperfirst=false,
            acronym,
            shortcuts,
            nonumberlist]{glossaries}

\NewDocumentCommand{\gSucc}{g}{\gradualstyle{\gtt{S}\IfValueT{#1}{\ #1}}}

\NewDocumentCommand{\gNil}{g}{\gradualstyle{{Nil}\IfValueT{#1}{\ #1}}}
\NewDocumentCommand{\sNil}{g}{\staticstyle{{Nil}\IfValueT{#1}{\ #1}}}

\NewDocumentCommand{\gCons}{gggg}{\gradualstyle{\gtt{Cons}\IfValueT{#1}{\ #1}\ \IfValueT{#2}{\ #2}\ \IfValueT{#3}{\ #3}\ \IfValueT{#4}{\ #4}}}

\NewDocumentCommand{\gVec}{gg}{\gradualstyle{{Vec}\IfValueT{#1}{\ #1}\ \IfValueT{#2}{\ #2}}}
\NewDocumentCommand{\sVec}{gg}{\staticstyle{{Vec}\IfValueT{#1}{\ #1}\ \IfValueT{#2}{\ #2}}}

\NewDocumentCommand{\conv}{gg}{\staticstyle{\stt{conv}\IfValueT{#1}{\ #1}\IfValueT{#2}{\ #2}}}

\usepackage{lstautogobble}
\usepackage{listings}

\lstdefinelanguage{Agda}%
  {morekeywords={let,in,as,data,record,import,infix,infixl,infixr,module,open,renaming,using,where,\_},%
   morekeywords=[2]{Set,Set1,Set2,Type},%
  literate=*%
     {?}{$\mathrm{\gqm}$}1
     {->}{$\mathrm{\to}$}2,
   otherkeywords={=,:,(,),\{,\},:=,;},
   sensitive=true,%
   morecomment=[n]{\{-}{-\}},%
   morecomment=[l]{--},%
   morestring=[b]{"}%
  }[keywords,comments,strings]%

\lstnewenvironment{gradualCode}{\lstset{language=Agda,
    basicstyle=\rmfamily\color{RoyalBlue},
    columns=flexible, mathescape=true}}{}

\lstnewenvironment{agdaCode}{\lstset{language=Agda,
    basicstyle=\rmfamily\color{Black},
    columns=flexible, mathescape=true}}{}

\lstnewenvironment{staticCode}{\lstset{language=Agda,
    basicstyle=\sffamily\color{BrickRed},
        literate=*%
     {?}{$\mathrm{\gqm}$}1
     {Type}{$\mathcal{U}$ }1
     {->}{$\mathrm{\to}$}2,
     columns=flexible, mathescape=true}}{}

\usepackage{sansmath}

\newcommand{\lang}{\ensuremath{\sf{GEq}}\xspace}
\newcommand{\clang}{\ensuremath{\sf{CastEq}}\xspace}

\newcommand{\cast}[2]{\g{\langle #2 <= #1 \rangle}}
\newcommand{\castnog}[2]{\g{\langle} #2 \g{<=} #1 \g{\rangle}}

\newcommand{\grefl}[3]{\g{refl_{#1 |- #2 \cong  #3}}}

\newcommand{\sind}[2]{\s{\mathsf{ind}_{#1}(}#2\s{)}}

\newcommand{\elabsto}{\rightarrowtriangle}
\newcommand{\echeck}[3]{#1 \elabsto #3 <= #2}
\newcommand{\esynth}[3]{#1 \elabsto #3 => #2}
\newcommand{\epsynth}[4]{#2 \elabsto #4 =>_{#1} #3}

\newcommand{\redsto}{\leadsto}

\newcommand{\sqube}{\sqsubseteq}
\newcommand{\squbr}{\sqsubseteq_{\mathsf{Surf}}}
\newcommand{\squbo}{\sqsubseteq^{\Vdash}}
\newcommand{\squbB}{\sqsubseteq_{\bB}}

\newcommand{\squbs}{\sqsubseteq_{\alpha}}

\newcommand{\qmat}[1]{\g{\gqm_{#1}}}

\crefformat{section}{\S#2#1#3}
\crefformat{subsection}{\S#2#1#3}
\crefformat{subsubsection}{\S#2#1#3}
\crefrangeformat{section}{\S\S#3#1#4 to~#5#2#6}
\crefmultiformat{section}{\S\S#2#1#3}{ and~#2#1#3}{, #2#1#3}{ and~#2#1#3}
\Crefformat{section}{Section #2#1#3}
\Crefformat{subsection}{Section #2#1#3}
\Crefformat{subsubsection}{Section #2#1#3}
\Crefrangeformat{section}{Sections #3#1#4 to~#5#2#6}
\Crefmultiformat{section}{Sections #2#1#3}{ and~#2#1#3}{, #2#1#3}{ and~#2#1#3}

\renewcommand{\slambda}[3]{\s{\lambda(#1 : #2)\ldotp #3}}

\renewcommand{\glambda}[3]{\g{\lambda(#1 : #2)\ldotp #3}}

\newcommand{\jform}[1]{\fbox{#1}\hspace{\fill}\\}

\newcommand{\psynth}[1]{{\,\mathbin{\Rightarrow_{#1}}}\,}

\newcommand{\psynthstar}[1]{{\,\mathbin{\Rightarrow^{*}_{#1}}}\,}
\newcommand{\ssorts}{\sType{ }}
\newcommand{\ind}{\mathtt{ind}}
\newcommand{\smatch}[4]{\s{\operatorname{\ind}_{#1}(#2,#3,#4)}}
\newcommand{\smatchnoarg}[1]{\s{\operatorname{\ind}_{#1}}}
\newcommand{\rmatch}[4]{\r{\operatorname{\ind}_{#1}(#2,#3,#4)}}
\newcommand{\gmatch}[4]{\g{\operatorname{\ind}_{#1}(#2,#3,#4)}}

\newcommand{\pars}{\operatorname{\mathbf{\color{black} Params}}}

\newcommand{\args}{\operatorname{\mathbf{\color{black} Args}}}
\newcommand{\parsub}[1]{[#1]}

\newcommand{\ulev}[1]{\scalebox{0.7}{@\{#1\}}}

\newcommand{\J}{\mathbf{J}}

\newcommand{\defprec}{{\sqube_{\stepsto}}}
\newcommand{\defsuprec}{{\sqube^{\longleftarrow}_{\stepsto}}}
\newcommand{\defcst}{{{\cong}_{\stepsto}}}
\newcommand{\acst}{{{\cong}_{\alpha}}}

\newcommand{\genprec}{\Gbox{\defprec}}
\newcommand{\gensuprec}{\Gbox{\defsuprec}}
\newcommand{\gencst}{\Gbox{\defcst}}

\usepackage{scalerel}
\DeclareMathOperator*{\bigamp}{\mathlarger{\&}}
\newcommand{\gcomp}[1]{\mathbin{\g{\&_{#1}}}}
\newcommand{\gcompop}{\g{\&}}
\usepackage{ifsym}

\newcommand{\proofappendix}{the appendix of the anonymized supplementary material}
\newcommand{\ruleappendix}{the appendix of the anonymized supplementary material}

\newcommand{\itercomp}{\seq{\bigamp}}

\newcommand{\germ}{\mathsf{\color{black} germ}}
\newcommand{\head}{\mathsf{\color{black} head}}

\acmYear{2022}

\ifdef{\reviewmode}{
  \settopmatter{printfolios=true,printccs=false,printacmref=false}
  \setcopyright{none}
  \renewcommand\footnotetextcopyrightpermission[1]{}
  \raggedbottom
}{
  \settopmatter{printfolios=true,printccs=false,printacmref=false}
  \setcopyright{none}
  \renewcommand\footnotetextcopyrightpermission[1]{}
  \pagestyle{plain}
}

\AtBeginDocument{%
  \providecommand\BibTeX{{%
    \normalfont B\kern-0.5em{\scshape i\kern-0.25em b}\kern-0.8em\TeX}}}

\begin{document}

\title{Propositional Equality for Gradual Dependently-Typed Programming}

  \author{Joseph Eremondi}

  \affiliation{
    \department{Department of Computer Science}              %
    \institution{University of British Columbia}            %
    \country{Canada}                    %
  }
  \email{{jeremond@cs.ubc.ca}}          %

  \author{Ronald Garcia}
  \affiliation{
    \institution{University of British Columbia}            %
    \country{Canada}                    %
  }
  \email{rxg@cs.ubc.ca}          %

    \author{\'{E}ric Tanter}
    \affiliation{
      \department{Computer Science Department (DCC)}              %
      \institution{University of Chile}            %
      \country{Chile}                    %
    }
    \email{etanter@dcc.uchile.cl}          %

\renewcommand{\shortauthors}{Eremondi, Garcia, and Tanter}

\begin{abstract}
  \textbf{Under Submission to ICFP 2022}

Gradual dependent types can help with the incremental adoption of dependently typed code by providing a principled semantics for \textit{imprecise} types and proofs, where some parts have been omitted.
Current theories of gradual dependent types, though,  lack a central feature of type theory: propositional equality.
Lennon-Bertrand et al. show that, when the reflexive proof $\mathit{refl}$ is the only closed
value of an equality type, a gradual extension of CIC with propositional equality violates static observational equivalences.
Extensionally-equal functions should be indistinguishable at run time, but
 the combination of equality and type imprecision
allows for contexts that distinguish extensionally-equal but syntactically-different functions.

This work presents a gradually typed language that supports propositional equality.
We avoid the above issues by devising an equality type where $\mathit{refl}$
is not the only closed inhabitant.
Instead, each equality proof is accompanied by a term that is at least as precise
as the equated terms, acting as a witness of their plausible equality.
These witnesses track partial type information as a program runs, raising errors
when that information shows that two equated terms are undeniably inconsistent.
Composition of type information is internalized as a construct of the language,
and is deferred for function bodies whose evaluation is blocked by variables.
By deferring, we ensure that extensionally equal functions compose without error,
thereby preventing contexts from distinguishing them.
We describe the challenges of designing consistency and precision relations
for this system, along with solutions to these challenges. Finally, we prove important metatheory: type-safety,
conservative embedding of CIC,
canonicity up to termination, and the gradual guarantees of Siek et al., which ensure that
reducing a program's precision introduces no new static or dynamic errors.

\end{abstract}

\ifdef{\reviewmode}{
}{

}

\def\acmBooktitle#1{\gdef\@acmBooktitle{#1}}
\acmBooktitle{Proceedings of \acmConference@name
       \ifx\acmConference@name\acmConference@shortname\else
         \ (\acmConference@shortname)\fi}

\maketitle

\section{Introduction}
\label{sec:intro}

\textit{Gradual dependent types} relax the discipline
of dependent types, so that programmers can write, type check and run programs with partial type information
and omit yet-to-be-devised terms or proofs.
These capabilities have the potential to help migrate code from non-dependently typed languages,
and reduce the learning curve for newcomers to this rich but complex type discipline.
Gradual languages~\citep{gradualTypeInitial} check programs against the type information statically available,
comparing types via \textit{consistency} $\cong$, i.e., equality up to missing type information.
Static checks skipped due to partial type information
are instead performed at run time when the actual
values are known.
Programs missing type information might not fail, nor are errors deferred until an unsafe operation is attempted.
Rather, partial type information is exploited at run time, so an error occurs when computation's result has a type incompatible with the context in which it arises.
Gradual dependent types let programmers use type-driven programming with holes~\citep{tdd-book} while still
running code and executing tests, even when missing parts of types, terms or proofs.

However, existing gradual dependent languages do not support
\textit{propositional equality}~\citep{MARTINLOF1982153}.
The propositional equality type $\s{t_{1} ==_{T} t_{2}}$ expresses that $\s{t_{1} }$ and $\s{t_{2}}$
are equal inhabitants of type $\sT$. Its only constructor is $\s{refl_{t}: t ==_{T} t}$, the proof that every term is equal to itself.
Equality is useful for practical dependently typed programming,
since it lets a function express pre- and post-conditions by taking
or returning equality proofs.
Likewise, a programmer can use an equality proof to rewrite the type
of the expression they are trying to produce.
Propositional equality even lays a path to support GADT-style inductive families,
since constructors with different return types
can be encoded with non-indexed inductive types and propositional equality~\citep{dtfpp}.

Limited means of representing and reasoning about equality have been used in existing gradual languages.
GCIC~\citep{bertrand:gcic} supports \textit{decidable equality}, where a type is
computed by pattern-matching on the equated terms. The programmer
must manually write a function and elimination principle for each type used,
and cannot equate terms of most function types.
Gradual Refinement Types~\citep{lehmannTanter:popl2017} support first-order constraints in linear integer arithmetic.
By contrast, propositional equality is general and lightweight: it works for every type, provides its own construction and elimination principles,
and can be used with
quantifiers.

Until now, the challenge with gradual equality has been propagating and enforcing equality constraints at run time. The problem is that the terms we equate may contain functions or dependent function types,
both of which bind variables. For example, $\s{(\lambda x \ldotp x + 0) ==_{\bN} (\lambda x \ldotp x)}$
and $\s{((x : \bN) -> Vec\ \bN\ (x+0)) ==_{\sType{}} ((x : \bN) -> Vec\ \bN\ x)}$ are both well-formed types.
Determining whether functions are extensionally equal (up to partial information) is undecidable.
Comparing functions syntactically, by directly comparing bound variables, is decidable.
Such a notion works for compile-time consistency checks,
but is problematic during run-time checks, since it destroys static reasoning principles.
Observationally equivalent terms in the static language may be distinguishable in
the gradual language. %
\Citet{bertrand:gcic} show that when $\r{refl}$ is the only constructor for propositional equality,
it cannot be included in a gradual language without violating static equivalences.
Moreover, code that compares bound variables cannot be easily compiled, since every function now needs a syntactic representation.

This paper presents the language \lang (pronounced ``geek''), which adds propositional equality to GCIC, allowing $\r{==}$,
$\r{refl}$ and elimination to be used like in the Calculus of Inductive Constructions~(CIC),
but with a dynamic semantics that is meaningful for gradual types.
Our \textbf{key insight} is to \textbf{represent an equality proof using a \textit{witness}
that captures equality constraints discovered at run time}.
 Taking inspiration from
 evidence in Abstracting Gradual Typing~\citep{agt} and middle-types
 in threesomes~\citep{Siek:2009:TWB:1570506.1570511},
 we represent witnesses with a term that is as precise as both equated terms.
 As a consequence, $\r{refl}$ and $\rqm$ are \textit{not} the only inhabitants of the equality type,
 avoiding the above impossibility result.
Our \textbf{contributions} are as follows:

\begin{itemize}[leftmargin=*]
  \item We {demonstrate how equality proofs between imprecise terms} are {useful} for {discovering bugs} in programs and for {guiding the development}  of static proofs (\cref{sec:example});

  \item We extend GCIC with {propositional equality} (\cref{sec:equality}) by
          typing equality using {consistency witnesses} between terms (\cref{subsec:cast-eq}).
          We give {operational semantics} via a cast calculus, where the eliminator for equality
                {uses casts going through the result type given by the witness} (\cref{subsec:cast-semantics}).
          To combine witnesses when casting between equality types, we add {witness composition} directly as a construct in \lang (\cref{subsec:meet-operator}).
        This operator delays the comparison of neutral terms until
        their variables are bound to values, so composing statically-equivalent functions
        does not raise an error;
  \item We prove {type safety}, {conservative extension of CIC}, {(weak) canonicity,} and the {gradual guarantees} for \lang (\cref{sec:generic}),
        so adding imprecision
        never introduces stuck states or
        (static or dynamic) errors, and \lang rejects ill-typed CIC programs.
        The proofs are
        parameterized over definitions of consistency and precision, revealing a sufficient set of  properties to prove the theorems;

  \item We define {precision and consistency} for the cast calculus (\cref{sec:prec}), showing that they
        {fulfill the previously-identified properties.} We separate \textit{static consistency},
        whether a term of some type can be used in a given context,
        from \textit{dynamic consistency}, whether two terms compose without error.
        These coincide for non-dependent gradual languages, but in \lang they must be separated
        to respect static equivalences while still rejecting ill-typed static programs.

\end{itemize}

\Cref{sec:static} reviews GCIC and its cast calculus CastCIC, upon which the other sections build. \Cref{sec:discussion} discusses
extensions enabled by \lang's features, along with related and future work.

\section{Setting The Stage}
\label{sec:example}

\newcommand{\boldpar}[1]{\textbf{#1:}}

\subsection{Programming vs. Proving and the Gradual Guarantees}

Though programming and proving are connected by the Curry-Howard
correspondence, the language features best supporting each task differ.
Our focus is dependently typed programming: we consider \lang
as a model of a programming language rather than as a type theory for mechanizing mathematics.
Nevertheless, we prove important metatheory about \lang that may aid in the
development of future gradual type theories.

One goal with \lang is proving
the \textit{gradual guarantees} of \citet{refinedCriteria}, which
state that a reduction in precision introduces no new static or dynamic errors.
These guarantees are useful for programming because of the contrapositive: if a program has a type error,
adding more type information does not remove the error. The types
are fundamentally inconsistent and must be changed.
By contrast, in current dependently typed languages holes block reduction, causing errors.

\subsection{Relationship to Existing Languages}

We wish to establish
how \lang, our gradual language with propositional equality, relates to the state-of-the-art for gradual dependent types.
\lang builds primarily off the work of two existing gradual dependent languages,
GDTL~\citep{Eremondi:2019:ANG:3352468.3341692} and GCIC~\citep{bertrand:gcic}.
\Cref{subsec:related} gives a broader discussion of related work.

\textbf{GDTL} is a \underline{G}radual \underline{D}ependently \underline{T}yped \underline{L}anguage
with dependent functions, a universe hierarchy,
and decidable type checking.
GDTL introduced the \textit{imprecise term} $\gqm$, which extended gradual typing
to allow imprecision not only in types, but type indices and proof terms.
\lang inherits $\gqm$ from GDTL.
Since it is based on AGT~\citep{agt,10.1145/3434342}, GDTL features some ideas similar to the witnesses
we use.
However, the authors only discuss equality and inductive types as an extension,
and do not include it in their metatheory.
Also, GDTL uses the naive syntactic notion of composition, and suffers from
the extensionality issues we discuss in the introduction.%

\textbf{GCIC} is a \underline{G}radual version of the \underline{C}alculus of \underline{I}nductive \underline{C}onstructions (CIC). It uses a cast calculus approach, extending a restricted version of
CIC with inductive types but no indexed inductive families or propositional equality.
\lang is a direct extension of GCIC.
The GCIC authors prove that no gradual language can simultaneously conservatively extend CIC,
have strong normalization, and have \textit{graduality}, a strengthening of the gradual guarantees
where decreasing then increasing precision
produces an equivalent term.
The authors give three variants of GCIC: $\mathsf{GCIC}^{\mathcal{G}}$, $\mathsf{GCIC}^{\mathcal{N}}$ and $\mathsf{GCIC}^{\uparrow}$
which respectively sacrifice one of strong normalization, graduality, and conservative extension of CIC,
while keeping the other two properties.
We build off $\mathsf{GCIC}^{\mathcal{G}}$.
$\mathsf{GCIC}^{\mathcal{N}}$ violates the gradual guarantees, and
$\mathsf{GCIC}^{\uparrow}$ is too restrictive for practical programming,
so we avoid them both.
Logical inconsistency
and non-terminating proofs are not as detrimental in programming as in mechanized mathematics,
so $\mathsf{GCIC^{\mathcal{G}}}$ suits our purposes best.

GCIC has no dedicated equality type, but decidable equality types are supported, as described in \cref{sec:intro}.
Though useful in a type theory context,
the programmer must manually construct an equality type for each type
about which they wish to write constraints,
along with the corresponding elimination principle.
We avoid placing this burden on the programmer by allowing
full propositional equality.

\subsection{A Motivating Example: Eagerly Enforcing Specifications}
\label{subsec:quicksort}

In this section, we motivate our development with examples of how gradual dependent types can catch errors
related to the lengths of lists.
A guiding principle of our work is that
{the types the programmer writes should, as much as possible, be treated as specifications to be checked, either statically or dynamically}, regardless of whether their enforcement is required for safety.

Throughout the paper, we write static terms using \staticdesc.
Terms from the gradual surface language use \surfdesc.
The theory is developed using a gradual cast calculus, which we write using \gradualdesc.

\boldpar{A Buggy Quicksort}
We begin by showing how gradual types help the migration of a sorting function from a non-dependently typed language to one with dependent types, and how this migration can help identify bugs.
Consider a flawed quicksort implementation:
\begin{flalign*}%
  &\s{sort : List\ Float -> List\ Float} &\mid \quad&\s{List\ A = Nil \mid Cons\ A\ (List\ A)}   \\
  &\s{sort\ Nil = Nil} &\mid\quad& \s{(++) : List\ A -> List\ A -> List\ A}   \textit{ (concatenation)}\\
  &\mathrlap{\s{sort\ (Cons\ h\ t) =}
    \s{(sort\ (filter\ (<\ h)\ t)) }
   \s{\ \doubleplus [h] \doubleplus (sort\ (filter\ (>\ h)\ t)) }}
 \end{flalign*}

\begin{wrapfigure}{r}{0.5\textwidth}
  \fbox{
\begin{minipage}{0.45\textwidth}
\begin{flalign*}%
  &\s{FList : \sType{} -> \bN -> \sType{}}\\
  &\underline{\s{FList\ A\ n = ((x : List\ A) \times length\ x ==_{\bN} n)}\qquad}\\
  &\s{sort : {(n : \bN)} -> FList\ Float\ n -> FList\ Float\ n} \\
  &\s{sort\ 0\ (Nil, p) = (Nil, p)} \\
  &\s{sort\ (1+n)\ (Cons\ h\ t, p) =} \\
  & \qquad \s{let\ lt = (filter\ (< h)\ t)}\\
  & \qquad \s{\quad (sortLt, plt) = sort\ (length\ lt)\ (lt, refl)}\\
  & \qquad \s{\quad gt = (filter\ (> h)\ t)}\\
  & \qquad \s{\quad (sortGt, pgt) = sort\ (length\ gt)\ (gt, refl)}\\
  & \qquad \s{in\ (sortLt \doubleplus [h] \doubleplus sortGt, {\color{black}\fbox{???}})}
\end{flalign*}
\end{minipage}
}
\caption{Sorting Fixed-Length Lists}
  \label{fig:sort}
  \end{wrapfigure}

  \noindent
 Since $\s{<}$ is used instead of $\s{\leq}$, duplicates are erroneously removed
 from the list.
The programmer may have a suspicion that they have made a mistake in their code,
or may have observed incorrect behavior while testing.
Their dependent type enthusiast friends have repeatedly assured them that dependently typed languages
can help eliminate bugs, so they try migrating their code to a dependently typed language
with propositional equality.

An approach to reasoning about the correctness of $\s{sort}$
is to use \textit{\underline{F}ixed-length lists}, called $\s{FList}$s. Dependent pairs and propositional equality
allow for a type
of lists indexed by their length.%
\footnote{A more conventional approach would be to use an indexed type family,
  which we discuss in \cref{sec:inductives}.
}
\Cref{fig:sort} shows this type, and how it can be used to express that $\s{sort}$
should preserve the length of the produced list.
Here, $\s{refl}$ is the \underline{refl}exive proof that $\st$ is equal to itself.
That $\s{sort}$ function behaves like the non-dependent version, except it must extract the $\s{List}$s from
the $\s{FLists}$ produced by the recursive calls, and produce a $\s{FList}$ with the appropriate proof that the length
is the same as the input.

At this stage, the programmer must fill the \textit{hole} $\fbox{???}$
by constructing a proof of type \\$\s{length\ (Cons\ h\ t) ==_{\bN} length\ (sortLt \doubleplus [h] \doubleplus sortGt)}$.
This task is difficult
for a newcomer, since they must use associativity of addition and
$\s{length\ (list_{1} \!\doubleplus\! list_{2}) ==_{\bN} length\ list_{1} \!+\! length\ list_{2}}$.
Moreover, they must prove that
$\s{length\ t ==_{\bN} (length\ lt) + (length\ gt)}$,
but such a proof is impossible, due to the bug. Even if they had such a proof, they would need to then use
the proofs from the recursive calls, $\s{plt}$ and $\s{pgt}$, to relate the lengths of $\s{lt}$ and $\s{gt}$
to the lengths of $\s{sortLt}$ and $\s{sortGt}$.

The programmer is now cursing their type-theorist friend.
For a non-buggy quicksort, one could construct the necessary proof,
but doing so is difficult, particularly for a newcomer.
The type checker does not detect the bug, so it does not inform the programmer that hole cannot be filled,
and it cannot say {which} aspect of the proof is impossible.
Also, development has now stopped: the programmer cannot run or test their code
without the missing proof.

\boldpar{Gradual Types to the Rescue}
\lang lets the programmer run and test $\r{sort}$ before writing
the missing proof, checking (within the limits of decidability) whether any static values could possibly replace
imprecise types and proofs.
In \cref{fig:sort}, replacing the hole with $\rqm$,
 the \textit{imprecise term,}%
 \footnote{Each  $\rqm$ is actually $\rqmat{0}$, i.e.,  annotated with its type's universe level. Our exposition omits levels; we explain them in \cref{subsec:gcic-surface-cast}.}
 yields a complete, well-typed \lang function
that can be called, tested, or used in other modules.%

The utility of gradual typing is shown in the run-time checks that let us identify bugs in code.
The run-time semantics of \lang are defined via type-directed elaboration to a
$\g{\text{cast calculus}}$  \clang, in which all implicit conversions
are replaced by explicit casts.
During type checking, $\rqm$ is elaborated into the \clang's
$\gqmat{ n ==_{\bN} length\ (sortLt \doubleplus [h] \doubleplus sortGt) }$
(the least precise term of type $\g{{n ==_{\bN} length\ (sortLt \doubleplus [h] \doubleplus sortGt)}}$),
which is not a value in \clang. Instead, it reduces to
the consistency witness
$\g{n \gcomp{\bN} length\ (sortLt \doubleplus [h] \doubleplus sortGt)}$.
The operator $\gcompop$ is the gradual composition operator, which combines
information statically known about its operands.
Because types depend on terms, composition is not limited to types, but can combine
terms of any type.
The $\gcompop$ operator is a syntactic construct of \clang,
not a meta-operation like it is in existing literature~\citep{Siek:2009:TWB:1570506.1570511,10.1145/2737924.2737968}.
Reifying $\gcompop$ into the object language is critical for composing functions (\cref{subsec:funcomp}).
Since $\g{n}$ is a variable, this composition expression does not
reduce further.

We can identify the bug once $\r{sort}$ is applied to a concrete list.
Consider the input $\r{[2.2, 1.1, 3.3, 2.2]}$, which elaborates to
 $\g{[2.2, 1.1, 3.3, 2.2]}$ in \clang. Applying $\g{sort}$ binds
$\g{lt := [1.1]}$ and $\g{gt := [3.3]}$, giving a result list of $\g{[2.2,1.1,3.3]}$.
Then $\g{n}$ is $\g{4}$ and $\g{length\ (sortLt \doubleplus [h] \doubleplus sortGt)}$ is $\g{3}$,
so the witness for the result is the composition
 $\g{4 \gcomp{\bN} 3}$, which reduces to a run-time error.

 \boldpar{Witness Composition}
 The key to finding the error above was tracking information with witnesses,
 and combining those witnesses using the composition operator.
While that composition was a simple equality check, in general
the composed values may be imprecise, and the result some value that is as precise as both inputs.
The information from the witness is used when eliminating an equality proof:
when using a witness $\g{t_{w}}$ of $\g{t_{1} ==_{T} t_{2}}$ to rewrite a term of type $\g{P(t_{1})}$ into $\g{P(t_{2})}$,
we first cast to $\g{P(t_{w})}$, then to $\g{P(t_{2})}$.
For a program with imprecise types or values,
the witness retains the information gained by running the program, preventing unsafe execution,
and informing the programmer when a counter-example to an imprecise equality is found.

Here we present an example of a bug that is found, not because of safety, but because
a remembered constraint was violated.
Consider the following functions:
\begin{flalign*}
  & \r{zip : (n : \bN) -> FList\ A\ n -> FList\ A\ n -> FList\ (A \times A)\ n}\\
  & \r{take : (n : \bN) -> (m : \bN) -> FList\ B\ (n + m) -> FList\ B\ n}
\end{flalign*}
Here, $\r{zip}$ takes two lists of exactly the same length, and produces a list of pairs
of their elements, while
$\r{take}$ takes a list with at least $\r{n}$ elements, and returns the first
$\r{n}$ elements of that list.
Each function constrains the size of its input, so by tracking equality witnesses, we can also track these constraints and detect where they are incompatible.
Now consider lists with imprecise types:
\begin{flalign*}
  & \r{list_{1} := { ([1.1, 2.2], refl_{\rqm})} : FList\ Float\ \rqm} & \qquad
  & \r{list_{2} := (Cons\ 1.1\ \rqm, refl_{\rqm})  : FList\ Float\ \rqm}
\end{flalign*}
For $\r{list_{1}}$, we are converting a list of length 2 to a fixed-length
list of unknown length, since $\r{2}$ is consistent with $\rqm$.
For $\r{list_{2}}$, however, the length is truly imprecise, since its tail is the unknown term.
We can zip these lists together as
$
  \r{zip\ \rqm\ list_{1}\ list_{2} : FList\ Float\ \rqm}$,
producing another list of unknown length, since recursively applying $\r{zip}$ to the unknown tail $\rqm$
produces an unknown result.
Applying $\r{take\ 3}$ to the result of $\r{zip}$ is well typed, since the length $\rqm$
is consistent with $\r{3 + \rqm}$,
i.e.,
 $ \r{take\ 3\ \rqm\ (zip\ \rqm\ list_{1}\ list_{2}) : FList\ Float\ (3 + \rqm)} $.
 However, computing the witness flags an error.

 This error represents something deeper than a simple safety check:
 it detects fundamental inconsistencies in statically-determined propositional equalities.
In the absence of equality proofs, the call could run safely:
$\r{Cons\ (1.1, 3.3)\ \rqm}$ would be a sensible result, having length consistent with $\r{3}$.
The witness composition is not
just checking if a list is empty before taking the head, or counting
the elements in the list before running $\r{take}$.
Rather, the information added by $\r{zip}$, that the list should have length $\r{2}$,
has been propagated using the $\r{list_{1}}$ witness
and composed with the conflicting information.
 \lang uses witnesses to enforce imprecise equality constraints at run time.

To understand how \lang detects this mismatch, we look the result of elaborating to \clang.
Initially, $\g{list_{2}}$ has $\g{1+\gqmat{\bN}}$ as the witness that $\gqmat{\bN}$ is equal to $\g{1+\gqmat{\bN}}$.
The result of $\g{zip}$ has $\g{2}$ as the witness of equality between
$\g{\gqmat{\bN}}$ and $\g{\gqmat{\bN}}$,
since that is the length of $\g{list_{1}}$.
This new witness was determined by composition:
since $\g{1+\gqmat{\bN}}$ is consistent with $\g{2}$,
this composition succeeds.
(Using a Peano representation of naturals, $\g{S(\gqm)}$ is consistent with $\g{S(S(0))}$)
Then, even though $\g{zip}$'s result has a type that is consistent with
what $\g{take}$ expects, the run-time type information remembers that $\g{zip}$
constrained the list to have length $\g{2}$.
The result of $\g{zip}$ is cast to $\g{3 + \gqmat{\bN}}$,
i.e., the type expected by $\g{take\ 3}$.
The $\g{zip}$ result has an equality proof
of type $\g{1+\gqmat{\bN} ==_{\bN} \gqmat{\bN} }$,
which is cast to type $\g{1+\gqmat{\bN} ==_{\bN} 3+\gqmat{\bN} }$.
During this cast,
the target value $\g{3+ \gqmat{\bN}}$ is composed with the witness $\g{2}$.
Despite the imprecision, these values are not consistent, and
composition produces an error: no value can replace $\gqm$ to make $\g{S(S(S(\gqm)))}$ equal to $\g{S(S(0))}$.
We detail the semantics enabling this in \cref{sec:equality}.

With equality witnesses, we achieve more than type safety.
From the gradual guarantees, we know the above code cannot possibly
be made static by replacing the $\rqm$ uses with static terms.
When a witness reduces to an error, the program is equating two terms that are
fundamentally not-equal. So the gradual guarantees now inform about
equality constraints, in addition to type constraints.
These constraints are expressed through types,
rather than an external language of assertions.

\subsection{Lazily Enforcing Specifications: Function Equalities and Extensionality}
\label{subsec:funcomp}

Propositional equality is neither restricted to first-order values
like numbers nor to types with decidable equality.
In particular, we can form equalities between functions,
for which equality is not in general decidable.
The following summarizes how \lang handles propositional equality for functions
without encountering the impossibility result of \citet{bertrand:gcic}.
Consider the example they use to
show the incompatibility between gradual typing and $\g{refl}$-based equality:
\begin{flalign*}
  & \r{id_{\bN} := (\lambda x \ldotp x) : \bN -> \bN}    \qquad\qquad \r{add0 := (\lambda x \ldotp x + 0) : \bN -> \bN}  \\
  & \mathrlap{\r{test := \lambda f \ldotp \J\ (\_.\bB)\ id_{\bN}\ true\ f\ (refl_{id_{\bN}}\ :: \rqm :: (id_{\bN} ==_{\bN -> \bN} f))  : (\bN -> \bN) -> \bB} }
\end{flalign*}
Here $\r{\J}$ is the eliminator for equality: we explain it fully in \cref{subsec:cast-eq},
but it suffices to know that in this case, it uses a proof of type $\r{id_{\bN} ==_{\bN -> \bN} f}$
to rewrite $\r{(\lambda \_ \ldotp \bB )\ id_{\bN}}$ to $\r{(\lambda \_ \ldotp \bB )\ f}$.
Both types reduce to $\r{\bB}$, but $\r{\J}$ only reduces if the equality
proof reduces without error.

Since $\r{id_{\bN}}$ and $\r{add0}$ agree on all inputs, they should be observationally equivalent,
producing the same result in any context in which we use them.
Violating this would mean that the embedding of CIC into GCIC or \lang
does not respect function extensionality, i.e., some statically-equivalent terms are distinguishable in the gradual language.
\citet{bertrand:gcic} offer $\r{test}$ as a context that distinguishes $\r{id_{\bN}}$
and $\r{add0}$. When $\r{id_{\bN}}$ is given as an argument, casting $\r{refl_{id_{\bN}}}$
to $\rqm$ then back to $\r{id_{\bN} ==_{\bN -> \bN} id_{\bN}}$ should produce $\r{refl_{id_{\bN}}}$.
However, when $\r{add0}$ is given for $\r{f}$, the cast must fail, since $\r{refl}$
cannot have type $\r{id_{\bN} ==_{\bN -> \bN} add0}$.

The first key piece that lets \lang avoid this inequivalence is the witness-based representation
of equality.
In \clang, $\g{refl}$ is the only constructor for equality, but it takes an argument:
the consistency witness for the equated terms. Moreover, it does not require the equated
terms to be syntactically identical, only that the witness be at least as precise as both of them.
So  $\g{refl}$ can have type $\g{id_{\bN} ==_{\bN -> \bN} add0}$.

The second key is the composition operator of \clang,
which builds the equality witness for $\g{id_{\bN}}$ and $\g{add_{0}}$.
Elaborating $\r{refl_{id_{\bN}}}$ creates witness $\g{id_{\bN}}$.
 The cast to $\g{id_{\bN} ==_{\bN -> \bN} f}$ composes that witness
with the destination endpoints, $\g{id_{\bN}}$ and $\g{f}$, yielding
$\g{id_{\bN} \gcomp{\bN -> \bN} id_{\bN} \gcomp{\bN -> \bN} f}$.
The semantics of $\gcompop$
reduce this composition to
$\g{\lambda x \ldotp x \gcomp{\bN} x \gcomp{\bN} (f\ x) }$,
similar to how a higher-order contract applied to a function produces a new function
that checks the input and result~\citep{Findler:2002:CHF:581478.581484}.
Since $\gcompop$ is an operator in the language, the composition does not need to reduce further,
but when the function is applied, it continues to reduce.
The same holds when we replace $\g{f}$ with $\g{id_{\bN}}$ or $\g{add0}$.
\Cref{sec:prec} defines precision such that $\g{x \gcomp{\bN} x \gcomp{\bN} f\ x}$
is more precise than both $\g{x}$ and $\g{f\ x}$,
so the above composition is a valid witness.
We define semantics for $\g{J}$ so that when it is given the equality proof with the above witness, it reduces, so $\g{test}$ reduces to $\g{true}$
for both $\g{id_{\bN}}$ and $\g{add0}$.

How can equating these functions be safe, since deferring composition means that we can prove an equality
between unequal functions?
As we saw with $\g{sort}$ above, $\g{\J}$ casts through
the witness, so when functions are extensionally non-equal,
trying to prove equality between their results dynamically fails.
Consider instead $\g{subadd1 := (\lambda (x : \bN) \ldotp x - 1 + 1)}$.
Since $\g{0 - t} = \g{0}$ in $\g{\bN}$,
 $\g{subadd1\ 0} = \g{1}$.
We can use the witness  $\g{\lambda x \ldotp (x \gcomp{\bN} x \gcomp{\bN} (subadd1\ x)) }$
to inhabit $\g{id_{\bN} ==_{\bN -> \bN} subadd1}$. However, if we try to use $\g{\J}$
to prove that $\g{id_{\bN}\ 0 ==_{\bN} subadd1\ 0}$, the result substitutes $\g{0}$ for $\g{x}$
in the witness, giving $\g{0 \gcomp{\bN} 0 \gcomp{\bN} 1}$, which reduces to an error.

The consequence of our approach is that \lang supports a limited form of extensionality.
\textit{Neutral terms}, i.e., variables or terms for which reduction is blocked by one or more variables,
always compose to a non-error,
so we can build a witness capturing the plausibility of equality between them,
given partial information. That witness makes an equality proof constructible.
Furthermore, any two functions with neutral bodies compose to a non-error.
If the functions agree on all inputs, eliminating their equality never fails
and the proof of equality can be freely used.
If the functions disagree on some input, an error is raised
when building a term that relies on the functions producing the same value for said input.
Since it is undecidable whether two functions agree on every input, this approach
finds a balance between decidability and flexibility.

\boldpar{Static vs. Dynamic Consistency}
For non-dependent gradual types, the successful composition of two types
usually implies that they are consistent.
However, for \lang, two neutrals always compose to a more-precise term.
To conservatively extend CIC, all ill-typed CIC programs must be ill-typed in \lang.
So \lang cannot have all
neutrals consistent, since this would yield a fully-static proof of $\g{(\lambda x \lambda y \ldotp x) ==_{T -> T -> T} (\lambda x \lambda y \ldotp y)}$.

We resolve this tension with separate static and dynamic notions of consistency (\cref{sec:prec}).
Terms are statically consistent if they are syntactically equal up to $\alpha$-equivalence,
reduction, and occurrences of $\gqmat{T}$.
Terms are dynamically consistent if they compose without error, or equivalently,
if there exists a non-error term as precise as both terms.
Essentially, terms are dynamically-consistent if they are statically consistent
in the non-neutral parts.
The type rules for \lang use static consistency.
Some pairs of terms are not statically consistent,
yet still compose to a non-error term.

To compare static and dynamic consistency,
consider the ill-typed CIC term $\s{refl : x ==_{\bN} y}$.
When embedded into \lang, $\r{refl : x ==_{\bN} y}$ is still ill-typed:
the expected type of $\r{x ==_{\bN} y}$ and the actual type of
$\r{x ==_{\bN} x}$ are not statically consistent, because the variables $\rx$ and $\ry$
are not identical. In \clang, $\gx$ and $\gy$ are neutral, and hence dynamically consistent,
meaning $\g{\gx \gcomp{\bN} \gy}$ witnesses $\g{x ==_{\bN} y}$.

Allowing neutrals to be dynamically consistent does not interfere with conservatively extending CIC.
For conservative extension, every ill-typed CIC program should be ill-typed in \lang.
In the absence of $\rqm$,
pairs of definitionally-unequal CIC terms are statically inconsistent.
While \lang gives $\r{refl}$ the same type as CIC, \clang lets
$\g{refl}$ prove equality for dynamically consistent terms.
However, dynamic consistency does not allow \clang to type ill-typed CIC terms,
because CIC programs are elaborated into a subset of \clang
where $\gt$ only witnesses $\g{t ==_T \gt}$ and all casts have the form $\cast{\gT}{\gT}$.
The type $\s{(x : \bN) -> (y : \bN) -> (x ==_{\bN} y)}$ is uninhabited in CIC,
and while it is inhabited in \clang using witness $\g{\gx \gcomp{\bN} \gy}$,
 the use of $\gcomp{\bN}$ puts the witness outside the static fragment of \clang.
It is not the elaboration of any \clang program, ill-typed or otherwise.

Static and dynamic consistency let us balance conflicting goals.
If all statically-inconsistent functions composed to an error,
then statically-equivalent terms would not be gradually equivalent,
making it harder to reason about program equivalence.
Using dynamic consistency during type checking would not conservatively extend CIC.
By separating these, we obtain conservative extension, and we conjecture
that all observational equivalences are preserved.

\section{The Static Language and GCIC}
\label{sec:static}

To begin, we review the state-of-the-art for handling
inductive types in a gradual language.
We describe the Bidirectional CIC (BCIC),
a modification of CIC whose bidirectional types are convenient
for gradual typing~\citep{LennonBertrand2021}.
We then describe the gradual surface language GCIC, along with the cast calculus, CastCIC,
and a translation from GCIC to CastCIC~\citep{bertrand:gcic}.
Specifically, we use $\mathsf{GCIC}^{\mathcal{G}}$, the variant of GCIC which satisfies the gradual
guarantees and embeds CIC, but sacrifices strong normalization.
We discuss options for decidable type checking in \cref{subsec:termination}.
Though GCIC is not a contribution of this paper, we use it as the starting point for our development,
making additions
to the surface language and cast calculus.

\subsection{Bidirectional CIC}

\subsubsection{Syntax}

\Cref{fig:bcic} gives
the \textit{bidirectional calculus of constructions} (BCIC) as
 originally presented by \citet{LennonBertrand2021},
though we modify their notation to maximize clarity for \lang's additions.
BCIC terms are denoted by metavariables $\st$ and $\sT$, loosely following the convention that
$\sT$ be reserved for types. Variables are denoted by $\sx, \sy, \sz$.
Like CIC, BCIC has variables, a predicative hierarchy of universes,
function types, functions, and applications.
Technically, BCIC extends the predicative, non-cumulative fragment of CIC: each function type is in a higher universe than its domain and codomain, and there is no subtyping between universe levels.
We assume a pre-existing set of inductive type constructors, denoted by the metavariable $\s{C}$, each of which has a fixed set of
data constructors $\s{D^{C}}$.
Type and data constructors are annotated with the level
of their type, though we omit these annotations when they are not relevant.

A combined form $\smatch{C}{\s{t_{1}}}{\sz.\s{T}}{x.\seq{y.t_{2}}}$
replaces CIC's $\s{fix}$ and $\s{match}$
This form branches on the scrutinee $\s{t_{1}}$ and has a parameterized  result type $\s{T}$, called the \textit{motive}~\citep{McBride2002}, that binds a variable of the scrutinee's type. The branches $\seq{\s{t_{2}}}$ correspond to the constructors $\s{\seq{D^{C}}}$ of $\s{C}$.
In each branch, the variables $\seq{\s{y}}$ are bound to the arguments to $\s{D^{C}}$,
and $\s{x}$ is replaced by the whole $\smatchnoarg{C}$ expression, to facilitate recursion.
The $\smatchnoarg{C}$ form expresses an induction principle: if each branch produces a result of type $\s{T}$ where $\s{z}$ is bound to $\s{D^{C}}$ applied to
$\seq{\sy}$,
 the  elimination has type $\s{T}$ where $\s{z}$ is bound to the scrutinee $\s{t_{1}}$. In essence, $\smatchnoarg{C}$
says that if we can build a $\s{T}$ for each constructor $\s{D^{C}}$ of $\s{C(\seq{t_{param}})}$, then we can build one for any value of $\s{C(\seq{t_{param}})}$.
Normally, a separate check ensures that recursive calls are only made on
structurally smaller arguments, but we omit this check, since it is orthogonal to gradual typing and \lang would not be strongly normalizing even with it.

BCIC also uses \textit{head tags}, denoted by $\s{h}$, which act as symbols to specify
a type constructor without specifying its arguments. We use these in typing, e.g.
for expressing that an applied function must synthesize a function type, even though we do not know
what the domain and codomain should be. Tags are also useful in GCIC for defining
the least precise type with a given head.

\subsubsection{Typing and Semantics}

\newcommand{\drawFigBcic}
{
  \ifapx{\begin{figure}[H]}
  \ifnotapx{\begin{figure}}
    \begin{boxedarray}{@{}l@{}}
      \boxed{
  \s{t}, \s{T}\defbnf \sx  \bnfalt{\sType{i}}
  \bnfalt \s{(x : T_{1}) -> T_{2}} \bnfalt \slambda{x}{T}{t} \bnfalt \s{t_{1}\ t_{2}}
  \bnfalt \s{C\ulev{i}}(\seq{\s{t}}) \bnfalt \s{D^C\ulev{i}(\seq{\st},\seq{\s{t'}})}
  \bnfalt \smatch{C}{\s{t_{1}}}{\sz.\s{T}}{x.\seq{y.t_{2}}}
}\\
\boxed{
  \s{h} \defbnf {\s{C}} \bnfalt \sType{} \bnfalt {\staticcolor{\Pi}}
  }
  \\\\
    \fbox{Synthesis: $\Gamma \vdash \st => \sT$\qquad
    $\Gamma \vdash \st <= \sT$ (Checking)\qquad
    $\Gamma \vdash \st \psynth{\s{h}} \sT$ (Constrained Synthesis)}\\
    \begin{inferbox}
    \inferrule[\mname{SCheck}]{\Gamma \vdash \st => \s{T'} \\
      \sT \stepstostar \s{T''} \\
    \s{T'} \stepstostar \s{T''}}
  {\Gamma \vdash \st <= \sT} \label{infrule:pts-check}

    \inferrule[\mname{SVar}]{(\sx : \sT) \in \Gamma}
    {\Gamma \vdash \sx => \sT} \label{infrule:pts-var}

    \inferrule[\mname{SType}]{ }
    {\Gamma \vdash \sType{i} => \sType{i+1}} \label{infrule:pts-sort}

    \inferrule[\mname{SFun}]{\Gamma \vdash \s{T_1} \psynth{\ssorts} \sType{i} \\\\
      \Gamma, \sx : \s{T_1} \vdash \s{T_2}  \psynth{\ssorts} \sType{j}}
      {\Gamma \vdash  \s{(x : T_1) -> T_2} => \sType{\max(i,j)}} \label{infrule:pts-prod}

      \inferrule[\mname{SApp}]{\Gamma \vdash \st \psynth{\Pi}  \s{(x : T_1) -> T_2} \\\\ \Gamma \vdash \st <= \sT}
      {\Gamma \vdash \s{t_0}\ \s{t_1} => [\s{t_1} / \sx]\s{T_2}} \label{infrule:pts-app}

      \inferrule[\mname{SAbs}]{\Gamma \vdash \s{T_1} \psynth{\ssorts} \sType{i} \\\\ \Gamma, \sx : \s{T_1} \vdash \st => \s{T_2}}
      {\Gamma \vdash \s{\slambda{x}{T_1}{t}} =>   \s{(x : \s{T_1}) -> \s{T_2}} } \label{infrule:pts-abs}

		\inferrule[\mname{SInd}]
    {\Gamma \vdash \s{t}_k <= \pars_k(\s{C},\s{i})\parsub{\seq{\st}}}
    {\Gamma \vdash \s{C\ulev{i}(\seq{\st})} => \sType{i}} \label{infrule:cic-ind}

      \inferrule[\mname{SCtor}]
    {\Gamma \vdash \st_k <= \pars_k(\s{C},\s{i})\parsub{\seq{\st}} \\
			\Gamma \vdash \s{t'}_m <= \args_m(\s{C},\s{i},\s{D})\parsub{\seq{\st}, \seq{\s{t'}}}}
    {\Gamma \vdash \s{D^C\ulev{i}(\seq{\st},\seq{\s{t'}})}
      => \s{C\ulev{i}(\seq{t})}}

		\inferrule[\mname{SMatch}]{
			\Gamma \vdash \s{t_{scrut}} \psynth{\s{C}}  \s{C\ulev{i}(\seq{t_{param}})} \\
			\Gamma, \sz : \s{C(\seq{t_{param}})} \vdash \s{T_P} \psynth{\ssorts} \sType{j} \\\\
			\seq{
        \Gamma, \s{x_{rec}} : (\s{ (z : C\ulev{i}(\seq{\s{t_{param}}}) ) -> T_P}), \seq{\sy} : \args(\s{C},\s{i},\s{D_k})\parsub{\seq{\s{t_{param}}},\seq{\sy}} \vdash \s{t_{rhs_k}} <= [\s{D^C_k\ulev{i} (\seq{t_{param}},\seq{\sy})} / \sz]\s{T_P}
        }
		}{\Gamma \vdash \smatch{C}{\s{t_{scrut}}}{\sz.\s{T_P}}{x_{rec}.\seq{\seq{y}.t_{rhs}}} => [\s{t_{scrut}}/\sz]\s{T_P}}

    \inferrule[\mname{SConstrSynthType}]{\Gamma \vdash \st => \sT \\\\ \sT \stepstostar \sType{i}}
    {\Gamma \vdash t \psynth{\ssorts} \sType{i}} \label{infrule:pts-sort-inf}

      \inferrule[\mname{SConstrSynthFun}]{\Gamma \vdash \st => \sT \\\\
        \sT \stepstostar \s{(x : T_1) -> T_2}}
      {\Gamma \vdash \st \psynth{{\staticcolor{\Pi}}}  \s{ (x : T_1) -> T_2}} \label{infrule:pts-prod-inf}

    \inferrule[\mname{SConstrSynthInd}]{\Gamma \vdash \st => \sT \\\\
      \sT \stepstostar \s {C\ulev{i}(\seq{\s{t_{p}}})}}
    {\Gamma \vdash \st \psynth{\s{C}} \s{ C\ulev{i}(\seq{\s{t_{p}}})}} \label{infrule:cic-ind-inf}
  \end{inferbox}
   \end{boxedarray}
  \ifapxCaption{Bidirectional CIC: Syntax and Typing}
  \ifnotapx{\label{fig:bcic}}
\end{figure}
}
\drawFigBcic

\newcommand{\drawFigBcicSem}
{
  \ifapx{\begin{figure}[H]}
  \ifnotapx{\begin{figure}}
    \begin{boxedarray}{@{}l@{}}
  \begin{inferbox}
    \jform{$\st \redsto \s{t'}$ (Notions of Reduction)  \qquad $\st \stepsto \s{t'}$ (Contextual reduction)}

    \inferrule[\mname{Red$\beta$}]{}{
      (\slambda{x}{T}{t})\ \s{t'} \redsto [\s{t'} / \sx] \s{t}
      }

    \inferrule{\st \redsto \s{t'}\\
      C \text{ an arbitrary context}
    }{C[\st] \stepsto C[\s{t'}]}

    \inferrule[\mname{RedMatch}]{}{
    \smatch{C}{D_k(\seq{t_{param}}, \seq{t_{arg}})}{z.T_P}{f.\seq{y.t_{rhs}}}
		\redsto [\slambda{x}{C(\seq{t_{param}})}{
      \smatch{C}{x}{z.T_{}}{x_{rec}.\seq{y.t_{rhs}}}
      / {f} }]     [\s{\seq{t_{arg}}} / {\seq{\sy}}] \s{t_{rhs_k}}
    }
  \end{inferbox}
\end{boxedarray}
\caption{BCIC: Dynamic Semantics}
\ifnotapx{\label{fig:bcic-sem}}
\end{figure}
}
\drawFigBcicSem

The typing (\cref{fig:bcic}) and semantics (\cref{fig:bcic-sem}) for BCIC
resemble the typical presentation of CIC, but typing
is divided into synthesis, which produces a type,
and checking, which consumes a type.
Semantics are given with primitive notions of reduction $\redsto$, contextual stepping $\stepsto$
where any sub-term reduces (even under binders), and multi-step contextual reduction
 $\stepstostar$, which allows zero or more steps with $\stepsto$.
Because function types bind their parameter in the codomain,
applications synthesize a type depending on the value of the argument,
        since it is substituted for $\sx$ in the codomain type.
A term checks against any type that reduces
to the same type as its synthesized type,
since an application may have produced a type
that must be reduced before comparing.
Constrained-synthesis, $\Gamma |- \st \psynth{\s{h}} \sT$,
        generalizes the pattern of synthesizing a type for a term after reducing it
        to a point that it has the desired head $\s{h}$.

The inductive elimination $\sind{C}$
        is essentially the dependent version of a case-expression:
         if $\s{T_{P}}$ is a type parameterized over a value $\sx$
        from the inductive type $\s{C}$, and we can (recursively) build a $\s{T_{P}}$
        for each constructor of $\s{C}$, then we can build a $\s{T_{P}}$ for any member or $\s{C}$. The $\sind{C}$ form gives an induction principle
        for $\s{C}$, hence the notation $\sind{C}{\ldots}$. The reduction rule essentially says
        that an $\sind{C}{\ldots}$ form given a value $\s{D^{C}(\ldots)}$ reduces to the branch corresponding to
        $\s{D^{C}}$.
Inductive types may be parameterized, but each constructor
has the same return type.

\subsection{GCIC: The Surface Language}
\label{subsec:gcic-surface-cast}

\Cref{fig:gcic-lemmas} extends BCIC into GCIC, the \underline{G}radual
CIC,
by adding the \textit{imprecise term} $\rqmat{i}$, which can be used at
any type in universe level $\r{i}$,
along with type ascriptions, which were not in BCIC because all forms
synthesized types.
We use $\rqm_{\rT}$ as sugar for $\rqmat{i}$ when $\rT : \rType{i}$.

\begin{figure}
  \centering
  \begin{boxedarray}{@{}l@{}}
    \begin{inferbox}
      \boxed{\r{t}, \r{T} \bnfadd \rqmat{i} \bnfalt \r{t :: T}}

    \inferrule
    { }
    {\Gamma |- \rqmat{i} => \rqmat{i+1} }

    \inferrule
    {\Gamma |- \r{t} <= \r{T} }
    {\Gamma |- \r{t :: T} => \rT }

      \inferrule
      {\Gamma |- \r{t} => \r{T'} \\
      \r{T'} \defcst \r{T}}
    {\Gamma |- \r{t} <= \r{T} }

    \inferrule
    {\Gamma |- \r{t} => \r{T}\\
      \r{T} \stepstostar \r{\rqmat{i}}}
      {\Gamma |- \r{t} \psynth{\surfcolor\Pi} \r{\rqmat{i} \!->\! \rqmat{i}} }

      \inferrule
      {\Gamma |- \r{t} => \r{T}\\
      \r{T} \stepstostar \r{\rqmat{i}}}
      {\Gamma |- \r{t} \psynth{\r{C}} \r{C}(\seq{\rqm_{\pars(\r{C},\r{i})}}) }

      \inferrule
      {\Gamma |- \r{t} => \r{T}\\
      \r{T} \stepstostar \r{\rqmat{i+1}}}
      {\Gamma |- \r{t} \psynth{\rType{  }} \rType{i} }
    \end{inferbox}
  \end{boxedarray}
  \ifapxCaption{GCIC: Syntax and Typing Lemmas}
    \ifnotapx{\label{fig:gcic-lemmas}}
  \end{figure}

Dependent types complicate the typing of GCIC. Because the dynamic semantics of GCIC
are defined using a cast calculus, and typing refers to reduction of terms,
\citet{bertrand:gcic} define GCIC typing with cast calculus types.
Nevertheless, we can establish lemmas (\cref{fig:gcic-lemmas}), phrased like rules, which provide intuition for how GCIC terms are typed against
GCIC types, helping GCIC be understood without diving into the details of the cast calculus.

The unknown term $\rqmat{i}$ synthesizes $\rqmat{i+1}$ i.e. the unknown term has the unknown type,
one level higher in the universe hierarchy.
A term checks against any type that is definitionally consistent with its synthesized type,
where the consistency relation $\defcst$ is understood to mean convertibility up to well-typed occurrences of $\rqm$
(explained fully in \cref{subsec:gcic-cst}).
An ascribed term synthesizes the given type if it checks against it,
relaxing or tightening the types
of gradual terms.
Finally, constrained synthesis accounts for terms synthesizing
$\rqm_{\rType{i}}$
by producing the \textit{germ} (called the \textit{ground type} in non-dependently typed literature).
The type $\germ_{i}(\g{h})$ is the least precise type with a given head in universe $i$. For function types,
the germ is $\r{\rqmat{i} -> \rqmat{i}}$, and for an inductive type, the germ
has  $\rqmat{i}$ with the appropriate $\r{i}$ for each parameter.
$\rType{i}$ is its own germ.

  \subsection{CastCIC: The Cast Calculus}
\label{subsec:gcic-surface-cast}

\subsubsection{Syntax, Typing and Reductions}
\Cref{fig:castcic-typing} presents CastCIC, the cast calculus for GCIC.
 CastCIC extends BCIC with
the unknown term $\gqm$, an error $\err$, and a cast $\cast{T_{1}}{T_{2}}\gt$
from type $\g{T_{1}}$ to $\g{T_{2}}$.
Forms $\gqmat{T}$
and $\errat{T}$ are ascribed with their type $\gT$, which affect the dynamic semantics of CastCIC.
The CastCIC type system contains all the rules from \cref{fig:bcic}
plus the rules of \cref{fig:castcic-typing}.
Terms $\gqm_{T}$ or $\err_{T}$ synthesize their ascribed type $\gT$, while
casts synthesize the destination type, given that the term being
cast checks against the source type, and that both types are well-formed.
Because casts are explicit, the checking rule uses definitional equality,
rather than consistency.

\newcommand{\drawCastCIC}{
\ifapx{\begin{figure}[H]}
\ifnotapx{\begin{figure}}
  \centering
  \begin{boxedarray}{@{}l@{}}
    \boxed{\g{t},\gT \bnfadd \qmat{\gT} \bnfalt \errat{T} \bnfalt \cast{T_{1}}{T_{2}}\gt}
      \\
      \begin{inferbox}
        \ifapx{
    \inferrule[\mname{CastCheck}]{\Gamma \vdash \gt => \g{T'} \\
      \gT \stepstostar \g{T''} \\
    \g{T'} \stepstostar \g{T''}}
  {\Gamma \vdash \gt <= \gT} \label{infrule:pts-check}

    \inferrule[\mname{CastVar}]{(\gx : \gT) \in \Gamma}
    {\Gamma \vdash \gx => \gT} \label{infrule:pts-var}

    \inferrule[\mname{CastType}]{ }
    {\Gamma \vdash \gType{i} => \gType{i+1}} \label{infrule:pts-sort}

    \inferrule[\mname{CastFun}]{\Gamma \vdash \g{T_1} \psynth{\gType{}} \gType{i} \\\\
      \Gamma, \gx : \g{T_1} \vdash \g{T_2}  \psynth{\gType{}} \gType{j}}
      {\Gamma \vdash  \g{(x : T_1) -> T_2} => \gType{\max(i,j)}} \label{infrule:pts-prod}

      \inferrule[\mname{CastApp}]{\Gamma \vdash \gt \psynth{\Pi}  \g{(x : T_1) -> T_2} \\\\ \Gamma \vdash \gt <= \gT}
      {\Gamma \vdash \g{t_0}\ \g{t_1} => [\g{t_1} / \gx]\g{T_2}} \label{infrule:pts-app}

      \inferrule[\mname{CastAbs}]{\Gamma \vdash \g{T_1} \psynth{\gType{}} \gType{i} \\\\ \Gamma, \gx : \g{T_1} \vdash \gt => \g{T_2}}
      {\Gamma \vdash \g{\glambda{x}{T_1}{t}} =>   \g{(x : \g{T_1}) -> \g{T_2}} } \label{infrule:pts-abs}

		\inferrule[\mname{CastInd}]
    {\Gamma \vdash \g{t}_k <= \pars_k(\g{C},\g{i})\parsub{\seq{\gt}}}
    {\Gamma \vdash \g{C\ulev{i}(\seq{\gt})} => \gType{i}} \label{infrule:cic-ind}

      \inferrule[\mname{CastCtor}]
    {\Gamma \vdash \gt_k <= \pars_k(\g{C},\g{i})\parsub{\seq{\gt}} \\
			\Gamma \vdash \g{t'}_m <= \args_m(\g{C},\g{i},\g{D})\parsub{\seq{\gt}, \seq{\g{t'}}}}
    {\Gamma \vdash \g{D^C\ulev{i}(\seq{\gt},\seq{\g{t'}})}
      => \g{C\ulev{i}(\seq{t})}}

		\inferrule[\mname{CastMatch}]{
			\Gamma \vdash \g{t_{scrut}} \psynth{\g{C}}  \g{C\ulev{i}(\seq{t_{param}})} \\
			\Gamma, \gz : \g{C(\seq{t_{param}})} \vdash \g{T_P} \psynth{\gType{}} \gType{j} \\\\
			\seq{
        \Gamma, \g{x_{rec}} : (\g{ (z : C\ulev{i}(\seq{\g{t_{param}}}) ) -> T_P}), \seq{\gy} : \args(\g{C},\g{i},\g{D_k})\parsub{\seq{\g{t_{param}}},\seq{\gy}} \vdash \g{t_{rhs_k}} <= [\g{D^C_k\ulev{i} (\seq{t_{param}},\seq{\gy})} / \gz]\g{T_P}
        }
		}{\Gamma \vdash \gmatch{C}{\g{t_{scrut}}}{\gz.\g{T_P}}{x_{rec}.\seq{\seq{y}.t_{rhs}}} => [\g{t_{scrut}}/\gz]\g{T_P}}

    }
      \inferrule[\mname{CastUnk}]
      {\Gamma |- \gT \psynth{\gType{ }} \gType{i} }
      {\Gamma |- \gqmat{T} => \gT}

      \inferrule[\mname{CastErr}]
      {\Gamma |- \gT \psynth{\gType{ }} \gType{i} }
      {\Gamma |- \errat{T} => \gT}

      \inferrule[\mname{CastCast}]
      {\Gamma |- \g{T}_j \psynth{\gType{  }} \gType{i} \text{ for } j \in \set{1,2} \\
      \Gamma |- \g{t} <= \g{T_1}
      }
      {\Gamma |- \cast{T_1}{T_2}\gt => \g{T_2} }
      \ifapx{

    \inferrule[\mname{CastConstrSynthType}]{\Gamma \vdash \gt => \gT \\\\ \gT \stepstostar \gType{i}}
    {\Gamma \vdash t \psynth{\gType{}} \gType{i}} \label{infrule:pts-sort-inf}

      \inferrule[\mname{CastConstrSynthFun}]{\Gamma \vdash \gt => \gT \\\\
        \gT \stepstostar \g{(x : T_1) -> T_2}}
      {\Gamma \vdash \gt \psynth{{\gradualcolor{\Pi}}}  \g{ (x : T_1) -> T_2}} \label{infrule:pts-prod-inf}

    \inferrule[\mname{CastConstrSynthInd}]{\Gamma \vdash \gt => \gT \\\\
      \gT \stepstostar \g {C\ulev{i}(\seq{\g{t_{p}}})}}
    {\Gamma \vdash \gt \psynth{\g{C}} \g{ C\ulev{i}(\seq{\g{t_{p}}})}} \label{infrule:cic-ind-inf}
      }
    \end{inferbox}
    \\
    \ifapx{
    \end{boxedarray}
    \ifapxCaption{CastCIC: Typing}
  \end{figure}
  \begin{figure}[H]
  \begin{boxedarray}{@{}l@{}}
    }
    \begin{inferbox}
      \textbf{Propogation Reductions:}\\
      \inferrule[\mname{RedPropFun(Unk,Err)}]
      {}
      {\gqmat{(x : T_1)->T_2} \redsto \glambda{x}{T_1}{\gqmat{T_2}}\\\\
\errat{(x : T_1)->T_2} \redsto \glambda{x}{T_1}{\errat{T_2}}
      }

      \inferrule[\mname{RedPropMatch(Unk,Err)}]
      {}
      {\gmatch{C}{\gqmat{C(\seq{\g{t_{param}}})}}{\g{z.T_P}}{\g{x_f.\seq{\gy}.\seq{\gt}}}
        \redsto
        \gqmat{[\gqmat{C(\seq{\g{t_{param}}})} / \gz]\g{T_P}}
        \\\\
        \gmatch{C}{\errat{C(\seq{\g{t_{param}}})}}{\g{z.T_P}}{\g{x_f.\seq{\gy}.\seq{\gt}}}
        \redsto
        \errat{[\errat{C(\seq{\g{t_{param}}})} / \gz]\g{T_P}}
      }

      \inferrule[\mname{RedPropCastInd(Unk,Err)}]
      {}
      {\cast{C(\seq{t_1})}{C(\seq{t_2})}\gqmat{\g{T_1}} \redsto \gqmat{T_2}
        \\\\
        \cast{C(\seq{t_1})}{C(\seq{t_2})}\errat{\g{T_1}} \redsto \errat{T_2}
      }

      \inferrule[\mname{RedPropCastDown(Unk,Err)}]{}{
        \cast{\gqmat{\gType{i}}}{\gT}\gqmat{\gqmat{\gType{i}}} \redsto \gqmat{\gT}\\\\
        \cast{\gqmat{\gType{i}}}{\gT}\errat{\gqmat{\gType{i}}} \redsto \errat{\gT}
      }
\\
\textbf{Cast Reductions:}\\

      \inferrule[\mname{RedCastFun}]
      {}
      {
        \cast{(x : T_1) -> T_2}{(x : T'_1) -> T'_2}\gt
        \redsto
        \g{\lambda y \ldotp \castnog{[\cast{T'_1}{T_1}\gy/\gx]T_1}{\g{T'_2}}(t\ \cast{T'_1}{T_1}y)}
      }

      \inferrule[\mname{RedCastInd}]
      {}
      {\cast{C(\seq{t_{param}})}{C(\seq{t'_{param}})}\g{D^C(\seq{t_{param}}, \seq{t_{arg}})} \redsto \g{D^C(\seq{t'_{param}}, \seq{t'_{arg}})}
\\\\ \textit{ where } \g{t'_{arg_{i}}} := \cast{\args_i(C,i,D)[\seq{t_{param}}, \seq{t_{args}}]}{\args_{i}(C,i,D)[\seq{t'_{param}}, \seq{t'_{args}}]}\g{t_{arg_{i}}}
}

      \inferrule[\mname{RedCastType}]
      {}
      {\cast{\gType{i}}{\gType{i}}\gT \redsto \gT}

      \inferrule[\mname{RedCastHeadErr}]
      {\head(\gT) \neq \head(\g{T'})}
      {\cast{T}{T'}\gt \redsto \errat{T'}}

      \inferrule[\mname{RedCastDomErr}]
      {}
      {\cast{\errat{\gType{i}}}{T}\gt \redsto \errat{\gT}}

      \inferrule[\mname{RedCastCodomErr}]
      {}
      {\cast{\gT}{\errat{\gType{i}}}\gt \redsto \errat{\errat{\gType{i}}}}

      \inferrule[\mname{RedCastUpDown}]
      {}
      {\cast{\gqmat{\gType{i}}}{\gT} \castnog{\germ_{i}(\g{h})}{\gqmat{\gType{i}}} \gt
        \redsto \castnog{\germ_{i}(\g{h})}{\gT}\gt
      }

      \inferrule[\mname{RedCastFunGerm}]
      {\g{(x : T_1)->T_2} \neq \germ_j({\gradualcolor{\Pi}}) \text{ for $j \geq i$ }}
      {\cast{(x : T_1)->T_2}{\gqmat{\gType{i}}} \gt
        \redsto
        \cast{\gqmat{\gType{i}} -> \gqmat{\gType{i}}}{\gqmat{\gType{i}}}
        \cast{(x : T_1)->T_2}{\gqmat{\gType{i}}->\gqmat{\gType{i}}}\gt
      }

      \inferrule[\mname{RedCastIndGerm}]
      {\g{C(\seq{t_{param}})} \neq \germ_j(\g{C}) \text{ for $j \geq i$ }}
      {\cast{C(\seq{t_{param}})}{\gqmat{\gType{i}}}\gt
        \redsto
        \cast{C(\seq{\gqmat{\pars(C)}})}{\gqmat{\gType{i}}}
        \cast{C(\seq{t_{param}})}{C(\seq{\gqmat{\pars(C)}})}\gt
      }
      \ifapx{

        \inferrule[RedCastSizeErr]{
          \min{\set{j \mid \exists \g{h} \ldotp \germ_j(\g{h}) = \g{T}}} > i
        }{
          \cast{\gT}{\gqmat{\gType{i}}}\gt \redsto \errat{\gType{i}}
        }
      }
    \end{inferbox}
  \end{boxedarray}
  \ifapxCaption{CastCIC: \ifnotapx{Typing and} Reduction rules \ifnotapx{that extend BCIC}}
  \ifnotapx{\label{fig:castcic-typing}}
\end{figure}
}
\drawCastCIC

The
CastCIC dynamic semantics includes all the reductions of BCIC, plus rules for casts, as well as what \citet{bertrand:gcic} call
``propagation rules'' that handle $\gqm$ and $\err$.
At type $\g{(x : T_{1})->T_{2}}$,
$\gqm$ and $\err$
expand to
$\glambda{x}{T_{1}}{\gqmat{T_{2}}}$
and $\glambda{x}{T_{1}}{\errat{T_{2}}}$.
Otherwise, eliminating
or casting $\gqm$ or $\err$
produces $\gqm$ or $\err$ at the target type.
The cast rules either convert between types with the same head,
cast to $\gqmat{\gType{\ell}}$, or produce an error.
A cast
 from $\gType{i}$ to itself reduces away.
For inductives, casts from $\g{C(\seq{t_{1}})}$ to $\g{C(\seq{t_{2}})}$
are reduced by casting the arguments to their new types.
Casting between types with mismatched heads produces an error,
as does casting to or from $\errat{\gType{i}}$.
Casting from the germ for a given head does \textit{not} reduce:
 $\cast{\germ(\gT)}{\gqmat{\gType{i}}}$
acts as a tag, injecting into $\gqmat{\gType{i}}$.
All casts from non-germ types to $\gqmat{\gType{i}}$ decompose into casts through the germ that
are then tagged with their injection into $\gqmat{\gType{i}}$.
Casting \textit{from} $\gqmat{\gType{i}}$ to $\gT$  reduces
when the value being cast originates from a type with a matching head,
and was accordingly tagged with a cast from $\head(\gT)$ to $\gqmat{\gType{i}}$.

\subsubsection{Elaboration}
Finally, elaboration (\cref{fig:gcic-elab}) defines the relationship between GCIC and CastCIC.
Like CastCIC, elaboration has synthesis, checking, and constrained synthesis, but each
produces the elaboration of the subject term as output.
\rrule{ElabUnk} synthesizes the unknown type for the unknown term at the given universe level.
\rrule{ElabApp} works like a normal dependent function application, but uses the elaboration
of the argument to replace the parameter in the return type.
\rrule{ElabCst}
checks a term against a type consistent with its synthesized type,
inserting the cast between these types into the elaboration.
\Cref{fig:gcic-elab} also defines new constrained synthesis rules.
Rule \rrule{ElabUnkFun} works like the corresponding lemma,
but adds the necessary cast to the elaborated term.
Rules \rrule{ElabUnkInd} and \rrule{ElabUnkUniv} work similarly.
We omit the elaboration rules corresponding to the remaining BCIC rules,
which homomorphically elaborate the sub-terms of a given term.

Elaboration defines GCIC typing:
 $\rt : \rT$ when $\cdot |- \epsynth{\gType{  }}{\rT}{\gType{i}}{\gT}$
and $\cdot |- \echeck{\rt}{\gT}{\gt}$.%

\newcommand{\drawFigElabGCIC}{
\ifapx{\begin{figure}[H]}
\ifnotapx{\begin{figure}}
  \centering
  \begin{boxedarray}{@{}l@{}}
    \jform{$\Gamma |-  \esynth{\r{t}}{\g{T}}{\gt}$\qquad $\Gamma |- \echeck{\rt}{\gT}{\gt} $ \qquad $\Gamma |- \epsynth{\g{h}}{\rt}{\gT}{\gt} $ \qquad (Elaboration)}
    \begin{inferbox}
 \inferrule[ElabUnk]
 { }
 {\Gamma |- \esynth{\rqmat{i}}{\gqmat{\gType{i}}}{\gqmat{\gqmat{\gType{i}}}}}

    \inferrule[ElabUnkFun]
      {\Gamma |- \esynth{\r{t}}{\g{\gqmat{\gType{i}}}}{\gt} }
      {\Gamma |- \epsynth{\gradualcolor{\Pi}}{\r{t}}{\g{\gqmat{\gType{i}} \!->\! \gqmat{\gType{i}}}}{\cast{\gqmat{\gType{i}}}{\gqmat{\gType{i}} \!->\! \gqmat{\gType{i}}}\g{t} } }

 \inferrule[ElabCst]
 {  \Gamma |- \esynth{\rt}{\g{T'}}{\gt}\\\\
   \gT' \defcst \g{V}\\
 }
 {\Gamma |- \echeck{\rt}{\gT}{\cast{\g{T'}}{\gT}\gt}}

 \inferrule[ElabApp]{
   \Gamma |- \epsynth{\surfcolor\Pi}{\r{t_0}}{\g{(x : T_1) -> T_2}}{\g{t_0}}\\\\
   \Gamma |- \echeck{\r{t_1}}{\g{T_1}}{\g{t_1}}
 }{
   \Gamma |- \esynth{\r{t_0\ t_1}}{[\g{t_1} /\gx]\g{T_2}}{\g{t_0\ t_1}}
 }

      \inferrule[ElabUnkInd]
      {\Gamma |- \esynth{\r{t}}{\g{\gqmat{i}}}{\gt} }
      {\Gamma |- \epsynth{\g{C}}{ \r{t}}{\g{C}(\seq{\gqm_{\pars(\g{C},\g{i})}})}{\cast{\gqmat{i}}{\g{C}(\seq{\gqm_{\pars(\g{C},\g{i})}})}\g{t}} }

      \inferrule[ElabUnkUniv]
      {\Gamma |- \esynth{\r{t}}{\g{\gqmat{i+1}}}{\gt} }
      {\Gamma |- \epsynth{\gType{  }}{\r{t}}{ \gType{i}}{\cast{\gqmat{i+1}}{\gType{i}}\g{t}} }
      \ifapx{

    \inferrule[\mname{ElabVar}]{(\gx : \gT) \in \Gamma}
    {\Gamma \vdash \esynth{\rx}{\gT}{\gx} }

    \inferrule[\mname{ElabType}]{ }
    {\Gamma \vdash \esynth{\rType{i}}{\gType{i+1}}{\gType{i}}
    } \label{infrule:pts-sort}

    \inferrule[\mname{ElabFun}]{\Gamma \vdash \epsynth{\gType{}}{\r{T_1}}{\gType{i}}{\g{T_1}}\\\\
      \Gamma, \gx : \g{T_1} \vdash \epsynth{\gType{  }}{\r{T_2}}{\gType{j}}{\g{T_2}}
    }
    {\Gamma \vdash  \esynth{\r{(x : T_1) -> T_2}}{\gType{\max(i,j)}}{\g{(x : T_1) -> T_2}}
      }

      \inferrule[\mname{ElabAbs}]{\Gamma \vdash \epsynth{\gType{}}{\r{T_1}}{ \gType{i} }{\g{T_1}} \\\\
        \Gamma, \gx : \g{T_1} \vdash \esynth{\rt}{\g{T_2}}{\gt}}
      {\Gamma \vdash \esynth{\r{\lambda(x : T_1)\ldotp t}}{\g{(x : \g{T_1}) -> \g{T_2}} }{\g{\glambda{x}{T_1}{t}}}   } \label{infrule:pts-abs}

		\inferrule[\mname{ElabInd}]
    {\Gamma \vdash \echeck{\r{t}_k}{\pars_k(\g{C},\g{i})\parsub{\seq{\gt}}}{\g{t}_k} }
    {\Gamma \vdash \esynth{\r{C\ulev{i}(\seq{\rt})}}{\gType{i}}{\g{C\ulev{i}(\seq{\gt})}}} \label{infrule:cic-ind}

      \inferrule[\mname{ElabCtor}]
    {\Gamma \vdash \echeck{\rt_k}{{\pars_k(\g{C},\g{i})\parsub{\seq{\gt}}}}{\gt_k} \\
			\Gamma \vdash \echeck{\r{t'}_m}{\args_m(\g{C},\g{i},\g{D})\parsub{\seq{\gt}, \seq{\g{t'}}}}{\g{t'}_m}
    }
    {\Gamma \vdash \esynth{\r{D^C\ulev{i}(\seq{\rt},\seq{\r{t'}})}}{ \g{C\ulev{i}(\seq{t})}}{\g{D^C\ulev{i}(\seq{\gt},\seq{\g{t'}})}}
      =>
    }

		\inferrule[\mname{ElabMatch}]{
			\Gamma \vdash \epsynth{\g{C}}{ \r{t_{scrut}}} {{\g{C\ulev{i}(\seq{t_{param}})}}}{\g{t_{scrut}}} \\\\
			\Gamma, \gz : \g{C(\seq{t_{param}})} \vdash \epsynth{\gType{  }}{\r{T_P}}{\gType{j}}{\g{T_P}} \\\\
			\seq{
        \Gamma, \g{x_{rec}} : (\g{ (z : C\ulev{i}(\seq{\g{t_{param}}}) ) -> T_P}), \seq{\gy} : {\args(\g{C},\g{i},\g{D_k})\parsub{\seq{\g{t_{param}}},\seq{\gy}} \vdash }}\\\\\qquad\seq{
         \echeck{\r{t_{rhs_k}}}
        {[\g{D^C_k\ulev{i} (\seq{t_{param}},\seq{\gy})} / \gz]\g{T_P}}{\g{t_{rhs_k}}}
        }
      }{\Gamma \vdash
        \esynth{\rmatch{C}{\r{t_{scrut}}}{\rz.\r{T_P}}{x_{rec}.\seq{\seq{y}.t_{rhs}}}}
        {[\g{t_{scrut}}/\gz]\g{T_P}}
        {\gmatch{C}{\g{t_{scrut}}}{\gz.\g{T_P}}{x_{rec}.\seq{\seq{y}.t_{rhs}}}}
      }

    \inferrule[\mname{ElabConstrSynthType}]{\Gamma \vdash  \esynth{\rt}{\gT}{\gt} \\\\ \gT \stepstostar \gType{i}}
    {\Gamma \vdash \epsynth{\gType{}}{\rt}{\gType{i}}{\gt} } \label{infrule:pts-sort-inf}

      \inferrule[\mname{ElabConstrSynthFun}]{\Gamma \vdash \esynth{\rt}{\gT}{\gt} \\\\
        \gT \stepstostar \g{(x : T_1) -> T_2}}
      {\Gamma \vdash \epsynth{{\gradualcolor{\Pi}}}{\rt}{{  \g{ (x : T_1) -> T_2}}}{\gt} } \label{infrule:pts-prod-inf}

    \inferrule[\mname{ElabConstrSynthInd}]{\Gamma \vdash \esynth{\rt}{\gT}{\gt} \\\\
      \gT \stepstostar \g {C\ulev{i}(\seq{\g{t_{p}}})}}
    {\Gamma \vdash \epsynth{\g{C}}{\rt}{ \g{ C\ulev{i}(\seq{\g{t_{p}}})}}{\gt} } \label{infrule:cic-ind-inf}
      }
    \end{inferbox}
  \end{boxedarray}
  \ifapxCaption{Elaboration from GCIC to CastCIC \ifnotapx{(homomorphic rules omitted)}}
  \ifnotapx{\label{fig:gcic-elab}}
\end{figure}
}
\drawFigElabGCIC

\section{Propositional Equality}
\label{sec:equality}

The main contribution of our paper is \lang:
an extension of GCIC with
propositional equality,
where the information about an imprecise value accumulates at run time
to detect inconsistencies.
We define \lang's semantics using a cast calculus \clang, which extends CastCIC with equality.

The core idea is that a surface-language proof of type $\r{ t_{1} ==_{T} t_{2}}$
is elaborated into a witness for the consistency of $\g{t_{1}}$ and $\g{t_{2}}$.
Much like the middle-type from threesomes~\citep{Siek:2009:TWB:1570506.1570511},
or evidence\footnote{
  Evidence is more complex in AGT, since it can witness subtyping. Evidence for plausible equality
  between types collapses to a single term as precise as the
  equated terms.
}
from AGT~\citep{agt}, the consistency witness between terms
is a term that is at least as precise as either term.
The standard equality proof, $\r{refl_{t} : t ==_{T} t}$, witnesses that  $\r{t}$
is consistent with itself,
while the imprecise proof $\rqm_{\r{t_{1} ==_{T} t_{2}}}$
is witnessed by the least precise term that is dynamically consistent with $\g{t_{1}}$ and $\g{t_{2}}$.
As a program runs, equality witnesses may take values between these extremes,
and when $\r{t}$ is imprecise, may be more precise than
the witness for $\g{refl}$.

The technical challenge with adding propositional equality is determining how to combine information
represented by the equality witnesses.
When casting between types $\g{t_{1} ==_{T} t_{2}}$ to $\g{t'_{1} ==_{T} t'_{2}}$,
both of which may be imprecise,
we must transform a witness $\g{t_{w}}$ for $t_{1} ==_{T} t_{2}$ to one for $t'_{1} ==_{T} t'_{2}$,
but even though $\g{t_{w}}$ is as precise as $\g{t_{1}}$ and $\g{t_{2}}$,
it may not be as precise as $\g{t'_{1}}$ and $\g{t'_{2}}$.
So we need a composition operator that can take $\g{t'_{1}}$, $\g{t'_{2}}$ and $\g{t_{w}}$
and produce a term that is as precise as all three.
However, to respect
static observational equivalences and avoid the problems of \cref{subsec:funcomp},
composition cannot be a syntactic meta-operation. The issue is with neutral terms,
i.e., variables, or terms whose reduction is blocked by applying or eliminating a variable.
Syntactic composition would require distinct
neutral terms to compose to an error, but that would violate extensionality.

Along with composition, we must define a notion of precision that determines valid witnesses of
consistency.
For the evolution of type information to be monotone, the operator $\gcompop$ should compute a lower
bound with respect to this notion of precision.
Computing the greatest lower bound prevents premature errors, although the proof that composition
is the greatest lower bound is left to future work.
With non-dependent gradual types, precision can be syntactically,
by adding structural rules to $\gt \sqube \gqmat{T}$, but structural rules are not flexible enough to handle composition.

The solutions to these two challenges are interdependent.
We avoid the issues with syntactic composition by adding it as a separate syntactic construct to \clang, so that
composition of neutral terms is itself a neutral term.
However, if composition is a construct in \clang, then precision must be
defined to accommodate terms that feature composition, so composing two neutral terms produces something
that is actually as precise as those two terms.
Precision must be defined to respect composition without
losing its other important properties, such as transitivity.

This section gives typing and semantics for gradual propositional equality,
where proofs of equality are represented by consistency witnesses,
but leave the exact definitions of consistency and precision unspecified.
In \cref{sec:generic}, we describe the properties that consistency and precision
should fulfill to ensure that \lang satisfies type safety and the gradual guarantees.
Finally, \cref{sec:prec} instantiates \lang with notions of precision and consistency
that fulfill our goals while ensuring decidable consistency-checking.
We separate our presentation in this way to motivate the choices we make in the design of precision,
and to avoid monolithic proofs when developing \lang's metatheory.

We write precision as
$\Gamma_{1} | \Gamma_{2} |- \g{t_{1}} \gensuprec \g{t_{2}}$
and consistency as $\g{t_{1}} \gencst \g{t_{2}}$,
highlighting the operators \Gbox{\text{in grey}} to indicate that their definitions are not yet specified.
The subscript $\stepsto$ on $\gencst$ indicates that it is \textit{definitional consistency},
whose name is chosen by analogy to definitional equality, since the operands can be reduced
before being compared structurally~\citep{MARTINLOUF197581}.
Precision is \textit{precision modulo conversion},
meaning it is closed under the equivalence relation given by convertibility.
Unlike consistency, precision modulo conversion can look backwards in time,
relating terms that are the \textit{results} of reducing syntactically-related terms,
in addition to relating terms that are syntactically-related after reducing.
We discuss the need for this in \cref{subsec:cast-eq}.
Precision takes two contexts, as its operands
must be typed in different contexts.

\subsection{\lang Syntax and Typing}
\label{subsec:geq-typing}

\begin{figure}
  \centering
  \begin{boxedarray}{@{}l@{}}
    \begin{inferbox}
      \genfrac{}{}{0pt}{2}{
      \fbox{ $\r{t} \bnfadd \r{t ==_{T} t} \bnfalt \r{refl_{t}} \bnfalt \r{\J(x.T,t_{1},t_{2},t_{3},t_{4})}$}
    }{
      \fbox{$ \r{h} \bnfadd \r{==} $ }
      }
    \ifapx{\\\\}

      \inferrule
      {\Gamma |-  \r{t_1} <= \rT\\
       \Gamma |- \r{t_2} <= \rT\\\\
       \Gamma |- \r{T} \psynth{\rType{  }} \rType{i}
      }
      {\Gamma |- \r{t_1 ==_T t_2} => \rType{i} }

      \inferrule
      {\Gamma |- \r{t} => \rT}
      {\Gamma |- \r{refl_t} => \r{t ==_T t} }

      \inferrule
      {
        \Gamma |- \r{t_{eq}} \psynth{\r{==}} \r{t_1 ==_{T} t_2}\\
        \Gamma, (\r{x} : \r{T}) |- \r{T_P} \psynth{\rType{  } } \rType{i}\\\\
        \Gamma |-  \r{t_1} <= \r{T}\\
        \Gamma |- \r{t_2} <= \r{T}\\
        \Gamma |- \r{t_{P1}} <= [ \r{t_1} / \rx ] \r{T_P}\\
      }
      {\Gamma |- \r{\J(x.T_P, t_1, t_2, t_{P1}, t_{eq} ) } => [\r{t_2} / \rx] \r{T_P} }
    \end{inferbox}
  \end{boxedarray}
  \ifapxCaption{\lang: Syntax and Typing Lemmas }
  \ifnotapx{\label{fig:geq-type-lemmas}}
\end{figure}

\cref{fig:geq-type-lemmas} extends GCIC to \lang by adding the equality type, introduction form, and eliminator.
Their types are identical to what is expected in the static setting.
Again, because surface typing is defined
by elaboration, the given rules are actually admissible lemmas.
An equality type  can describe an equality between any two values of consistent types (because each endpoint is checked against $\r{T}$).
The reflexive proof $\r{refl_{t}}$ synthesizes type $\r{t ==_{T} t}$,
so long as $\r{t}$ is well typed at type $\r{T}$.
The eliminator $\r{\J}$
takes a type $\r{T_{P}}$ parameterized over a value of type $\r{T}$,
\footnote{The full $\g{J}$  in type theory
  parameterizes $\g{T_{P}}$ over the equality proof. \Cref{subsec:axiomK} shows why this is not needed
  for \lang.
}
along with two values of type $\r{T}$. Then, given a value of
type $\r{[\g{t_{1}} /\gx]T_{P}}$,
and a proof $\r{t_{eq}}$ that $\r{t_{1}}$ and $\r{t_{2}}$ are equal, the elimination has type
$\r{[\g{t_{2}} /\gx]T_{P}}$.
That is, if two values are equal, we can take any term whose type refers to the first,
and transform it into a term whose type refers to the second.

\subsection{\clang Syntax and Typing}
\label{subsec:cast-eq}

\newcommand{\drawClangTyping}{
\ifapx{\begin{figure}[H]}
\ifnotapx{\begin{figure}}
  \centering
  \begin{boxedarray}{@{}l@{}}
    \begin{inferbox}
      \genfrac{}{}{0pt}{2}{
    \boxed{
      \g{t} \bnfadd \g{t_{1} ==_{T} t_{2}} \bnfalt \grefl{t}{t_{1}}{t_{2}} \bnfalt \g{\J(x.y.z.T,t_{1},t_{2},t_{3},t_{4})} \bnfalt \g{t_{1}} \gcomp{T} \g{t_{2}}
    } }
  {\boxed{\g{h} \bnfadd  \g{\lambda} \bnfalt \g{C^{D}} \bnfalt \g{refl} \bnfalt \g{==}}}
    \ifapx{\\\\}

      \inferrule[{ElabRefl}]{
        \Gamma |- \esynth{\rt}{\gT}{\gt}
      }{
        \Gamma |- \esynth{\r{refl_t}}{\g{t ==_T t}}{\grefl{t}{t}{t}}
      }

 \inferrule[{CastRefl}]
 { \Gamma |- \g{t_{w}} => \gT\\
   \Gamma |- \g{t_{1}} <= \gT\\
   \Gamma |- \g{t_{2}} <= \g{T}\\\\
   \Gamma | \Gamma |- \g{t_{w}} \gensuprec \g{t_{1}} \\
   \Gamma | \Gamma |- \g{t_{w}} \gensuprec \g{t_{2}} %
 }
 {\Gamma |- \grefl{t_{w}}{t_{1}}{t_{2}} => \g{t_{1} ==_{\gT} t_{2}}  }

 \inferrule[{CastComp}]
 {\Gamma |- \g{t_1} <= \gT\\\\
 \Gamma |- \g{t_2} <= \gT}
{\Gamma |- \g{t_1} \gcomp{T} \g{t_2} => \gT}
\ifapx{

      \inferrule[ElabEq]
      {
       \Gamma |- \epsynth{\gType{ }}{\rT}{\gType{i}}{\g{T}}\\\\
        \Gamma |-  \echeck{\r{t_1}}{\gT}{\g{t_1}}\\
        \Gamma |-  \echeck{\r{t_2}}{\gT}{\g{t_2}}
      }
      {\Gamma |- \esynth{\r{t_1 ==_T t_2}}{\gType{i}}{\g{t_1 ==_T t_2}} }

      \inferrule[CastEq]
      {\Gamma |-  \g{t_1} <= \g{T}\\
       \Gamma |- \g{t_2} <= \g{T}\\\\
       \Gamma |- \g{T} \psynth{\gType{  }} \gType{i}
      }
      {\Gamma |- \g{t_1 ==_T t_2} => \gType{i} }

      \inferrule[ElabJ]
      {
        \Gamma |- \epsynth{\g{=}}{\r{t_{eq}}}{\g{t_1 ==_{T_1} t_2}}{\g{t_{eq}} }\\
        \Gamma, (\g{x} : \g{T_1}) |- \epsynth{\gType{  }}{\r{T_P}}{\gType{i}}{ \g{T_P}}\\\\
        \Gamma |-  \echeck{\r{t_1}}{{\g{T_1}}}{\g{t_1}}\\
        \Gamma |- \echeck{\r{t_2}}{\g{T_1}}{\g{t_2}}\\
        \Gamma |- \echeck{\r{t_{P1}}}{[ \g{t_1} / \gx ] \g{T_P}}{\g{t_{P1}} }\\
      }
      {\Gamma |- \esynth{\r{\J(x.T_P, t_1, t_2, t_{P1}, t_{eq} ) }}{[ \g{t_2} / \gx ] \g{T_P}}{ \g{\J(x.T_P, t_1, t_2, t_{P1}, t_{eq} ) }}
      }

      \inferrule[CastJ]
      {
        \Gamma |- \g{t_{eq}} \psynth{\g{=}} \g{t_1 ==_{T} t_2}\\
        \Gamma, (\g{x} : \g{T}) |- \g{T_P} \psynth{\gType{  } } \gType{i}\\\\
        \Gamma |-  \g{t_1} <= \g{T}\\
        \Gamma |- \g{t_2} <= \g{T}\\
        \Gamma |- \g{t_{P1}} <= [ \g{t_1} / \gx ] \g{T_P}\\
      }
      {\Gamma |- \g{\J(x.T_P, t_1, t_2, t_{P1}, t_{eq} ) } => [ \g{t_2} / \gx ] \g{T_P} }
}
    \end{inferbox}
  \end{boxedarray}
  \ifapxCaption{\clang: Syntax, \ifnotapx{Key} Typing and Elaboration \ifnotapx{Rules} \ifapx{(Extends CastCIC)}}
  \ifnotapx{\label{fig:clang-typing}}
\end{figure}
}
\drawClangTyping

\cref{fig:clang-typing} extends CastCIC
to \clang by adding propositional equality and the gradual composition operator.
We extend the syntax for heads to include value constructors, not just types, which is useful
when defining the semantics of composition.
Proofs of reflexivity
take three arguments: the endpoints of the equality, and a witness of the (dynamic) consistency
of those endpoints.
We borrow the notation $\g{t_{w} |- t_{1} \cong t_{2}}$ from \citet{agt}
to indicate that $\g{t_{w}}$ contains information supporting the (dynamic) consistency of $\g{t_{1}}$ and $\g{t_{2}}$.
Composition is ascribed with the type of its arguments so that we can ascribe the
proper $\gT$ when the composition of two terms steps to $\errat{T}$.

For typing, \rrule{CastComp} synthesizes a composition's ascribed type when both arguments
check against that type.
In \rrule{CastRefl}, $\grefl{t_{w}}{t_{1}}{t_{2}}$ synthesizes $\g{t_{1} ==_{T} t_{2}}$ if the witness $\g{t_{w}}$
is as precise as both $\g{t_{1}}$ and $\g{t_{2}}$.
In \rrule{ElabRefl}, $\r{refl_{t}}$
is elaborated into $\grefl{t}{t}{t}$, i.e., a term serves as the initial witness that it is
equal to itself. If $\r{t}$ is imprecise, casts applied to the equality proof
may produce more precise witnesses,
but the programmer never constructs a witness directly.
We omit typing rules for $\g{==_{T}}$ and $\g\J$,
as they mirror the lemmas in \cref{fig:geq-type-lemmas},
as do their elaboration rules.

Precision must be closed under convertibility because,
as \citet{bertrand:gcic} note, syntactic precision is not preserved by stepping
the less precise term. Since $\gqmat{T}$ is less precise than $\gx$, the less precise term
may reduce in a way that is blocked for the other term.
So for contextual steps to preserve \rrule{CastRefl},
 the results of stepping related
terms must be related.

\subsection{Cast Semantics}
\label{subsec:cast-semantics}

\begin{figure}
  \centering
  \begin{boxedarray}{@{}l@{}}
    \begin{inferbox}
       \inferrule[RedJ]
      {}
      {\g{\J(x.T_{P},t_{1},t_{2},t_{P1},\grefl{t_{w}}{t_{1}}{t_{2}})}
        \redsto \cast{[\g{t_w}/\gx]T_{P}}{[\g{t_2}/\gx]T_{P}}\cast{[\g{t_1}/\gx]T_{P}}{[\g{t_w}/\gx]T_{P}}\g{t_{P1}}
      }

       \inferrule[RedEqGerm]
      {\g{t_{1} ==_{T} t_{2} \neq \germ_\ell(\g{=})}} %
      {\cast{t_{1} ==_{T} t_{2}}{\gqmat{\gType{\ell}}}\gt
        \redsto
        \cast{\gqmat{\gqmat{\gType{\ell}}} ==_{\gqmat{\gType{\ell}}} \gqmat{\gqmat{\gType{\ell}}} }{\gqmat{\gType{\ell}}}
        \cast{t_{1} ==_{T} t_{2}}{\gqmat{\gqmat{\gType{\ell}}} ==_{\gqmat{\gType{\ell}}} \gqmat{\gqmat{\gType{\ell}}} }\gt
      }

       \inferrule[RedCastEq]
      {} %
      {\cast{t_{1} ==_{T} t_{2}}{t'_{1} ==_{T'} t'_{2}}\grefl{t_{w}}{t_{1}}{t_{2}}
        \\\\ \redsto
      \grefl{(\cast{T}{T'}\g{t_{w}} \gcomp{T'} \g{t'_{1}} \gcomp{T'} \g{t'_{2}})}{t'_{1}}{t'_{2}}}

       \inferrule[PropEqUnk]{}{
        \gqmat{t_{1} ==_{T} t_{2}} \redsto \grefl{t_{1} \gcomp{T} t_{2}}{t_{1}}{t_{2}}
        }

      \inferrule[PropEq(Unk,Err)]{}{
        \err_{\g{t_{1} ==_{T} t_{2}}} \redsto \grefl{\err_{\gT}}{t_{1}}{t_{2}}
        }
    \end{inferbox}
  \end{boxedarray}
  \ifapxCaption{\clang: Reductions for Casts and $\g{J}$}
  \ifnotapx{\label{fig:j-cast-semantics}}
\end{figure}

A challenge with gradual equality is designing its dynamic semantics.
In a fully static language, $\s{refl}$ is only inhabited for identical values,
so $\s{J}$ performs no computation other than pattern matching on the proof of equality.
In the presence of type imprecision, $\g{J}$ must perform casts.
We also need reductions for casts between equality types.
\Cref{fig:j-cast-semantics} gives reductions for $\g{J}$ and casts.
The \rrule{RedJ} rule reduces by casting \textit{through} the motive $\g{T_{P}}$ with $\g{x}$
bound to the witness $\g{t_{w}}$. The typing of equality guarantees that this witness $\g{t_{w}}$ is as precise as either $\g{t_{1}}$ and $\g{t_{2}}$. So $[\g{t_{w}}/\gx]\g{T_{P}}$ is like a middle type,
since it is  more precise than
$[\g{t_{1}}/\gx]\g{T_{P}}$ (the type of $\g{t_{P1}}$) and $[\g{t_{2}}/\gx]\g{T_{P}}$
(the type of the result).

Why cast through the middle, and not directly from $[\g{t_{1}}/\gx]\g{T_{P}}$ to $[\g{t_{2}}/\gx]\g{T_{P}}$?
As \cref{subsec:quicksort} showed, the witness
tracks constraints as the program runs, and since composition is monotone,
its precision only increases. So constraints are remembered,
and the programmer can see when a static constraint has
been dynamically violated.
  Also, the witness ensures that equalities between inconsistent values
cannot be used without flagging an error. Without a witness,
one could have $\gqmat{2 ==_{\bN} 5}$, despite the type being statically uninhabited.
Then $\g{\J}$ could use this equality to convert from $\g{Vec\ Float\ (2 \mod 3)}$\
to $\g{Vec\ Float\ (5 \mod 3)}$: the cast would succeed, despite the absurdity of
the initial equality. Going through the middle type catches such absurd cases.

For casting $\g{refl}$ between equality types, the \rrule{RedCastEq} rule casts the witness to the correct type. The typing rule \rrule{CastRefl}
requires the witness to be as precise as the endpoints,
but the result of casting the witness is not guaranteed to fulfill this!
So the witness is composed with both endpoints, producing a precision-related result.
These casts are precisely why we need a composition operator.

The propagation rules \rrule{PropEqUnk} and \rrule{PropEqErr} reduce $\gqm$ and $\err$ at equality types
to $\g{refl}$ with the least and most precise witnesses, respectively.
\rrule{RedEqGerm} casts an equality proof to $\gqmat{\gType{\ell}}$ by casting through
the germ type, just like with functions and constructors.

\subsection{Semantics of Composition}
\label{subsec:meet-operator}

\begin{figure}
  \centering
  \begin{boxedarray}{@{}l@{}}
    \begin{inferbox}
      \inferrule[RedCompGerm]
      {}
      {\g{(\cast{\germ(h)}{\gqmat{\gType{\ell}}}\g{t_1})} \gcomp{\gqmat{\gType{\ell}}} \g{(\cast{\germ(h)}{\gqmat{\gType{\ell}}}\g{t_2})}
        \redsto
        \cast{\germ(h)}{\gqmat{\gType{\ell}}}\g{(t_1 \gcomp{\germ(h)} t_2)}
      }

      \inferrule[RedComp(Unk,Err)(L,R)]{
        \g{T} \text{ matches } $\g{\gType{\ell}}$ \text{,} $\g{C(\seq{t'})}$ \text{ or } $\gqmat{\gType{\ell}}$
      }{
        \gqmat{T} \gcomp{\gT} \gt \redsto \gt
        \qquad
        \gt \gcomp{\gT} \gqmat{\gT} \redsto \gt \\\\
        \errat{T} \gcomp{\gT} \gt \redsto \errat{\gT}
        \qquad
        \gt \gcomp{\gT} \errat{\gT} \redsto \errat{\gT}
      }

      \inferrule[RedCompGermErr]
      {\g{h_1} \neq \g{h_2}}
      {\g{(\cast{\germ(h_1)}{\gqmat{\gType{\ell}}}\g{t_1})} \!\gcomp{\gqmat{\gType{\ell}}}\! \g{(\cast{\germ(h_2)}{\gqmat{\gType{\ell}}}\g{t_2})}
        \\\\ \redsto
        \errat{\gqmat{\gType{\ell}}}
      }

      \inferrule[RedCompHeadErr]{
        \head(\g{t_1}) = \g{h_1}\quad
        \head(\g{t_2}) = \g{h_2}\\\\
        \g{h_1} \neq \g{h_2}
      }{
        \g{t_1} \gcomp{\gT} \g{t_2} \redsto \errat{T}
      }

      \inferrule[RedCompEq]{
        \g{t''_1} := \g{
          \cast{T_1}{T_1 \!\gcomp{\gType{\ell}}\! T_2}\g{t_1} \!\gcomp{(T_1 \gcomp{\gType{\ell}}\! T_2) } \!\cast{T_2}{T_1 \!\gcomp{\gType{\ell}}\! T_2}\g{t_2}} \\
        \g{t''_2} := \g{
          \cast{T_1}{T_1 \!\gcomp{\gType{\ell}}\! T_2}\g{t'_1} \!\gcomp{(T_1 \gcomp{\gType{\ell}} T_2) }\! \cast{T_2}{T_1 \!\gcomp{\gType{\ell}}\! T_2}\g{t'_2}
        }
      }
      {
        \g{(t_1 ==_{T_1} t'_1)} \!\gcomp{\gType{\ell}}\! \g{(t_2 ==_{T_2} t'_2)}
        \redsto
        \g{t''_1 ==_{(T_1 \gcomp{\gType{\ell}} T_2)} t''_2 }
      }

      \inferrule[RedCompRefl]{}
      {\grefl{t_1}{t}{t'} \gcomp{t ==_T t' } \grefl{t_2}{t}{t'}
        \redsto \grefl{t_1 \gcomp{T} t_2}{t}{t'}
      }

      \inferrule[RedCompInd]
      {}
      {\g{C}(\seq{\g{t_1}}) \gcomp{\gType{\ell}} \g{C}(\seq{\g{t_2}})
         \redsto \g{C}(\itercomp(\pars(\g{C}, \g{i}), \seq{\g{t_1}}, \seq{\g{t_2}})})

      \inferrule[RedCompLam]
      {}
      {\g{\lambda (x : T_1) \ldotp t_1} \gcomp{(x : T_1) -> T_2} \g{\lambda (x : T_1) \ldotp t_2 } \\\\ \redsto \g{\lambda (x : T_1) \ldotp (t_1 \gcomp{T_2} t_2)}}

      \inferrule[RedCompCon]
      {}
      {\g{D^C(} \seq{\g{t}}, \seq{\g{t_1}} \g{)} \gcomp{C(\seq{t})} \g{D^C(} \seq{\g{t}}, \seq{\g{t_2}} \g{)}
         \\\\\redsto \g{D^C}( \seq{\gt}, \itercomp(\args(\g{C},\g{i},\g{D_k}), \seq{\g{t_1}}, \seq{\g{t_2}}) ) }

      \inferrule[RedCompPi]{}
      {\
        \g{(x : T_1) -> T'_1} \gcomp{\gType{\ell}} \g{(x : T_2) -> T'_2}
        \redsto \\
        \g{(x : (T_1 \gcomp{\gType{\ell}} T_2)) ->} [\cast{T_1 \gcomp{\gType{\ell}} T_2}{T_1}\gx / \gx]\g{T'_1} \gcomp{\gType{\ell}} [\cast{T_1 \gcomp{\gType{\ell}} T_2}{T_2}\gx / \gx]\g{T'_2}
      }
    \end{inferbox}
    \\
    \jform{$\itercomp(\seq{(\gx : \gT)}, \seq{\g{t_{1}}}, \seq{\g{t_{2}}})$ (\textit{Iterative Dependent Type-Preserving Composition})}
    \begin{inferbox}
      \inferrule
      {}
      {\itercomp(\cdot, \cdot, \cdot) := \cdot
      \\\\}

      \inferrule
      {}
      {\itercomp((\gx : \g{T_1})\cdot \seq{(\gy : \g{T_2})}, \g{t_1}\cdot\seq{\g{t'_1}}, \g{t_2}\cdot\seq{\g{t'_2}}) := \\\\
        \g{(t_1 \gcomp{T_1} t_2)}\cdot\itercomp(
        \seq{(\gy : [\g{(t_1 \gcomp{T_1} t_2)} / \gx ]\g{T_2} )}, seq_1, seq_2) \textit{ where}\\\\
        \hspace{11cm}\mathllap{seq_1 := \seq{\castnog{[\g{t_1} / \gx ]\g{T_2}}{[\g{(t_1 \gcomp{T_1} t_2)} / \gx]\g{T_2}}\g{t'_1} } \qquad
        seq_2 := \seq{\castnog{[\g{t_2} / \gx ]\g{T_2}}{[\g{(t_1 \gcomp{T_1} t_2)} / \gx]\g{T_2}}\g{t'_2} }
        )}
      }
    \end{inferbox}
  \end{boxedarray}
  \ifapxCaption{\clang: Semantics for $\gcompop$ }
  \ifnotapx{\label{fig:clang-meet-semantics}}
\end{figure}

\Cref{fig:clang-meet-semantics}
gives the semantics of composition.
Technically, we do not need composition as an operator in \clang itself,
but only for witnesses and cast type ascriptions.
However, because dependent types
remove the separation between terms and types, witnesses and cast types need dynamic semantics.
So for simplicity, we let
witnesses and cast-types be any \clang terms, and add composition to \clang's semantics,
rather than duplicating \clang's semantics for a witness-specific language.

When designing reductions for composition, it is essential that when $\g{t_{1} \gcomp{T} t_{2}}$
reduces, the result is a term that is as precise as both $\g{t_{1}}$ and $\g{t_{2}}$.
For $\gqmat{T} \gcomp{T} \gt$, we produce
$\gt$ (\rrule{RedCompUnk(L,R)}), since $\gt$ is always as precise
as itself and $\gqmat{T}$.
Likewise, the rules that produce $\err$ satisfy this, since it is the most precise term.
We see this in \rrule{RedCompUnk(L,R)} , which composes with $\err$,
and in  \rrule{RedCompHeadErr} and \rrule{RedCompGermErr}, where composing non-neutral terms with distinct heads
reduces to $\err$.

The remaining rules compose terms with the same head $\g{h}$
e.g. both are functions, built with the same $\g{D^{C}}$, etc.
In these cases, the head $\g{h}$ is applied to the respective
composition of the arguments, e.g., the composition of functions is a function returning the composition of the bodies.
For functions, equality proofs, and inhabitants of $\gqmat{\gType{\ell}},$ the
head can be applied directly (\rrule{RedCompLam, RedCompRefl, RedCompGerm}).
In the remaining cases, we must account for how types of later arguments depend on
the values of earlier arguments.
\rrule{RedCompPi} produces a domain by composing the argument domains,
which is the type of the parameter $\gx$. The codomains $\gx$'s each have their own domain types,
so we cast all uses of $\gx$ from the composed type to the expected type.
\rrule{RedCompEq} adds casts the equated terms
by composing the element types.

The most complex rules are \rrule{RedCompInd} and \rrule{RedCompCon}.
Because type and data constructors have dependent function types,
the type of later parameters and arguments may depend on the \textit{values}
of previous parameters and arguments.
To compose type or data constructor applications, we compose the parameters and arguments element-wise,
but composing two arguments changes the type of later arguments.
The metafunction $\itercomp$ traverses the types
of type and data constructors, composing arguments element-wise and adding
casts to the bound variables in later arguments.

To see why composing needs casts, consider dependent pairs
 formulated as inductive types.
\begin{flalign*}
  &\g{data\ DPair\ : (X : \gType{}) -> (P : X -> \gType{}) -> \gType{}\ \textbf{where}}
  \\&\qquad
 \g{mkDPair : (x : X) -> P\ x -> DPair\ X\ P}
\end{flalign*}
One example of a dependent pair type is $\g{DPair\ (\bN \times \bN)\ (\lambda x \ldotp (\pi_{1}\ x) + (\pi_{2}\ x) ==_{\bN} 3)}$,
i.e., the Curry-Howard equivalent of ``there exists a pair of numbers such that adding them yields 3.''
Suppose we want to compose two inhabitants of this type, say
$\g{mkDPair\ (1,\gqmat{\bN})\ \grefl{3}{1 + \gqmat{\bN}}{3}}$
and $\g{mkDPair\ (\gqmat{\bN}, 2)\ \grefl{3}{\gqmat{\bN} + 2}{3}}$.
To compose the first element, we can produce $\g{(1 \gcomp{\bN} \gqmat{\bN}, \gqmat{\bN} \gcomp{\bN} 2)}$,
which reduces to $\g{(1,2)}$.
However, for the second element, the proofs $\grefl{3}{1 + \gqmat{\bN}}{3}$
and $\grefl{3}{ \gqmat{\bN} + 2}{3}$ do not have the same type:
they equate different terms, so we cannot compose them!
Instead, we must first cast each of them to the type
$\g{(1 \gcomp{\bN} \gqmat{\bN}) + ( \gqmat{\bN} \gcomp{\bN} 2) ==_{\bN} 3 }$,
i.e., the value
obtained by replacing $\gx$ with the composition of the pairs' first elements
in $\g{ ( (\pi_{1}\ x) + (\pi_{2}\ x) ==_{\bN} 3)}$.

\section{Parameterized Metatheory: Criteria for Precision and Consistency}
\label{sec:generic}

\newcommand{\lemReason}[1]{#1}

\lang is now defined except for \clang's precision and consistency relations.
For non-dependent languages, the semantics of precision can be justified
either in terms of sets of static terms~\citep{agt}
or in terms of semantic precision~\citep{10.1145/3236768}.
Such justifications are difficult with dependent types.
Our approach is different:
we define the important metatheoretic criteria for \lang
without referring to precision and consistency,
 then describe the criteria precision and consistency must fulfill
to prove the desired metatheoretic properties.
Meeting these criteria guides and justifies our definition of precision and consistency (\cref{sec:prec}).
We see precision and consistency as a means to the end of the desired metatheory.

\subsection{Stating the Gradual Guarantees}
\label{subsec:guarantees}

The gradual guarantees state that reducing the precision of a surface term introduces
no new static or dynamic errors. However, to state them formally, we must define
what precision means for surface terms. We follow \citet{bertrand:gcic} and define
surface precision as the relation generated by $\r{t} \squbr \rqmat{i}$,
plus all the usual structural rules.
Essentially, $\r{t} \squbr \r{t'}$ holds if we can obtain $\r{t'}$ by replacing some parts of $\r{t}$
with some $\rqmat{i}$.
To guarantee preservation of typing, we also need such replacements to be
\textit{universe adequate}~\citep{bertrand:gcic}.
We say that the judgment $\r{t} \squbr \r{t'}$ is universe adequate if,
for every subterm $\r{r}$ of $\r{t}$, when $\Gamma |- \esynth{\r{r}}{\gT}{\g{r}}$ and
$\Gamma |- \gT \psynth{\gType{ }} {\gType{i}}$, then any uses of
$\r{r} \squbr \rqmat{j}$ have $\r{i} = \r{j}$.
This essentially says $\r{t'}$ is $\r{t'}$
with some subterms replaced by $\rqmat{i}$ \textit{for the right $\r{i}$}.
We can now state the static gradual guarantee:

\begin{definition}[Static Gradual Guarantee]
  If $\cdot |- \rt : \rT$ and $\r{t} \squbr \r{t'}$ universe-adequately,
  then $\cdot |- \r{t'} : \rT$.
\end{definition}
\noindent
That is, reducing the precision of a program causes no new type errors.

To state the dynamic guarantee without referring to \clang precision, we must formalize what it
means to introduce no new dynamic errors.
We follow \citet{10.1145/3236768} and do this
with \textit{semantic precision},
which compares terms by quantifying over all possible boolean contexts.
We use booleans because of their simplicity:
gradual booleans have only the four values $\g{true}$, $\g{false}$, $\gqmat{\bB}$
and $\errat{\bB}$.
If a context exists such that
reducing a term's precision changes the result from $\g{true}$ to
$\g{false}$, then we have violated the guarantee that precision only affects
behavior via errors. Likewise, if a context exists such that reducing
precision turns $\g{true}$ to $\errat{\bB}$, then reducing precision introduced a new error.
By defining semantic precision in terms of \textit{all} contexts, we capture the idea
that the above behaviors are impossible for precision-related terms.
We formalize this as follows:

\begin{definition}[Semantic Precision]
  \textit{Boolean precision} $\squbB$ is defined by
  ${\g{true} \squbB \g{true}}$,
  ${\g{false} \squbB \g{false}}$,
  ${\errat{\bB} \squbB \g{b}}$,
  and
  ${\g{b} \squbB \gqmat{\bB}}$
  for all $\g{b : \bB}$.
  Then two closed terms are related by \textit{semantic precision},
  written ${ |= \g{t} \squbo \g{t'} : {\gT}}$ if, for all
  $C : \g{T -> \bB}$, whenever $C[\g{t}] \stepstostar {\g{v}}$,
  then $C[\g{t'}] \stepstostar {\g{v'}}$
  and $\g{v} \squbB \g{v'}$.

\end{definition}

Then the gradual guarantee states that reducing a surface term's precision
causes a corresponding reduction in the semantic precision
of the surface terms' elaborations.

\begin{definition}[Dynamic Gradual Guarantee]
  Suppose $\cdot |- \echeck{\rt}{\gT}{\gt}$ and $\cdot |- \echeck{\r{t'}}{\gT}{\g{t'}}$.
  If $\r{t} \squbr \r{t'}$ universe-adequately, then $\Gamma |= \g{t} \squbo \g{t'}$.
\end{definition}

\subsection{Necessary Properties of Precision and Consistency}
\label{subsec:criteria}

Next, we list properties that, if satisfied by $\gensuprec$ and $\gencst$,
suffice to prove type safety, conservative extension of CIC, well-typedness of elaboration,
and the gradual guarantees.
Each criterion is accompanied by a specific case of the safety or
gradual guarantee proofs that motivate its inclusion.
We also include criteria that composition should satisfy.
While the semantics appear in a prior section (\cref{subsec:meet-operator}),
the criteria are new from
GCIC, so we list them to highlight our contribution.

Though $\gensuprec$ is ideal for typing witnesses, it is too lenient to express the monotonicity properties
of \clang.
In particular, if the monotonicity of reduction is phrased with $\gensuprec$,
then consistency must also be closed under convertibility, which would make it undecidable
even when its operands terminate.
So we introduce a strictly stronger relation $\genprec$ which, like $\gencst$,
only compares after reductions, and not before.
This distinction is acceptable because the precision side-condition of \rrule{CastRefl} is never
used to prove safety or monotonicity. Rather, the side-condition ensures that witnesses always entail at least as much information as the equated terms. Including the side-condition in \clang's type  conveniently
captures the invariant that, when $\r{refl_{t}}$ is elaborated with initial witness $\g{t}$,
future witnesses are never less precise than $\gt$.

\paragraph{Safety and Elaboration}
For elaboration to preserve types, precision must be reflexive
so the initial witness for $\r{refl}$ is valid.
For safety,
progress requires that each well-typed non-value can step. So
composition must step for each non-value.
For preservation, each reduction must preserve types,
including composition reductions.
If composition yields a precision lower-bound and precision is transitive,
the side-conditions of \rrule{CastRefl} can be preserved.

\begin{theoremEnd}[apxproof, restate command = lemOne]{lemma}[Precision Reflexive]
  \label{lem:reflexive}
  If $\Gamma |- \gt <= \gT$ then $\Gamma | \Gamma |- \gt \genprec \gt $
        \lemReason{(\textit{For \rrule{ElabRefl} to produce an elaboration that satisfies the $\genprec$ side-condition of \rrule{CastRefl}}).}
      \end{theoremEnd}
      \begin{proofEnd}
        Straightforward induction on the definition of precision. Works because every form
        has a diagonal rule.
      \end{proofEnd}

      {\vspace{-1em}}
      \begin{theoremEnd}[apxproof, restate command = lemTwo]{lemma}[Composition Safety]
        \label{lem:comp-safety}
   If\ $\g{t_{1}} \gcomp{T} \g{t_{2}}$ is not a value and $\Gamma |- \g{t_{1}} \gcomp{T} \g{t_{2}} <= \g{T}$, then
        $\g{t_{1}} \gcomp{T} \g{t_{2}} \stepsto \g{t_{3}}$ for some $\g{t_{3}}$ and
         $\Gamma |- \g{t_{3}} <= \g{T}$
        \lemReason{(\textit{For progress and preservation})}
      \end{theoremEnd}
      \begin{proofEnd}
 Each composition of two canonical forms of the same type has a reduction.
If one of the composed terms is not a canonical form, then either
(1) one of the composed terms can reduce,
(2) one term is a $\gqm_{\gT}$ or $\err_{\gT}$ where $\g{T}$ is not a function or equality type,
and we can reduce with \rrule{RedCompUnk} or \rrule{RedCompErr},
or (3) one of the composed terms is neutral, and hence the composition is neutral.
      \end{proofEnd}

      \vspace{-1em}
      \begin{theoremEnd}[apxproof, restate command = lemThree]{lemma} [Composition Confluence]
        \label{lem:comp-confluence}
        If\ $\g{t_{1}} \gcomp{T} \g{t_{2}} \Rrightarrow \g{t_{3}} $ and $\g{t_{1}} \gcomp{T} \g{t_{2}} \Rrightarrow \g{t'_{3}} $
        maximally, then $\g{t_{3}} \Rrightarrow \g{t'_{3}} $, where $\Rrightarrow$ is the parallel reduction
        operator, standard in confluence proofs~\citep{TAKAHASHI1995120}
        \lemReason{\textit{(For confluence, which is needed to show that $\beta$-reductions preserve types)};}
      \end{theoremEnd}
      \begin{proofEnd}
        The proof is straightforward. We provide the case for composition that
        fits into the overall induction proof of confluence.
        If $\g{t'_{3}}$ is obtained only by stepping within $\g{t_{1}}$ and $\g{t_{2}}$
        then the result follows from IH. If $\g{t'_{3}}$ and $\g{t'_{3}}$ are both the results a \rrule{RedComp*} rule,
        then the result follows from IH, plus the preservation of parallel reduction
        under substitution (since some rules use $\itercomp$).
        Finally, if $\g{t'_{3}}$ is the result of a \rrule{RedComp*} rule but $\g{t_{3}}$ is not,
        then we can step $\g{t_{3}}$ with that same rule, and apply the IH
        and preservation under substitution to get the result.
        The key is that in each case, there is only one possible non-contextual reduction.
      \end{proofEnd}
      \vspace{-1em}

      \begin{theoremEnd}[apxproof, restate command = lemFour]{lemma} [Composition Lower Bound]
        \label{lem:comp-lower}
        If\ $\Gamma |- \g{t_{1}} \gcomp{T} \g{t_{2}} <= \g{T}$,
        then
        $\Gamma | \Gamma |- \g{t_{1}} \gcomp{T} \g{t_{2}} \genprec \g{t_{1}} $
        and $\Gamma | \Gamma |- \g{t_{1}} \gcomp{T} \g{t_{2}} \genprec \g{t_{2}} $
        \lemReason{(\textit{Preserving the $\gensuprec$ condition of \rrule{CastRefl} for reduction \rrule{RedCastEq}})};
      \end{theoremEnd}
      \begin{proofEnd}
        Given by \rrule{PrecComp(L,R)}.
      \end{proofEnd}
      \vspace{-1em}

      \begin{theoremEnd}[apxlem]{lemma}[Presynthesis is Monotone]
        \label{lem:presynth-mono}
        If $\Gamma |- \g{t_{1}} =>* \g{T_{1}}$, $\Gamma |- \g{t_{2}} =>* \g{T_{2}}$
        and $\Gamma_{1} | \Gamma_2 |- \g{t_{1}} \squbs \g{t_{2}}$, then $\Gamma_{1} | \Gamma_{2} |- g{T_{1}} \squbs \g{T_{2}}$.
      \end{theoremEnd}
      \begin{proofEnd}
        Straightforward induction on the type derivation. \Cref{lem:sub-mono} is used
        for \rrule{CastApp}, and \cref{lem:precision-step} and the catch-up lemmas are used to show that the results of constrained synthesis are precision related.
      \end{proofEnd}

      \begin{theoremEnd}[apxproof, restate command = lemFive]{lemma} [Precision Transitive]
        \label{lem:trans}
        If\ $\Gamma_{1} | \Gamma_{2} |- \g{t_{1}} \genprec \g{t_{2}} $ and
        $\Gamma_{2} | \Gamma_{3} |- \g{t_{2}} \genprec \g{t_{3}} : \gT$ then\\
        ${\Gamma_{1} | \Gamma_{3} |- \g{t_{1}} \genprec \g{t_{3}} }$
        \lemReason{(\textit{Preserving the $\gensuprec$ side-condition of \rrule{CastRefl} for reduction \rrule{RedCastEq}})};
      \end{theoremEnd}
      \begin{proofEnd}
        We prove for by mutual induction for structural and definitional precision.
        For definitional precision, it follows from the inductive hypothesis, plus confluence.

        For structural precision, we proceed
  by induction on the combined depths of the derivations $\mathcal{D_{1}} :: \Gamma_{1} | \Gamma_{2} |- \g{t_{1}} \squbs \g{t_{2}}$ and $\mathcal{D_{2}} :: \Gamma_{2} | \Gamma_{3} |- \g{t_{2}} \squbs \g{t_{3}}$.
  Cases where both use the same diagonal rule are straightforward, as are any with $\err$ on the left or $\gqm$ on the right. We show a few examples for remaining cases, the reasoning is similar in those we omit.

  \case{\rrule{DiagComp}, \rrule{PrecCompL} (\rrule{PrecCompR} symmetric)}{
    Then $\g{t_{1}} = \g{t_{1L} \gcomp{T_{1}} \g{t_{1R}}}$,
    and $\g{t_{2}} = \g{t_{2L} \gcomp{T_{2}} \g{t_{2R}}}$,
    where  $\Gamma_{1} | \Gamma_{2} |- \g{t_{1L}} \squbs \g{t_{2L}}$,
    $\Gamma_{1} | \Gamma_{2} |- \g{t_{1R}} \squbs \g{t_{2R}}$,
    and $\Gamma_{2} | \Gamma_{3} |- \g{t_{2L}} \squbs \g{t_{3}}$.
    IH gives that $\Gamma_{1} | \Gamma_{3} |- \g{t_{1L}} \squbs \g{t_{3}}$,
    so \rrule{PrecCompL} yields our result.
  }
  \case{$\mathcal{D_{1}}$ with \rrule{CastL} or $\mathcal{D_{2}}$ with \rrule{CastR}}
  {
    Follows from IH, plus \cref{lem:presynth-mono} to obtain the type precision premises
    of \rrule{CastL} or \rrule{CastR}.
  }
  \case{$\mathcal{D_{1}}$ with \rrule{CastR} or \rrule{DiagCast}, $\mathcal{D_{2}}$ with \rrule{CastL}}
  {
    Similar reasoning to above, IH gives us a precision relation between the terms being cast, and then
    we can apply \rrule{CastL} to obtain result.
  }

\end{proofEnd}
\vspace{-1em}

\begin{theoremEnd}[apxproof, restate command = lemSix]{lemma} [Precision Modulo Conversion]
  \label{lem:super-prec}
        If $\Gamma_{1} | \Gamma_{2} |- \g{t_{1}} \gensuprec \g{t_{2}} $,
        where $\g{t_{2}} \stepstostar \g{t'_{2}}$, then $\Gamma_{1} | \Gamma_{2} |- \g{t_{1}} \gensuprec \g{t'_{2}}$
        \lemReason{(Preservation of \rrule{CastRefl} under contextual reduction)}
      \end{theoremEnd}
      \begin{proofEnd}
        Immediate from the definition of precision modulo conversion.
      \end{proofEnd}

\paragraph{Conservativity}

If \lang is to conservatively extend CIC, then a fully-static program should be well-typed in CIC if and only if it is  well-typed in \lang. For the most part, the rules only differ when $\rqm$
is involved, but the major exception is \rrule{ElabCst}, which let us replace a
type with any consistent type (after conversion). So for fully static terms, consistency should coincide
with syntactic equality.

\begin{theoremEnd}[apxproof, restate command = lemSeven]{lemma}[Static Consistency]
  \label{lem:static-cst}
  For any static terms $\s{t_{1}}$ and $\s{t_{2}}$,
        let $\g{t_{1}}$ and $\g{t_{2}}$ be their embedding in \clang.
        Then $\g{t_{1}} \gencst \g{t_{2}}$ iff $\s{t_{1}} =_{\alpha\beta} \s{t_{2}}$,
        i.e., if they are statically definitionally equal
        \lemReason{(\textit{For \lang to conservatively extend CIC})}.
      \end{theoremEnd}
      \begin{proofEnd}
        In the static fragment of GCIC, $\gqm$, $\err$ and $\gcompop$ are absent,
        and all closed equality proofs have the form $\grefl{t_{w}}{t}{t'}$ where $\g{t_{2}}$,
        $\g{t}$ and $\g{t'}$ are all definitionally equal.
        So mutual induction on the derivations of $\alpha$-consistency and definitional consistency shows the result:
        rules \rrule{CstCompDiag}, \rrule{CstComp(L,R)}, and \rrule{CstUnk(L,R)}
        can never occur.
        All other rules are a head constructor with consistency premises between the arguments.
      \end{proofEnd}

\paragraph{Monotonicity}

The last group of properties  relate to the gradual guarantees.
The dynamic gradual guarantee requires
that evaluating precision-related terms produces precision-related results.
Because of the dependency in dependent types, proving the static guarantee
relies on the proof of the dynamic guarantee: \rrule{ElabCst} reduces types
before comparing for consistency, so precision of types before reduction
should be preserved, and reducing the precision of a type
should make it consistent with no fewer types.
Likewise, to show the static guarantee,
elaboration must be monotone in both synthesized types and elaborated terms, since
dependent application uses the argument's elaboration in the return type.

\begin{theoremEnd}[apxproof, restate command = lemEight]{lemma}[Cast Monotonicity]
  \label{lem:cast-mono}
  Suppose $\Gamma_{1} | \Gamma_{2} |- \g{t_{1}} \genprec \g{t_{2}} $,
        and that $\Gamma _{1}|- \g{t_{1}} => \g{T_{1}}$ and $\Gamma_{2} |- \g{t_{2}} => \g{T_{2}}$
        where $\Gamma_{1} | \Gamma_{1} |- \g{T_{1}} \genprec \g{T'_{1}} $
        and $\Gamma_{2} | \Gamma_{2} |- \g{T_{2}} \genprec \g{T'_{2}} $.
        Then $\Gamma_{1} | \Gamma_{2} |- \cast{T_{1}}{T'_{1}}\g{t_{1}} \genprec \cast{T_{2}}{T'_{2}}\g{t_{2}}$
        \lemReason{\textit{(For \rrule{ElabCst} to produce $\genprec$-related elaborations for $\squbr$-related inputs)}}
      \end{theoremEnd}
      \begin{proofEnd}
        Given by \rrule{DiagCast}.
      \end{proofEnd}

      \begin{theoremEnd}[apxlem]{lemma}[Substitution Monotone for Structural Precision]
        \label{lem:struct-sub-mono}
        Suppose ${\Gamma_{1} | \Gamma_{2} |- \g{t_{1}} \squbs \g{t_{2}}} $,
        where ${\Gamma_{1} |- \g{t_{1}} => \g{T_{1}}}$
        and ${\Gamma_{2} |- \g{t_{2}} => \g{T_{2}}}$.
        If $\Gamma_{1},(\gx : \g{T_{1}}),\Delta_{1} | \Gamma_{2},(\gx : \g{T_{2}}),\Delta_{2} |- \g{t'_{1}} \squbs \g{t'_{2}} $, then
        $\Gamma_{1} [\g{t_{1}} / \gx]\Delta_{1} | \Gamma_{2} [\g{t_{2}} / \gx] \Delta_{2} |- [\g{t_{1}} / \gx] \g{t'_{1}} \squbs [\g{t_{2}} / \gx] \g{t'_{2}} $
        \lemReason{(\textit{Needed for \rrule{ElabApp} to be monotone in the return type})}
      \end{theoremEnd}
      \begin{proofEnd}
        As \citet{bertrand:gcic} say, the proof follows from weakening of typing,
        with induction on the precision derivation. The only non-trivial new case is
        for \rrule{PrecComp(L)} (\rrule{PrecCompR} is symmetric).
        In this case, we have $\Gamma_{1}(\gx : \g{T_{1}})\Delta_{1} | \Gamma_{2}(\gx : \g{T_{2}})\Delta_{2} |- \g{t'_{L1} \gcomp{T} \g{t'_{R1}}} \squbs \g{t'_{2}}$, where
        $\Gamma_{1}(\gx : \g{T_{1}})\Delta_{1} | \Gamma_{2}(\gx : \g{T_{2}})\Delta_{2} |- \g{t'_{L1}} \squbs \g{t'_{2}} $.
        IH $\Gamma_{1} [\g{t_{1}} / \gx]\Delta_{1} | \Gamma_{2} [\g{t_{2}} / \gx] \Delta_{2} |- [\g{t_{1}} / \gx] \g{t'_{L1}} \squbs [\g{t_{2}} / \gx] \g{t'_{2}} $, so we apply \rrule{PrecCompL} to get our result.
      \end{proofEnd}

\vspace{-1em}
      \begin{theoremEnd}[apxproof, restate command = lemNine]{lemma}[Substitution Monotone]
        \label{lem:sub-mono}
        Suppose ${\Gamma_{1} | \Gamma_{2} |- \g{t_{1}} \genprec \g{t_{2}}} $,
        where ${\Gamma_{1} |- \g{t_{1}} => \g{T_{1}}}$
        and ${\Gamma_{2} |- \g{t_{2}} => \g{T_{2}}}$.
        If $\Gamma_{1}(\gx : \g{T_{1}})\Delta_{1} | \Gamma_{2}(\gx : \g{T_{2}})\Delta_{2} |- \g{t'_{1}} \genprec \g{t'_{2}} $, then
        $\Gamma_{1} [\g{t_{1}} / \gx]\Delta_{1} | \Gamma_{2} [\g{t_{2}} / \gx] \Delta_{2} |- [\g{t_{1}} / \gx] \g{t'_{1}} \genprec [\g{t_{2}} / \gx] \g{t'_{2}} $
        \lemReason{(\textit{For \rrule{ElabApp} to be monotone in the return type})}
      \end{theoremEnd}
      \begin{proofEnd}
        Follows from \cref{lem:struct-sub-mono} for $\squbs$, plus the preservation of reduction
        under substitution.
      \end{proofEnd}

      \begin{theoremEnd}[apxlem]{lemma}[Catch-up Lemmas]
        \label{lem:catchup}
        Suppose $\Gamma_{1} \squbs \Gamma_{2}$, and $\Gamma_{1} | \Gamma_{2} |- \g{t_{1}} \squbs \g{t_{2}}$ where $\g{t_{1}}$ has head $\g{h}$, and $\Gamma_{1} |- \g{t_{1}} => \gT$.
        Then $\g{t_{2}} \stepstostar \g{t'_{2}}$ where either (1) $\g{t'_{2}}$ has head $\g{h}$
        and $\Gamma_{1} | \Gamma_{2} |- \g{t_{1}} \squbs \g{t'_{2}}$,
        or (2)
        $\g{t'_{2}}$ is $\gqmat{T'}$ where $\Gamma_{1} | \Gamma_{2} |- \g{T_{1}} \squbs \g{T_{2}}$.
        The same holds for $\defprec$.
      \end{theoremEnd}
      \begin{proofEnd}
        The proof is by induction on the precision derivation, and is identical to GCIC, except for the cases regarding equality. First, we note
        that if $\g{t_{1}}$ has head $\g{h}$, it cannot be a composition or a cast expression. The only rules that allow
        composition on the right require that the term on the left either be a cast or a composition,
        so $\g{t_{2}}$ cannot be a composition.

        \rcase{DiagEq}{
          Identical to the GCIC case for function types.
        }
        \rcase{DiagRefl}{
          Then $\g{t_{2}} = \cast{T'_{1}}{T'_{2}}\ldots\cast{T'_{k-1}}{T'_{k}}\g{t'_{2}}$,
          since \rrule{CastL} and \rrule{DiagRefl} are the only rules that apply.
          All of the $\g{T'_{i}}$s are definitionally less precise than
          $\g{T}$, but $\g{T}$ must be some $\g{t_{L} ==_{T_{elem}} t_{R}}$ by \rrule{CastEq}.
          So each $\g{T'_{i}}$ must either reduce to either $\g{t_{iL} ==_{T_{elem_{i}}} t_{iR}}$
          or $\gqmat{\gType{\ell}}$ (using the previous case for head $\gType{ }$).
          And the typing premise of \rrule{CastL}, $\g{T'_{k}}$ must be $\g{t_{kL} ==_{T_{elem_{k}}} t_{kR}}$.
          If $\g{t'_{2}}$ is $\gqmat{t_{kL} ==_{T_{elem_{k}}} t_{kR}}$ (because the first non-cast precision rule was \rrule{GenUnk}) then it reduces with \rrule{PropEqUnk} to
          $\grefl{t_{kL} \gcomp{T_{elem_{k}}} t_{kR}}{t_{kL}}{t_{kR}}$. Otherwise it already has the form
          $\grefl{t_{kw}}{t_{kL}}{t_{kR}}$ (because the first non-cast precision rule was \rrule{DiagRefl}).
          All casts can then be reduced using one of \rrule{RedCastEq}, \rrule{RedUpDown}, \rrule{RedEqUnk} or \rrule{RedEqGerm}. Since we begin and end with an equality type, the resulting term
          must have the form $\grefl{t''_{w}}{t''_{1}}{t''_{2}}$.
        }
      \end{proofEnd}

\vspace{-1em}
      \begin{theoremEnd}[apxproof, restate command = lemTen]{lemma}[Reduction Monotone]
        \label{lem:precision-step}
        If $\Gamma_{1} | \Gamma_{2} |- \g{t_{1}} \genprec \g{t_{2}} $ and $\g{t_{1}} \stepstostar \g{t'_{1}}$,
        then $\g{t_{2}} \stepstostar \g{t'_{2}}$ for some $\g{t'_{2}}$ where $\Gamma_{1} | \Gamma_{2} |- \g{t'_{1}} \genprec \g{t'_{2}}$
        \lemReason{(\textit{For DGG, to preserve \rrule{ElabCst} when reducing precision, and to preserve typing under contextual reduction of $\grefl{t_{w}}{t_{1}}{t_{2}}$ })}
      \end{theoremEnd}
      \begin{proofEnd}
        We simultaneously prove the result for $\defprec$ and $\squbs$ by mutual induction on the precision derivation.
        Because \rrule{PrecComp(L,R)} have composition on the left-hand side,
        it is never a possibility in the inversions performed in the GCIC proof.
        So the GCIC case for each of its reductions still holds for \lang. We show
        the cases for the rules added to \lang.
        If we ever take a contextual step, the result is immediate from the IH
        and the corresponding diagonal rule.

        \case{\rrule{CastL}}{
          The only case now from GCIC is the cast for equality:

          \rcase{RedCastEq}{
            Then $\g{t_{1}} = \cast{t_{startL} ==_{Tstart} t_{endR}}{t_{endL} ==_{Tend} t_{endR}}\grefl{t_{w}}{t_{startL}}{t_{endR}} $,
            and $\Gamma_{2} |- \g{t_{2}} =>* \g{T_{2}} $,
            where $\Gamma_{1} | \Gamma_{2} |- \g{t_{startL} ==_{Tstart} t_{endR}} \genprec \g{T_{2}}$,
          $\Gamma_{1} | \Gamma_{2} |- \g{t_{endL} ==_{Tend} t_{endR}} \genprec \g{T_{2}}$,
        and $\Gamma_{1} | \Gamma_{2} |- \grefl{t_{w}}{t_{startL}}{t_{endR}} \squbs \g{t_{2}} $.
        Our goal is to show that
        $\Gamma_{1} | \Gamma_{2} |- \grefl{\cast{T_{start}}{T_{end}}\gt_{w} \gcomp{T_{end}} t_{endL} \gcomp{T_{end}} t_{endR}}{t_{endL}}{t_{endR}} \squbs \g{t'_{2}} $ and $\g{t_{2} \stepstostar \g{t'_{2}}}$.
        By \cref{lem:catchup}, we have (1) $\g{t_{2}}\stepstostar \grefl{t'_{2w}}{t'_{2L}}{t'_{2R}}$
        with $\Gamma_{1} | \Gamma_{2} |- \grefl{t_{w}}{t_{startL}}{t_{endR}} \squbs \grefl{t'_{2w}}{t'_{2L}}{t'_{2R}} $,
        or  (2) $\g{t_{2}} \stepstostar \gqmat{T_{2}}$. The result is immediate for (2) by \rrule{GenPrecUnk}.

        For (1),
        inversion on typing gives that $\g{T_{2}} = \g{t'_{2L} ==_{T'_{2}} t'_{2R}}$,
        so inversion on  $\g{t_{startL} ==_{Tstart} t_{startR}} \genprec \g{T_{2}}$
        gives $\g{\Gamma_{1} |- \Gamma_{2} |- \g{t_{startL}} \squbs \g{t'_{2L}}}$,
        $\g{\Gamma_{1} |- \Gamma_{2} |- \g{t_{startR}} \squbs \g{t'_{2R}}}$,
        and $\Gamma_{1} | \Gamma_{2} |- \g{T_{start}} \squbs \g{T'_{2}}$.
        We get the same for $\g{t_{endL}}$, $\g{t_{endR}}$, and $\g{T_{end}}$.
        Inversion on the precision from \cref{lem:catchup} gives $\Gamma_{1} | \Gamma_{2} |- \g{t_{w}} \squbs \g{t'_{2w}}$,
        so then \rrule{CastL} gives $\Gamma_{1} | \Gamma_{2} |- \cast{T_{start}}{T_{end}}\g{t_{w}} \squbs \g{t'_{2w}}$.
        Two applications of \rrule{PrecCompR} give $\Gamma_{1} | \Gamma_{2} |- \cast{T_{start}}{T_{end}}\g{t_{w}} \gcomp{T_{end}} \g{t_{endL}} \gcomp{T_{end}} \g{t_{endR}} \squbs \g{t'_{2w}}$,
        so then our result is built with \rrule{DiagRefl}.
          }
        }
        \case{\rrule{PrecCompL} (\rrule{PrecCompR} is symmetric)}{
            Then $\g{t_{1}} = \g{t_{1L} \gcomp{T} t_{1R}}$, with
            $\mathcal{D} :: \Gamma_{1} | \Gamma_{2} |- \g{t_{1L}} \squbs \g{t_{2}}$.
            We proceed by cases on the reduction rule used on $\g{t_{1L} \gcomp{T} t_{1R}}$.

            \rcase{RedCompGerm}{
              Then $\g{t_{1L}} = \g{\castnog{\germ(\g{h})}{\gqmat{\gType{\ell}}} \g{t'_{1L}}}$
              and $\g{t_{1R}} = \g{\castnog{\germ(\g{h})}{\gqmat{\gType{\ell}}} \g{t'_{1R}}}$.
              We proceed by inversion on $\mathcal{D}$.

              \rcase{DiagCast}{
                Then $\g{t_{2}} = \cast{T_{2start}}{T_{2end}}\g{t'_{2}}$,
            with $\Gamma_{1} | \Gamma_{2} |- \g{t'_{1L}} \squbs \g{t'_{2}}$,
            $\Gamma_{1} | \Gamma_{2} |- \germ(\g{h}) \squbs \g{T_{2start}}$
            and $\Gamma_{1} | \Gamma_{2} |- \gqmat{\gType{\ell}} \squbs \g{T_{2end}}$.
            Then $\g{T_{2end}}$ is $\gqmat{\gType{\ell}}$ with some number of casts,
            so we can apply \rrule{GenUnk} and \rrule{CastL} to get
            $\Gamma_{1} | \Gamma_{2} |- \germ(\g{h}) \squbs \g{T_{2end}}$.
            So by \rrule{PrecCompL} we have
            $\Gamma_{1} | \Gamma_{2} |- \g{t'_{1L} \gcomp{\germ(\g{h})} t'_{1R}} \squbs \g{t'_{2}}$.
            Then by \rrule{DiagCast} we have our result.
              }
              \rcase{CastL}{
                Then we have $\Gamma_{1} | \Gamma_{2} |- \g{t'_{1L}} \squbs \g{t_{2}}$, so
                by \rrule{PrecCompL} we have $\Gamma_{1} | \Gamma_{2} |- \g{t'_{1L}} \gcomp{\germ(\g{h})} \g{t'_{1R}} \squbs \g{t_{2}}$.
                The result then follows from \rrule{CastL}.
              }
          }
          \rcase{RedCompUnkL}{
            Then $\g{t_{2}} \stepstostar \gqmat{\gT}$, giving our result.
          }
          \rcase{RedCompUnkR}{
            Follows from premise of \rrule{PrecCompL}.
          }
          \rcase{RedCompErr, RedCompHeadErr, RedCompGermErr}{
            Trivial since $\err$ is least.
          }
          \rcase{RedCompEq,RedCompRefl,RedCompInd,RedCompLam,RedCompCon,RedCompPi}{
            In each case, $\g{t_{2}}$ has the same head as $\g{t_{1L}}$ and $\g{t_{1R}}$,
            so we can push the use of \rrule{PrecCompL} deeper, applying the necessary
            diagonal rules, along with \cref{lem:sub-mono}.
          }
        }
        \rcase{DiagComp}{

          By cases on the reduction rule used on the left.

          \rcase{RedCompGerm}{
            Then we have
            $\Gamma_{1} | \Gamma_{2} |- \castnog{\germ(\g{h})}{\gqmat{\gType{\ell}}}\g{t_{L}} \gcomp{\gType{\ell}}
            \castnog{\germ(\g{h})}{\gqmat{\gType{\ell}}}\g{t_{L}} \squbs \g{t_{2L} \gcomp{T_{2}} \g{t_{2R}}}$.
            Then by \cref{lem:catchup} $\g{T_{2}}$ must reduce to $\gqmat{\gType{\ell}}$, since it is less precise than $\gqmat{\gType{\ell}}$. The precision relation between
            $\castnog{\germ(\g{h})}{\gqmat{\gType{\ell}}}\g{t_{L}}$ and $\g{t_{2L}}$
            must either use \rrule{DiagCast} or \rrule{CastL}, likewise for on the RHS.
            If both use \rrule{DiagCast}, then we can step $ \g{t_{2L} \gcomp{T_{2}} \g{t_{2R}}}$
            with \rrule{RedCompGerm} and use \rrule{DiagCast} and \rrule{DiagComp} for our result.

            Consider then the case where the precision relation with $\g{t_{2L}}$ uses \rrule{CastL},
            and with $\g{t_{2R}}$ uses \rrule{DiagCast} (the case for vice versa is symmetric).
            We will get our result in zero steps of the RHS.
            Here, $\Gamma_{1} | \Gamma_{2} |- \g{t_{1L} \squbs \g{t_{2L}}}$, with
            and $\g{t_{2R}} = \cast{T_{2start}}{T_{2end}}\g{t'_{2R}}$
            with $\Gamma_{1} | \Gamma_{2} |- \g{t_{1R}} \squbs \g{t'_{2R}}$,
            $\Gamma_{1} | \Gamma_{2} |- \germ(\g{h}) \squbs \g{T_{2start}}$
            and $\Gamma_{1} | \Gamma_{2} |- \gqmat{\gType{\ell}} \squbs \g{T_{2end}}$.
            Then $\g{T_{2end}}$ is $\gqmat{\gType{\ell}}$ with some number of casts,
            so we can apply \rrule{GenUnk} and \rrule{CastL} to get
            $\Gamma_{1} | \Gamma_{2} |- \germ(\g{h}) \squbs \g{T_{2end}}$.
             Then by \rrule{CastR}, we have $\Gamma_{1} | \Gamma_{2} |- \g{t_{1R}} \squbs \cast{T_{2start}}{T_{2end}}\g{t'_{2R}}$.
             So by \rrule{DiagComp} we have
             $\Gamma |- \g{t_{1L} \gcomp{\germ(\g{h})} t_{1R} } \squbs \g{t_{2L}} \comp{T_{2end}} \cast{T_{2start}}{T_{2end}}\g{t'_{2R}}$.
             Finally, our result comes from applying \rrule{CastL}.
             The case where both use \rrule{CastL} is similar, but without the need to use \rrule{CastR}.
          }
          \rcase{RedCompUnk}{
            We handle \rrule{RedCompUnk(L)}, the other case is symmetric.
            In this case, we have $\Gamma_{1} | \Gamma{2} |- \gqmat{T_{1}} \gcomp{T_{1}} \g{t_{1R}} \squbs \g{t_{2L}} \gcomp{T_{2}} \g{T_{2R}} $.
            Inversion gives that $\g{t_{2L}}$ must be some sequence of casts applied to $\gqmat{T'_{2}}$,
            where each cast type and $\g{T'_{2}}$ are all less precise than $\g{T_{1}}$.
            Because $\g{T}$ is one of $\gType{\ell}$, $\g{C(\seq{t})}$ or $\gqmat{\gType{\ell}}$,
            and because each cast type is less precise,
            $\g{t_{2L}} \stepstostar \gqmat{T_{2}}$. So we can step with \rrule{RedCompUnk(L)}
            and get our result from inversion on the original precision derivation.
          }
          \rcase{RedCompErr, RedCompHeadErr, RedCompGermErr}{
            Trivial since $\err$ is least.
          }
          \rcase{RedCompEq,RedCompRefl,RedCompInd,RedCompLam,RedCompCon,RedCompPi}{
            Then we have $\Gamma_{1} | \Gamma_{2} |- \g{ t_{1L} \gcomp{T} \g{t_{1R}} } \squbs \g{t_{2L} \gcomp{T_{2}} \g{t_{2R}}}$,
            where $\g{t_{1L}}$ and $\g{t_{2L}}$ have the same head $\g{h}$.
            We have
            $t_{1L} \gcomp{T} \g{t_{1R}} \stepstostar \g{t'_{1}}$ and want to find $\g{t'_{2}}$
            such that $\g{t_{2L} \gcomp{T_{2}} t_{2R}} \stepstostar \g{t'_{2}}$ and $\Gamma_{1} | \Gamma_{2} |- \g{t'_{1}} \squbs \g{t'_{2}}$.
            \Cref{lem:catchup} on each side of $\gcompop$ gives $\g{t_{2L} \gcomp{T_{2}} t_{2R}} \stepstostar \g{t'_{2L}} \gcomp{T_{2}} \g{t'_{2R}}$ that is less precise than $\g{t_{1L} \gcomp{T} \g{t_{1R}}}$,
            where $\g{t'_{2L}}$ and $\g{t'_{2R}}$ both have head $\g{h}$.
            This term then steps with the same reduction as $\g{t_{1L} \gcomp{T} \g{t_{1R}}}$,
            and using the appropriate diagonal rules and \cref{lem:prec-sub} gives our result.
          }
        }
        \rcase{DiagJ}{
          Then the step is with \rrule{RedJ}. So we can step the RHS with \rrule{RedJ},
          and obtain our result with \rrule{DiagCast} and the monotonicity of substitution.
        }
        \rcase{DiagEq}{
          Then only a contextual step is possible, and the result follows from IH, since the type-comparisons in \rrule{DiagEq} are all modulo reduction.
        }
        \rcase{DiagRefl}{
          Then only a contextual step is possible, so we take the same step in the RHS and use \rrule{DiagRefl}
          for our result.
        }
      \end{proofEnd}

\begin{theoremEnd}[apxlem]{lemma}[$\alpha$-Consistency Upward Closed for Precision]
  \label{lem:acst-monotone}
  If $\Gamma | \Gamma |- \g{t_{lowL}} \squbs \g{t_{highL}}$
        and  $\g{t_{lowL}} \acst \g{t_{R}}$,
        then $\g{t_{highL}} \acst \g{t_R}$
      \end{theoremEnd}
      \begin{proofEnd}

        By induction on the derivations of $\Gamma_{1} | \Gamma_{2} |- \g{t_{lowL}} \squbs \g{t'_{highL}}$.

        \case{$\acst$ derived with \rrule{CstCompR} or \rrule{CstCastR}}{
          If some sequence of \rrule{CstCompR} and \rrule{CstCastR} was used,
          we can unwrap the derivation from these until there is a use of a rule that constrains
          the syntax of the LHS.
          We then
          use the derivation from the corresponding case below, then re-apply the same sequence of \rrule{CstCompR} and \rrule{CstCastR}
          to obtain our result.
        }
        \case{\rrule{PrecCompL} (\rrule{PrecCompR} is symmetric)}{
          So $\g{t_{lowL}} =\g{t_{1} \gcomp{T} \g{t_{2}}}$, and $\Gamma_{1} | \Gamma_{2} |- \g{t_{1}} \squbs \g{t_{highL}}$.
          Then by IH we have $\g{t_{highL}} \acst \g{t_{R}}$.
        }
        \rcase{DiagComp}{
          Then $\g{t_{lowL}} = \g{t_{1} \gcomp{T} \g{t_{2}}}$ and $\g{highL} = \g{t'_{1} \gcomp{T} \g{t'_{2}}}$,
          where $\Gamma_{1} | \Gamma_{2} |- \g{t_{1}} \squbs \g{t'_{1}}$
          and $\Gamma_{1} | \Gamma_{2} |- \g{t_{2}} \squbs \g{t'_{2}}$.
          If $\acst$ was derived with \rrule{CstCompDiag},
          inversion gives $\g{t_{R}} = \g{t_{R1} \gcomp{T_{R}} \g(t_{R2})}$, so we can apply IH and \rrule{CstCompDiag}.
          If $\acst$ was derived with \rrule{CstCompL}, we can use IH and \rrule{CstCompL}.
        }
        \rcase{DiagRefl}{
          Follows from IH plus \rrule{CompRefl}.
        }
        \case{Remaining cases}{Same as GCIC}
      \end{proofEnd}

\begin{theoremEnd}[apxlem]{lemma}[$\alpha$-Consistency Symmetric]
  \label{lem:acst-symm}
  $\acst$ Is symmetric.
\end{theoremEnd}
\begin{proofEnd}
  Straightforward induction. Rules \rrule{CstVar} and \rrule{CstType} have identical terms
  on each side.
  Any uses of \rrule{CstCompL} can be turned into \rrule{CstCompR} and vice versa,
  likewise for \rrule{CstCastL} and \rrule{CstCastR}.
  The other cases follow easily from IH.
\end{proofEnd}

\vspace{-1em}
\begin{theoremEnd}[apxproof, restate command = lemEleven]{lemma}[Consistency Upward Closed for Precision]
  \label{lem:cst-up}
  If $\Gamma | \Gamma |- \g{t_{1}} \genprec \g{t'_{1}}$
  and
  \\ $\Gamma | \Gamma |- \g{t_{2}} \genprec \g{t'_{2}}$, and $\g{t_{1}} \gencst \g{t_{2}}$,
        then $\g{t'_{1}} \gencst \g{t'_{2}}$
        \lemReason{\textit{(So reducing precision of $\g{V}$ and $\g{V'}$ preserves \rrule{ElabCst})}}
      \end{theoremEnd}
      \begin{proofEnd}
        Follows from \cref{lem:acst-symm}, \cref{lem:acst-monotone} and \cref{lem:precision-step}.
      \end{proofEnd}
      \vspace{-1em}
      \begin{theoremEnd}[apxproof, restate command = lemTwelve]{lemma}[Structural Precision]
        \label{lem:prec-struct}
        $\genprec$ contains all structural rules
        \lemReason{(\textit{For homomorphic elaboration rules to produce $\genprec$-related elaboration for
          $\squbr$-related inputs})}
      \end{theoremEnd}
      \begin{proofEnd}
        Given by the diagonal rules of precision.
      \end{proofEnd}

\subsection{Metatheory: Proving Safety and the Gradual Guarantees}
\label{subsec:ggs}

Finally, we summarize the properties that we can prove by assuming \lang
satisfies the criteria of \cref{subsec:criteria}.
The general idea is that each case
in the proofs either (1) is the same as the proof for GCIC~\citep{bertrand:gcic}
or (2) follows directly from one of our criteria.
Full proofs can be found in \proofappendix.

\subsubsection{Type Safety}

Type safety is shown in the usual way for operational semantics, via progress and preservation~\citep{WRIGHT199438}.
Each well-typed \clang term is either a value, or can step to a well-typed term.
Confluence is necessary to prove preservation for dependent types.
Space restrictions mean that the formalization of values $\g{v}$
in \lang are in \ruleappendix,
but the idea is to follow \citet{bertrand:gcic}, adding  $\g{t_{1}} \gcomp{T} \g{t_{2}}$ as value when $\g{t_{1}}$ and $\g{t_{2}}$
are values, neither of $\g{t_{1}}$ and $\g{t_{2}}$ is $\gqm$ or $\err$, and $\g{T}$ is not a function type.

\begin{theoremEnd}[apxlem]{lemma}[Canonical Forms]
  \label{lem:canonical-forms}
  If $\Gamma |- \gv <= \gT$ or $\Gamma |- \gv => \gT$, and $\gv$ is not neutral, then $\g{v}$ is either
  $\gqmat{\gT'}$ or $\errat{\gT'}$ where $\g{T'}$ is convertible with $\g{T}$, or one of the
  following holds:
  \begin{itemize}
    \item $\g{T} = \g{(x : T_{1}) -> T_{2}}$ and $\g{v} = \g{\lambda (x : T'_{1}) -> T'_{2}}$
    \item $\g{T} = \g{t_{1} ==_{T'} t_{2}}$ and $\g{v} = \grefl{t_{w}}{t'_1}{t'_{2}}$
    \item $\g{T} = \g{C(\seq{t})}$ and $\g{v} = \g{D^{C}(\seq{t'})}$
    \item $\g{T} = \g{\gType{\ell}}$ and $\g{v} = \g{(x : T_{1}) -> T_{2}}$
          or $\g{v} = \g{C(\seq{i})}$
          or $\g{v} = \g{t_{1} ==_{T} t_{2}}$
          or $\g{v} = \gType{\ell'}$ for $\ell' < \ell$.
  \end{itemize}
\end{theoremEnd}
\begin{proofEnd}
  By induction on the typing derivation.

  \rcase{CastCheck}{Follows from IH on the synthesis derivation}
  \rcase{CastVar,CastApp,CastMatch,CastCast,CastJ,CastComp}{
    If $\g{v}$ is not neutral, then it cannot be a value, since the only
    value forms for eliminations are neutrals.
  }
  \rcase{CastFun,CastType,CastFun,CastApp,CastInd,CastCtor,CastUnk,CastErr,CastEq,CastRefl}{
    Immediate from the form of the typing derivation.
  }
\end{proofEnd}

\begin{theoremEnd}[apxproof]{proposition}[Confluence, Progress, Preservation and Elaboration]
  \
  \begin{itemize}
  \item $\stepsto$ is confluent.
    \item If\ $\Gamma |- \gt <= \gT$, then $\gt$ is a value or $\gt \stepsto \g{t'}$ for some $\g{t'}$.

    \item If\ $\Gamma |- \g{t_{1}} <= \gT$ and $\g{t_{1}}\stepsto \g{t_{2}}$
  then $\Gamma |- \g{t_{2}} <= \gT$.
    \item   If\ $\Gamma |- \echeck{\r{t}}{\gT}{\gt}$, then $\Gamma |- \gt <= \gT$.
  \end{itemize}
\end{theoremEnd}
\begin{proofE}
  (1) As in \citep{bertrand:gcic}. The proof uses the usual parallel reduction strategy.
  The additional cases are for \rrule{RedJ}, \rrule{RedCastEq}, \rrule{PropEqUnk} and \rrule{PropEqEq}, all of which are straightforward,
  and for $\gcompop$, which is given by our criteria.

  (2)
  As in \citep{bertrand:gcic}. We perform mutual induction on the typing derivation with the corresponding propositions
  for synthesis and constrained synthesis. The gradual criteria guarantee that the composition
  does not interfere with the canonical-forms
  property for functions, types (in cast statements) and members of inductive types,
  so all cases from GCIC are readily adapted to \clang.
  Additional cases are for \rrule{CastUnk}, \rrule{CastErr}, \rrule{CastEq}, \rrule{CastRefl}, \rrule{CastJ}, and possibly $\gcompop$.
  For \rrule{CastUnk} and \rrule{CastErr}, there is a new case because of the new type former $\g{==}$,
  but we always step with \rrule{PropEqUnk} or \rrule{PropEqErr}.
  For \rrule{CastEq} and \rrule{CastRefl}, by our hypothesis, either we can take a step in a sub-term
  or all sub-terms are normal, meaning the entire term is normal. For \rrule{CastJ}, the arguments to $\g{J}$
  must all be normal, or a step can be taken in one of them. If they are all normal, then the canonical forms
  lemma gives that the scrutinee must be $\g{refl}$, enabling a reduction with \rrule{RedJ}.
  Our gradual criteria ensure that any non-value $\gcompop$ terms can step.

  (3) As in \citep{bertrand:gcic}. We perform mutual induction on the typing derivation with the corresponding propositions
  for synthesis and constrained synthesis.
  The new cases are for \rrule{RedJ}, \rrule{RedCastEq}, \rrule{PropEqUnk} and \rrule{PropEqErr},
  as well as $\gcompop$.
  For \rrule{RedJ}, the cast types are well-typed by preservation of typing under substitution,
  and the whole expression can be typed with two applications of \rrule{CastCast}.
  For \rrule{RedCastEq}, the witness and cast-types are well-typed by the premise, so casting
  the witness is well-typed at the destination type by \rrule{CastCast}. Our gradual criteria give that
  composing terms of the same type preserves that type. Finally, we have that the witness is
  more precise than both endpoints, by transitivity and the lower-bound property from the gradual criteria,
  allowing us to apply \rrule{CastRefl}. The cases for \rrule{PropEqUnk} and \rrule{PropEqErr} are trivial,
  and the case for $\gcompop$ follows from the gradual criteria.
  Finally, we have contextual steps. These are all straightforward, or identical to GCIC, except
  for preserving the precision side-conditions of \rrule{CastRefl}, where it follows from our definition of  precision modulo convertibility.

  (4) As in \citep{bertrand:gcic}. We perform mutual induction on the typing derivation with the corresponding propositions
  for synthesis and constrained synthesis.
  The new cases are \rrule{ElabEq}, which is trivial, along with \rrule{ElabRefl}.
  For \rrule{ElabRefl}, our criteria give reflexivity of $\genprec$, so this combined with our premises
  allow us to apply \rrule{CastRefl} to type the elaboration.
\end{proofE}

These together yield the main safety theorem.

\begin{theorem}[Type Safety]
  If $\cdot |- \rt : \rT$, then $\rt$ has an elaboration that either steps to a normal form
  or steps indefinitely.
  \end{theorem}

  As a corollary, we can perform inversion on the typing derivations to obtain weak canonicity.
  That is, every well-typed closed term that terminates steps to a canonical term of its type.

  \begin{theoremEnd}[apxproof]{corollary}[Weak Canonicity]
    \label{thm:canonicity}
    Suppose $\cdot |- \g{t} : \g{T}$. Then either $\g{t}$ diverges, or $\g{t} \stepstostar \g{v}$ where:
    \begin{itemize}
      \item If\ $\g{T}$ is $\g{(x : T_{1}) -> T_{2}}$ then $\gv$ is $\g{\lambda x \ldotp \g{t'}}$
      \item If\ $\g{T}$ is $\g{C\ulev{i}(\seq{\g{t_{1}}})}$ then $\g{v}$ is $\g{D^{C}\ulev{i}(\seq{t_{2}})}$ for some $\g{D}$.
      \item If\ $\g{T}$ is $\g{t_{1} ==_{T'} t_{2}}$ then $\g{v}$ is $\grefl{t'}{t_{1}}{t_{2}}$
      \item If\ $\g{T}$ is $\gType{i}$ then $\g{v}$ is one of $\g{C\ulev{i}(\seq{\g{t_{1}}})}$,
            $\g{(x : T_{1}) -> T_{2}}$, $\gType{i-1}$ or $\g{t_{1} ==_{T'} t_{2}}$.
    \end{itemize}
  \end{theoremEnd}
  \begin{proofEnd}
   Follows from \cref{lem:canonical-forms} with type safety.
  \end{proofEnd}

  \subsubsection{Conservatively Extending CIC}

  Each CIC rule has a direct analogue in \clang, so it is clear that it extends CIC.
  Since most of the gradual-specific rules refer to $\gqm$ or $\err$, knowing that consistency
  collapses to $\alpha$-equivalence on static terms is enough to show that
  said extension is conservative.

  \begin{theoremEnd}[apxproof]{theorem}[Conservativity]
    For any BCIC-terms $\s{t}$ and $\s{T}$,
    let $\rt$ and $\rT$ be the \lang terms corresponding to $\st$ and $\sT$
    by mapping BCIC $\s{\lambda}$ to \lang $\r{\lambda}$, etc.
    Then $\cdot |- \st <= \sT$ iff $\cdot |- \rt : \rT $.
  \end{theoremEnd}
  \begin{proofEnd}
    BCIC as given by \citet{LennonBertrand2021} does not have propositional equality,
    but it is easily added with the types given in \cref{fig:geq-type-lemmas}.
    We identify $\s{refl_{t} : t ==_{T} t}$ with $\grefl{t}{t}{t}$.

    The ``only if'' direction is straightforward, since each BCIC rule has an analogue
    in \lang.
    For the other direction, we need to transform each GCIC type derivation into a BCIC derivation.
    However, this is straightforward induction: all typing rules have corresponding BCIC rules,
    except for \rrule{CastComp}, \rrule{CastUnk} and \rrule{CastErr}, but these all use gradual
    features not present in the embedding of a GCIC term into BCIC.
    For \rrule{CastCheck},
    the result follows from \cref{lem:static-cst}.
  \end{proofEnd}

\subsubsection{Gradual Guarantees}

To prove the gradual guarantees, we use the gradual criteria to show that
elaboration is monotone. This, when combined with the monotonicity of $\redsto$
with respect to semantic precision, gives us both the static and gradual guarantees.

\begin{theoremEnd}[apxproof]{proposition}[Elaboration Gradual Guarantee]
  Suppose $\r{t_{1}} \squbr \r{t_{2}}$ and $\Gamma_{1} \genprec \Gamma_{2}$
  (i.e. entries in $\Gamma_{1}$ and $\Gamma_2$ are respectively related by $\genprec$). Then:
  \begin{itemize}
    \item If\ $\Gamma_{1} |- \echeck{\r{t_{1}}}{\gT}{\g{t_{1}}}$  then $\Gamma_{2} |- \echeck{\r{t_{2}}}{\gT}{\g{t_{2}}}$
          for some $\g{t_{2}}$ where $\Gamma_{1} | \Gamma_{2} |- \g{t_{1}} \genprec \g{t_{2}}$.

    \item If\ $\Gamma_{1} |- \esynth{\r{t_{1}}}{\g{T_{1}}}{\g{t_{1}}}$ then $\Gamma_{2} |- \esynth{\r{t_{2}}}{\g{T_{2}}}{\g{t_{2}}}$
          for some $\g{T_{2}},\g{t_{2}}$ where
          $\Gamma_{2} |- \g{T_{}}$
          and $\Gamma_{1} | \Gamma_{2} |- \g{t_{1}} \genprec \g{t_{2}}$.
  \end{itemize}
\end{theoremEnd}
\begin{proofEnd}
  Our gradual criteria ensure that every rule of structural precision from \citep{bertrand:gcic} is admissible
  for $\genprec$, so our proof follows the same format. The new cases are for \rrule{ElabEq}, and \rrule{ElabRefl}, which are immediate from the IH.
  We reiterate the case for \rrule{ElabCst} to show how it fits with the gradual criteria.
  Suppose that
  $\Gamma_{1} \vdash {\r{t_{1}}} \elabsto \g{t_{1}} => \g{T_{1}}$,
  that
  $\Gamma_{2} \vdash {\r{t_{2}}} \elabsto \g{t_{2}} => \g{T_{2}}$,
  and that $\r{t_{1}} \squbr \r{t_{2}}$.
  We know that $\g{T} \defcst \g{T_{1}}$,
  and our goal is to show that
  $\Gamma_{1} | \Gamma_{2} \vdash \cast{T_{1}}{T}\g{t_{1}} \genprec \cast{T_{2}}{T}\g{t_{2}} $
  and that
  $\g{T} \defcst \g{T_{2}}$.
  The former is immediate from \cref{lem:cast-mono}.
  The latter follows from \cref{lem:cst-up}.
\end{proofEnd}

This, combined with the preservation of precision under evaluation, is enough to prove the
static and dynamic gradual guarantees as stated in \cref{subsec:guarantees}.
The hard work lies in proving that reduction preserves precision, which we leave to \cref{subsec:fulfill-criteria}.

\section{Consistency and Precision}
\label{sec:prec}

Motivated by the criteria of
\cref{subsec:criteria}, in this section we extend GCIC's precision and consistency relations
to accommodate propositional equality and composition.
We show that our relations fulfill the laws of \cref{subsec:criteria},
thus showing that \lang fulfills type safety and the gradual guarantees,
justifying the design of precision and consistency.

\subsection{Review: Precision and Consistency in GCIC}

\subsubsection{Structural Precision}

\Cref{fig:gcic-prec} recalls \textit{structural precision}
from GCIC~\citep{bertrand:gcic},
written  as $\Gamma |- \g{t_{1}} \squbs \g{t_{2}}$.
Structural precision is the syntactic relation out of which
definitional precision $\defprec$ is build.
The \textit{generating rules} \rrule{GenUnk} and \rrule{GenErr} establish $\qmat{\gT}$
and $\errat{\gT}$ as the least and most precise terms of type $\g{T}$.
For technical reasons, \rrule{GenUnkUniv} allows some cumulativity for $\qmat{\gType{j}}$,
while \rrule{GenErrLam} encodes a version of $\eta$-expansion for errors.
The \textit{diagonal rules} (named \rrule{Diag*})
are structural: terms are precision related if they are built with the same
syntactic construct and the corresponding sub-terms are precision-related.
We show a few examples, but omit most diagonal rules for space reasons.
Finally, \textit{cast rules} capture non-structural properties of casts.
Rule \rrule{Cast-L} states that a casting $\g{t}$ is more precise that $\g{t'}$ if
the cast's source and destination types are both more precise than the type of $\g{t'}$,
and if $\g{t}$ is more precise that $\g{t'}$.
The rule \rrule{Cast-R} says the opposite: casting $\g{t}$ is less precise than $\g{t}$
is the source and destination are both less precise than the type of $\g{t}$ and $\g{t}$ itself
is less precise than $\g{t'}$.

\newcommand{\drawGCICPrec}{
  \ifapx{\begin{figure}[H]}
    \ifnotapx{\begin{figure}}
  \centering
  \begin{boxedarray}{@{}l@{}}
    \jform{$\Gamma |- \g{t_1} \squbs \g{t_2}$ \textit{(Precision: \ifnotapx{Key} Rules from GCIC)}}
    \begin{inferbox}
      \inferrule[GenUnk]
      {\Gamma_1  |- \gt =>* \g{T'}\\\\
      \Gamma_1 | \Gamma_2 |- \g{T'} \defprec \g{T} }
      {\Gamma_1 | \Gamma_2 |- \gt \squbs \qmat{T}}

      \inferrule[GenErr]
      {\Gamma_1  |- \gt =>* \g{T'}\\\\
      \Gamma_1 | \Gamma_2 |- \g{T} \defprec \g{T'} }
      {\Gamma_1 | \Gamma_2 |-  \errat{T} \squbs \gt}

      \inferrule[GenUnkUniv]
      {\Gamma_1  |- \gT \psynthstar{\gType{  }} \gType{i}\\\\
      \g{i} < \g{j} }
      {\Gamma_1 | \Gamma_2 |- \gT \squbs \qmat{\gType{j}}}

  \inferrule[DiagVar]
  { }
  {\Gamma_1 | \Gamma_2 |- \g{x} \squbs \g{x}}

      \inferrule[GenErrLam]
      {\Gamma_2 |- \g{t_2} \psynthstar{\gradualcolor{\Pi}} \g{(x : T_2) -> T'_2} \\\\
      \Gamma_1 | \Gamma_2 |- \g{(x : T_1) -> T'_1} \defprec \g{(x : T_2)-> T'_2}}
      {\Gamma_1 | \Gamma_2 |- \g{\lambda(x : T_1)\ldotp \errat{T'_1}} \squbs \g{t_2}}

      \inferrule[DiagAbs]
      {\Gamma_1 | \Gamma_2 |- \g{T_1} \defprec \g{T_2} \\\\
        \Gamma_1, (\gx : \g{T_1}) | \Gamma_2, (\gx : \g{T_2}) |- \g{t_1} \squbs \g{t_2}
      }
    {\Gamma_1 | \Gamma_2 |- \g{\lambda(x : T_1)\ldotp t_1}  \squbs \g{\lambda(x : T_2)\ldotp t_2}}

  \inferrule[DiagCast]
  {
    \Gamma_1 | \Gamma_2 |- \g{T_1} \squbs \g{T_2}\\
    \Gamma_1 | \Gamma_2 |- \g{T'_1} \squbs \g{T'_2}\\
    \Gamma_1 | \Gamma_2 |- \g{t_1} \squbs \g{t_2}
  }
  {\Gamma_1 | \Gamma_2 |- \cast{T_1}{T'_1}\g{t_1} \squbs \cast{T_2}{T'_2}\g{t_2}}

  \inferrule[CastL]
  {\Gamma_2 |- \g{t_2} =>* \g{T_2}\\
    \Gamma_1 | \Gamma_2 |- \g{T_1} \defprec \g{T_2}\\\\
    \Gamma_1 | \Gamma_2 |- \g{T'_1} \defprec \g{T_2}\\
    \Gamma_1 | \Gamma_2 |- \g{t_1} \squbs \g{t_2}
  }
  {\Gamma_1 | \Gamma_2 |- \cast{T_1}{T'_1}\g{t_1} \squbs \g{t_2}}

  \inferrule[CastR]
  {\Gamma_1 |- \g{t_1} =>* \g{T_1}\\
    \Gamma_1 | \Gamma_2 |- \g{T_1} \defprec \g{T_2}\\\\
    \Gamma_1 | \Gamma_2 |- \g{T_1} \defprec \g{T'_2}\\
    \Gamma_1 | \Gamma_2 |- \g{t_1} \squbs \g{t_2}
  }
  {\Gamma_1 | \Gamma_2 |- \g{t_1} \squbs \cast{T_2}{T'_2}\g{t_2} }
  \ifapx{

      \inferrule[DiagEq]
      {
        \Gamma_1 | \Gamma_2 |- \g{T_1} \defprec \g{T_2}\\\\
        \Gamma_1 | \Gamma_2 |- \g{t_1} \squbs \g{t_2}\\
        \Gamma_1 | \Gamma_2 |- \g{t'_1} \squbs \g{t'_2}
      }
      {\Gamma_1 | \Gamma_2 |- \g{t_1 ==_{T_1} t'_1} \squbs \g{t_2 ==_{T_2} t'_2}}

    \inferrule[DiagInd]
    {\Gamma_1 | \Gamma_2 |- \seq{\g{t_1} \squbs \g{t_2}} }
    {\Gamma_1 | \Gamma_2 |- \g{C\ulev{i}}\seq{\g{t_1}} \squbs \g{C\ulev{i}}\seq{\g{t_2}} }

    \inferrule[DiagCons]
    {\Gamma_1 | \Gamma_2 |- \seq{\g{t_1} \squbs \g{t_2}}\\
    \Gamma_1 | \Gamma_2 |- \seq{\g{t'_1} \squbs \g{t'_2}}}
  {\Gamma_1 | \Gamma_2 |- \g{D^C(}\ \seq{\g{t_1}}, \seq{\g{t'_1}}\g{)} \squbs \g{D^C(}\ \seq{\g{t_2}}, \seq{\g{t'_2}} \g{)} }

    \inferrule[DiagApp]
    {\Gamma_1 | \Gamma_2 |- \g{t_1} \squbs \g{t_2}\\\\
    \Gamma_1 | \Gamma_2 |- \g{t'_1} \squbs \g{t'_2}}
  {\Gamma_1 | \Gamma_2 |- \g{t_1\ t'_1} \squbs \g{t_2\ t'_2}}

      \inferrule[DiagUniv]
      { }
      {\Gamma_1 | \Gamma_2 |- \gType{\ell} \squbs \gType{\ell} }

      \inferrule[DiagProd]
      {\Gamma_1 | \Gamma_2 |- \g{T_1} \squbs \g{T_2}\\\\
        \Gamma_1, (\gx : \g{T_1}) | \Gamma_2, (\gx : \g{T_2}) |- \g{T'_1} \squbs \g{T'_2}
      }
      {\Gamma_1 | \Gamma_2 |- \g{(x : T_1) -> T'_1} \squbs \g{(x : T_2) -> T'_2}
      }

  \inferrule[DiagFix]
  {\Gamma_1 | \Gamma_2 |- \g{s} \squbs \g{s'}\\
    \Gamma_1 |- \gs {\psynthstar{\g{C}}} \g{C(\seq{a})}\\
    \Gamma_2 |- \gs' {\psynthstar{\g{C}}} \g{C(\seq{a'})}\\
    \Gamma_1,(\gz : \g{C(\seq{a})}) | \Gamma_2, (\gz : \g{C(\seq{a'})}) |- \g{P} \squbs \g{P'}\\
    \Gamma_1, (\g{f} : \g{(z : C(\seq{a}))}), \seq{(\g{y} : \g{Y_k}[\g{a}/\g{x}])}
    | \Gamma_2, (\g{f} : \g{(z : C(\seq{a'}))}), \seq{(\g{y} : \g{Y_k}[\g{a'}/\g{x}])}
    |- \g{t_k} \squbs \g{t'_k}
  }
  {\Gamma_1 | \Gamma_2 |- \g{\ind_C(s,z.P,f.y.t)} \squbs \g{ind_C(s',z.P', f.y.t')}  }
    }
\end{inferbox}
\ifapx{
\end{boxedarray}
\ifapxCaption{GCIC Structural Precision}
\end{figure}
\begin{figure}[H]
  \begin{boxedarray}{@{}l@{}}
}
\ifnotapx{\\\\}
    \jform{$\g{t_1} \acst \g{t_2}$ \textit{(Consistency\ifnotapx{: Non-structural Rules})}}
    \begin{inferbox}
      \inferrule[CstVar]
      { }
      {\gx \acst \gx}

      \inferrule[CstUnkL]
      { }
      {\qmat{\gT} \acst \gT}

      \inferrule[CstUnkR]
      { }
      {\g{t} \acst \qmat{\gT}}

      \inferrule[CstCastL]
      {\gt \acst \g{t'}}
      {\cast{T_1}{T_2}\gt \acst \g{t'}}

      \inferrule[CstCastR]
      {\gt \acst \g{t'}}
      {\gt \acst \cast{T_1}{T_2}\g{t'}}
      \ifapx{

      \inferrule[CstRefl]
      {
         \g{t_1} \acst \g{t_2}\\\\
         \g{t'_1} \acst \g{t'_2}\\
         \g{t''_1} \acst \g{t''_2}
      }
      { \grefl{t_1}{t'_1}{t''_1} \acst \grefl{t_2}{t'_2}{t''_2}}

      \inferrule[CstAbs]
      { \g{T_1} \acst \g{T_2} \\\\
        \g{t_1} \acst \g{t_2}
      }
    { \g{\lambda(x : T_1)\ldotp t_1}  \acst \g{\lambda(x : T_2)\ldotp t_2}}

  \inferrule[CstVar]
  { }
  { \g{x} \acst \g{x}}

  \inferrule[CstCast]
  {
     \g{T_1} \acst \g{T_2}\\
     \g{T'_1} \acst \g{T'_2}\\\\
     \g{t_1} \acst \g{t_2}
  }
  { \cast{T_1}{T'_1}\g{t_1} \acst \cast{T_2}{T'_2}\g{t_2}}

      \inferrule[CstEq]
      {
         \g{T_1} \acst \g{T_2}\\\\
         \g{t_1} \acst \g{t_2}\\
         \g{t'_1} \acst \g{t'_2}
      }
      { \g{t_1 ==_{T_1} t'_1} \acst \g{t_2 ==_{T_2} t'_2}}

    \inferrule[CstInd]
    { \seq{\g{t_1} \acst \g{t_2}} }
    { \g{C\ulev{i}}\seq{\g{t_1}} \acst \g{C\ulev{i}}\seq{\g{t_2}} }

    \inferrule[CstCons]
    { \seq{\g{t_1} \acst \g{t_2}}\\
     \seq{\g{t'_1} \acst \g{t'_2}}}
  { \g{D^C(}\ \seq{\g{t_1}}, \seq{\g{t'_1}}\g{)} \acst \g{D^C(}\ \seq{\g{t_2}}, \seq{\g{t'_2}} \g{)} }

    \inferrule[CstApp]
    { \g{t_1} \acst \g{t_2}\\\\
     \g{t'_1} \acst \g{t'_2}}
  { \g{t_1\ t'_1} \acst \g{t_2\ t'_2}}

      \inferrule[CstUniv]
      { }
      { \gType{\ell} \acst \gType{\ell} }

      \inferrule[CstProd]
      { \g{T_1} \acst \g{T_2}\\\\
        \g{T'_1} \acst \g{T'_2}
      }
      { \g{(x : T_1) -> T'_1} \acst \g{(x : T_2) -> T'_2}
      }

  \inferrule[CstFix]
  {\seq{\gt \acst \g{t'}}}
  {\g{\ind_C(\seq{\gt})} \acst \g{\ind_C(\seq{\g{t'}})}}
  }
    \end{inferbox}
  \end{boxedarray}
  \ifapxCaption{\ifnotapx{Structural Precision and } Consistency for GCIC\ifnotapx{: Key Rules}}
  \ifnotapx{\label{fig:gcic-prec}}
\end{figure}
}
\drawGCICPrec

Structural precision uses an auxiliary type judgment:
\textit{presynthesis } $\Gamma |- \gt =>* \gT $ is defined to be
exactly the type synthesis relation without the $\squbs$ side-condition in \rrule{CastRefl}.
Presynthesis types strictly more terms than synthesis, and both produce the same type,
since they differ only in side-conditions.
The side-condition is not used in the type-safety proof,
so any run-time terms that presynthesize a type are safe.
Unlike GCIC, \lang uses precision to type equality witnesses,
so presynthesis avoids a circular dependency between typing and precision.

Structural precision is defined mutually with definitional precision (\cref{fig:gcic-prec}) $\defprec$,
which acts as $\genprec$ from \cref{sec:generic}.
Definitional precision allows reducing before comparing, and is used with type ascriptions, such as for functions, equality proofs
and casts.
Since the checking rule for CastCIC allowed arbitrary reductions, a term may be well-typed
even if its type ascriptions are not fully reduced. Type ascriptions on a term may need to be reduced before structural precision is apparent.
This definition is due to \citet{bertrand:gcic}.

\subsubsection{Syntactic Consistency}
\label{subsec:gcic-cst}

\Cref{fig:gcic-prec} defines consistency for GCIC.
All terms are consistent with $\gqm$ (\rrule{CstUnkL, CstUnkR}).
Each syntactic construct also has an (omitted) structural rule.
Unlike precision, consistency between terms ignores type ascriptions,
and casts are also ignored (\rrule{CstCastL, CstCastR}).
We follow GCIC and let $\errat{T}$ be consistent with $\gqmat{T}$.

\subsection{Precision and Consistency for \lang}
\label{subsec:comp-prec}

The structural precision laws are not sufficient for handling composition.
In particular, we want $\Gamma | \Gamma' |- \g{t_{1} \gcomp{T} t_{2}} \sqube \g{t_{1}}$,
with the same holding for $\g{t_{2}}$. However, this fact is not derivable
from the diagonal rule for composition. Instead, we must add rules to ensure that composing produces a lower
bound.
However, once we start adding non-structural rules, we must be careful not to disrupt the other
properties we need from precision. For example, \cref{subsec:criteria} states that precision must be transitive.
If $\g{(x \gcomp{T} y) \gcomp{T} z} \sqube \g{x \gcomp{T} y} $ and
$\g{x \gcomp{T} y} \sqube \g{x}$, but we also want to have $\g{(x \gcomp{T} y) \gcomp{T} z} \sqube \g{x}$,
i.e., we need to be able to transitively apply the fact that composing produces a lower bound.

\Cref{fig:clang-prec-cst} shows the added rules.
\rrule{DiagRefl} and \rrule{DiagComp}, along with the omitted \rrule{DiagEq} and \rrule{DiagJ}
are like the other diagonal rules.
The rules \rrule{PrecCompL} and \rrule{PrecCompR} encode that the composition is a precision-lower bound
in a way that preserves transitivity.
The rules for $\defsuprec$ are also shown: like $\defprec$ they allow
for reductions before comparing with structural precision, but they also allow backwards steps,
fulfilling \cref{lem:super-prec}. We only allow backwards-steps for the less precise
term, since backwards steps on the left-hand are admissible by \cref{lem:precision-step}.

Last are the rules for (static) consistency for composition.
Recall that \cref{subsec:criteria} required that reducing precision preserved consistency.
Since composition is as precise as both its arguments, $\g{t_{1} \gcomp{T} \g{t_{2}}} \acst \g{t_{3}}$
should imply that $\g{t_{1}}$ and $\g{t_{2}}$ are both consistent with $\g{t_{3}}$.
We conjecture that composition is a (semantic) greatest lower bound,
which would mean that errors in composing witnesses are never flagged earlier than necessary.
For this to hold, the composition of two terms must be consistent with everything that is
consistent with both of those two terms.
Our composition consistency rules in \cref{fig:clang-prec-cst} express this: \rrule{CstCompL} and \rrule{CstCompR}
ensure that $\g{t_{1}} \gcomp{T} \g{t_{2}}$ is consistent with exactly the terms that are consistent with
both $\g{t_{1}}$ and $\g{t_{2}}$.

With consistency fully defined, the difference between static and dynamic consistency is now clear:
two terms that share a non-error lower-bound may be statically inconsistent
if they differ only in neutral terms.
Variables are only statically consistent with themselves (\rrule{CstVar})
or $\gqm$ (\rrule{CstUnk{L,R}}).
However, for any two variables $\g{x}$ and $\g{y}$, $\g{x} \gcomp{T} \gy$ is a non-error
term that is as precise as both, as given by \rrule{PrecComp(L,R)}.
This disconnect between precision and consistency is justified
by the criteria of \cref{subsec:criteria}: we show below that reducing precision
preserves consistency.
The separation of static and dynamic consistency enables the gradual guarantees and
conservatively embedding CIC while maintaining static equivalences.

\newcommand{\drawGeqPrec}{
\begin{figure}
  \centering
  \begin{boxedarray}{@{}l@{}}
    \begin{inferbox}
      \inferrule[DiagRefl]
      {
        \Gamma_1 | \Gamma_2 |- \g{t_1} \squbs \g{t_2}\\\\
        \Gamma_1 | \Gamma_2 |- \g{t'_1} \defprec \g{t'_2}\\
        \Gamma_1 | \Gamma_2 |- \g{t''_1} \defprec \g{t''_2}
      }
      {\Gamma_1 | \Gamma_2 |- \grefl{t_1}{t'_1}{t''_1} \squbs \grefl{t_2}{t'_2}{t''_2}}

      \inferrule[DiagComp]
      {
        \Gamma_1 | \Gamma_2 |- \g{T_1} \squbs \g{T_2}\\
        \Gamma_1 | \Gamma_2 |- \g{t_1} \squbs \g{t_2}\\
        \Gamma_1 | \Gamma_2 |- \g{t'_1} \squbs \g{t'_2}
      }
      {\Gamma_1 | \Gamma_2 |- \g{t_1} \gcomp{T_1} \g{t'_1} \squbs \g{t_2} \gcomp{T_2} \g{t'_2}}

      \inferrule[PrecCompL]
      {\Gamma_1 | \Gamma_2 |- \g{t_1} \squbs \g{t_3}}
      {\Gamma_1 | \Gamma_2 |- \g{t_1} \gcomp{T} \g{t_2} \squbs \g{t_3} }

      \inferrule[PrecCompR]
      {\Gamma_1 | \Gamma_2 |- \g{t_2} \squbs \g{t_3}}
      {\Gamma_1 | \Gamma_2 |- \g{t_1} \gcomp{T} \g{t_2} \squbs \g{t_3} }

      \inferrule[CstComp(L,R)]{\g{t_1} \acst \g{t_3}\\
      \g{t_2} \acst \g{t_3}}
    {\g{t_1} \gcomp{T} \g{t_2} \acst \g{t_3}\\\\
      \g{t_3} \acst \g{t_1} \gcomp{T} \g{t_2}
    }

    \inferrule[CstCompDiag]
    {\g{t_1} \acst \g{t'_1}\\
    \g{t_2} \acst \g{t'_2}}
  {\g{t_1} \gcomp{T} \g{t_2} \acst \g{t'_1} \gcomp{T'} \g{t'_2}}

  \inferrule[PrecModStruct]{
    \Gamma_1 | \Gamma_2 |- \g{t_1} \squbs \g{t_2}
  }{
    \Gamma_1 | \Gamma_2 |- \g{t_1} \defsuprec \g{t_2}
  }

  \inferrule[PrecModStepL]{
    \g{t_1} \stepstostar \g{t'_1}\\\\
    \Gamma_1 | \Gamma_2 |- \g{t'_1} \defsuprec \g{t_2}
  }{
    \Gamma_1 | \Gamma_2 |- \g{t_1} \defsuprec \g{t_2}
  }

  \inferrule[PrecModStepR]{
    \g{t_2} \stepstostar \g{t'_2}\\\\
    \Gamma_1 | \Gamma_2 |- \g{t_1} \defsuprec \g{t'_2}
  }{
    \Gamma_1 | \Gamma_2 |- \g{t_1} \defsuprec \g{t_2}
  }

  \inferrule[PrecModStepBack]{
    \g{t'_2} \stepstostar \g{t_2}\\\\
    \Gamma_1 | \Gamma_2 |- \g{t_1} \defsuprec \g{t'_2}
  }{
    \Gamma_1 | \Gamma_2 |- \g{t_1} \defsuprec \g{t_2}
  }
  \ifapx{

  \inferrule[DiagJ]
  {
    \Gamma_1 |- \g{t_1} =>* \g{T}\\
    \Gamma_2 |- \g{t'_1} =>* \g{T'}\\
    \Gamma_1, (\gz : \g{T}) | \Gamma_2, (\gz : \g{T'}) |- \g{T_P} \squbs \g{T'_P}\\
    \Gamma_1 | \Gamma_2 |- \g{t_1} \squbs \g{t'_1}\\
    \Gamma_1 | \Gamma_2 |- \g{t_2} \squbs \g{t'_2}\\
    \Gamma_1 | \Gamma_2 |- \g{t_3} \squbs \g{t'_3}\\
    \Gamma_1 | \Gamma_2 |- \g{t_4} \squbs \g{t'_4}
  }
  {\Gamma_1 | \Gamma_2 |- \g{\J(z.T_P, t_1, t_2, t_3, t_4)} \squbs \g{\J(z.T'_P, t'_1, t'_2, t'_3, t'_4)}}

  \inferrule[CstJ]
  {\seq{\gt \acst \g{t'}}}
  { \g{\J(\seq{\gt})} \acst \g{\J(\seq{\g{t'}})} }
  }
    \end{inferbox}
  \end{boxedarray}
  \caption{\clang Precision and Consistency rules for composition}
  \ifnotapx{\label{fig:clang-prec-cst}}
\end{figure}
}
\drawGeqPrec

\subsection{Fulfilling The Critera}
\label{subsec:fulfill-criteria}

\lang is now fully defined: we have defined the precision relations $\defprec$ and $\defsuprec$ and the consistency relation $\defcst$
to instantiate $\genprec$, $\gensuprec$ and $\gencst$.
We now establish that these relations fulfill
the criteria of \cref{subsec:criteria}.
We give the intuition
behind some of the cases that are new compared to GCIC.
Full proofs can be found in \proofappendix.

\begin{itemize}[leftmargin=*]

  \item \textbf{Immediate Results:}
        Proving reflexivity of $\defprec$ (\cref{lem:reflexive}) is a straightforward induction.
        The rules \rrule{PrecCompL} and \rrule{PrecCompR} make the composition of terms as precise
        as either term, proving \cref{lem:comp-lower}.
        The closure of $\defsuprec$ under convertibility is built into its definition, proving
        \cref{lem:super-prec}.
        \rrule{DiagCast}
        gives that casts are monotone, proving \cref{lem:cast-mono}.
        The monotonicity of substitution (\cref{lem:sub-mono}) is proved with a straightforward induction,
        relying on presynthesis preserving types under substitution.
        The remaining diagonal rules give that
        $\defprec$ has all structural rules, fulfilling \cref{lem:prec-struct}.

\item \textbf{Composition Safety} (\cref{lem:comp-safety})
 For progress, each composition of two canonical forms of the same type has a reduction.
        If one of the composed terms is not a canonical form, then either
(1) one of the composed terms can reduce,
(2) one term is a $\gqm_{\gT}$ or $\err_{\gT}$ where $\g{T}$ is not a function or equality type,
and we can reduce with \rrule{RedCompUnk} or \rrule{RedCompErr},
or (3) one of the composed terms is neutral, and hence the composition is neutral.
        For preservation, either the result is immediate, or casts are inserted to ensure that
        types are preserved.

\item \textbf{Composition Confluence} (\cref{lem:comp-confluence})
 \rrule{RedCompUnk(L,R)} ensures that composing with $\gqmat{T}$ only reduces
when $\gqmat{T}$ cannot reduce, avoiding a ``diamond'' problem.

  \item \textbf{Precision Transitive} (\cref{lem:trans}):
        We actually prove this \textit{after} monotonicity of reduction,
        which lets us prove that precision-related types have precision-related terms,
        which is necessary to fulfill premises on term's types, such as in \rrule{CastL}
        and \rrule{CastR}.
        The rest is straightforward induction.

  \item \textbf{Static Consistency} (\cref{lem:static-cst})
  The \lang rules not present in GCIC are for equality, which are trivially handled, and consistency rules for composition, for which the result vacuously holds since composition is
  not present in the static language.

  \item \textbf{Monotonicity of Reduction} (\cref{lem:precision-step}):
  The key fact is that, since \rrule{PrecCompL} and \rrule{PrecCompR} only have composition on the left,
  all the inversions in the GCIC proofs are still valid for \lang.
  The interesting case is when precision is derived using \rrule{PrecCompL} (\rrule{PrecCompR} is symmetric),
  and the composition reduces. The result is trivial for \rrule{RedCompUnk(L,R)} and \rrule{RedCompErr(L,R)}.
  For the remaining cases, two terms with the same head are being composed, and the result is either $\err$
  or another term with the same head. When $\err$ is produced the result is trivial. When a term with
  the same head is produced, the  \rrule{PrecCompL} can be used with the appropriate diagonal rule.
  In the case that casts are present in the result of composition, \rrule{CastL} is used.
  The other notable case is when $\g{J}$ reduces, where the result is derived using \rrule{DiagCast}.

  \item \textbf{Consistency Upward Closed} (\cref{lem:cst-up})
        We first show that it consistency is upward closed on the left, then prove that it is
        symmetric to obtain upward closure for both arguments.
        The case when $\acst$ is derived with \rrule{CstCompR} or \rrule{CstCastR}
        must be handled specially, since they each take an operand that can be any term.
        The trick is to unwrap the chain of \rrule{CstCompR} and \rrule{CstCastR}
        uses, use the induction hypothesis on the contained derivation, then re-apply
        \rrule{CstCompR} and \rrule{CstCastR} in the same order to obtain the result.
        When precision is derived with \rrule{PrecCompL} or \rrule{PrecCompR}, then consistency
        must have either been derived with \rrule{CstCompDiag}, in which case the result
        follows from the induction hypothesis, or with \rrule{CstCompL} or (symmetrically) \rrule{CstCompR}.
        For \rrule{CstCompL}, the premise gives that both composed terms are consistent with the
        right-hand term, yielding our result. The remaining cases are straightforward.

\end{itemize}

\section{Discussion}
\label{sec:discussion}

\subsection{Extensions Enabled by Equality}
\label{sec:inductives}

In addition to catching the kinds of bugs discussed in \cref{sec:example},
we show some benefits of having propositional equality in \lang.
Three new language features can be encoded using propositional equality,
without augmenting the cast calculus: empty types, Axiom K, and indexed inductive type families.
For type families, we discuss some limitations of the approach and workarounds for these limitations,
showing how our
cast calculus is expressive enough to pave the way for future improvements.

\subsubsection{The Empty Type}

Just as the gradual $\g{J}$ needed computation,
eliminating the empty type has computational content in a gradual language.
In static languages, the empty type $\s{Empty}$
has no closed values,
so either $\s{Empty}$ contains no terms, or
(for logically inconsistent languages) any such terms
are non-terminating.
The elimination function $\s{exfalso : (X : \sType{i}) -> Empty -> X}$
produces a result of any type, given a value of the empty type.
In a gradual language, however, $\rqm$ and $\err$ can be used at any type,
including the empty type. So a gradual $\g{exfalso}$ must produce a value of type $\gX$.

We  again follow the goal of dynamically tracking constraints expressed by types.
For the empty type, a value of type $\g{f : T -> Empty}$ encodes the constraint that $\gT$ should be impossible,
and a branch built using $\g{exfalso}$ should be unreachable.
If $\g{f}$ is applied to $\g{t} : \g{Empty}$, created using $\gqm$ or casts,
then the constraint has been violated, and an error should be raised.

We can encode this behavior by defining $\g{Empty}$ to be $\g{true ==_{\bB} false}$,
and $\g{exfalso} $ to be
\\$\g{\lambda X \ldotp \lambda t \ldotp J\ (b \ldotp if\ b\ \bB\ X)\ true\ t}$.
The key is that  $\g{Empty}$ and $\errat{Empty}$ both evaluate to
$\grefl{\errat{\bB}}{true}{false}$. So the only value of type $\g{Empty}$ is a dynamic type error.
Likewise, the eliminator $\g{exfalso}$ casts $\g{t}$ to type $\errat{\gType{\ell}}$ before casting it
to type $\gX$, so the result is always $\errat{\gX}$.
Without adding any features to \clang, the bug-finding described in \cref{sec:example}
handles constraints encoded as logical negation.

\subsubsection{Axiom K}
\label{subsec:axiomK}
Because $\gqmat{t_1 ==_T t_2}$ steps to $\grefl{t_1 \gcomp{T} g_2}{t_1}{t_2}$,
\lang is in the class of dependently typed languages where $\g{refl}$ is the only constructor
for equality. Composition can be used to derive a (gradual) proof of this uniqueness, even though
no such proof can be derived in most static type theories~\citep{hofmann1998groupoid}:
\begin{flalign*}
& \g{K : (x : T) -> (pf : x ==_{T} x) -> pf ==_{x ==_{T} x} \grefl{x}{x}{x}}\\
& \g{K\ x\ pf = \grefl{pf \gcomp{x ==_{T} x} \grefl{x}{x}{x}}{pf}{\grefl{x}{x}{x}} }
\end{flalign*}
Axiom K can be used to prove that all equality proofs of a given type are equal ~\citep{streicher1993investigations}, so
our proof-irrelevant $\g{\J}$ principle does not lose any expressivity,
since any types parameterized by an equality proof can be rewritten with $\g{K}$.
Also, Axiom K allows for conventional dependent pattern matching to be elaborated
into inductive eliminators~\citep{Goguen2006},
providing a lightweight alternative to the cumbersome $\g{\ind}$ form.
The combination of Axiom K and function extensionality suggests a connection to
Observational Type Theory~\citep{Altenkirch:2007:OE:1292597.1292608,10.1145/3498693}
that warrants future exploration.

\subsubsection{Inductive Types}

\Citet{dtfpp} describes how, using propositional equality, indexed inductive families
can be encoded. The main idea is, instead of having each constructor return different indices,
each index is a parameter, and each constructor takes an equality proof that the parameter
has the desired value.
In the elimination principle, the $\g{J}$ is used to rewrite the type of the returned value
using the stored equality.
Consider how the classic vector type is transformed:
\begin{flalign*}
  \scriptstyle
  & \g{data\ Vec\ (X \!:\! \gType{}) \!:\! (n :\!\bN) \!->\! \gType{}\ \textit{where}} &
  & \g{data\ Vec' (X : \gType{}) (n : \bN) \!:\! \gType{}\ \textit{where}}\\
  & \qquad \g{Nil : Vec\ X\ 0}&
  & \qquad \g{Nil' : (n ==_{\bN} 0) -> Vec\ X\ y}\\
  & \qquad \g{Cons : X \!->\! Vec\ X\ n  \!->\! Vec\ X\ (1\!+\!n)}&
  & \qquad \g{Cons' : (z \!:\! \bN) \!->\! X \!->\! Vec\ X\ z } \\
  &&& \qquad\qquad \g{-> n \!==_{\bN}\! (1\!+\!z) \!->\! Vec\ X\ n}
\end{flalign*}
This transformation gives a low-overhead way to incorporate indexed inductive families
with gradual dependent types. Since no extensions to \clang are required, the safety
and gradual guarantee results from \cref{sec:generic} apply.
The constructors take equality proofs, so violations of those equalities
raise dynamic type errors.

However, the approach is limited in its ability to eagerly detect errors.
The problem is that dynamic consistency is fundamentally not transitive,
since
otherwise all types are consistent through $\gqm$.
Members of inductive types are essentially trees,
and equality constraints track constraints at each level of the tree,
but consistency across the entire tree is not ensured. The witnesses track the evolution
of type information across time, but not across space.
Consider
$
  \g{Cons'\ \gqmat{\bN}\ true\ (Nil'\ \gqmat{\bN}\ \grefl{0}{\gqmat{\bN}{0}})\ \grefl{2}{2}{\gqmat{\bN}}  }
  : \g{Vec'\ \bB\ 2}
  $,
a vector with one element, whose type says it has length 2. Constructing this vector raises
no run-time type errors. At each level, the equality proof is correct: $\g{0}$ is consistent with $\gqmat{\bN}$,
and $\g{2}$ is consistent with $\g{1 + \gqmat{\bN}}$.
Gradually,
the non-transitivity means that imprecision at each level can cause disconnects between levels.

Thankfully, \clang is expressive enough to encode a solution to this problem.
By having composition as an operator in the language, one can define
so-called ``smart constructors'' that have the same types as the normal constructors,
but that access the equality proofs stored in the previous level of the tree when constructing new ones.
For example, using $\g{J}$ we can write $\g{cong1 : (m\ n : \bN) -> m ==_{\bN} n -> 1 + m == 1 + n }$,
which can be used in a ``smart''  $\g{Cons}$:%
\begin{flalign*}
  &\g{smartCons\ z\ h\ t\ eq = Cons\ z\ h\ t\ (eq\ \gcomp{(1 + z) ==_{\bN} n}\ (\cast{1\!+\!z ==_{\bN} 1\!+\!z}{(1 \!+\! z) ==_{\bN} n}(cong1\ (wit\ t))) )}\\
  &\textit{where } \g{wit : Vec\ X\ z -> z ==_{\bN} z}\\
  &\qquad\ \  \g{wit\ (Nil\ eq) = \cast{0 ==_{\bN} z}{z ==_{\bN} z}eq} \quad \mid \quad
  \g{wit\ (Cons\ x\ h\ t\ eq) = \cast{1\!+\!x ==_{\bN} z}{z ==_{\bN} z}eq}
\end{flalign*}%
When $\g{smartCons}$ is used in place of $\g{Cons'}$, the witness $\grefl{0}{\gqmat{\bN}}{0}$
is transformed to $\grefl{1}{\gqmat{\bN}}{\gqmat{\bN}}$, which produces an error when cast to
$\g{1 \!+\! \gqmat{\bN} ==_{\bN} 2}$, since $\g{1 \gcomp{\bN} 2} \redsto \errat{\bN}$.
Formalizing the general version of this approach is beyond the scope of this paper, but it shows how
having composition as an operator enables more detailed
manipulation of run-time type information.

\subsection{Future Work}

\subsubsection{Termination and Approximate Normalization}
\label{subsec:termination}

As presented, \lang has undecidable type checking, since some terms do not
terminate and consistency compares modulo reduction.
In GCIC, \citet{bertrand:gcic} show that termination can be obtained by sacrificing
the gradual guarantees, or by restricting universes so that they are not closed under function types. While useful for type theory, these sacrifices remove reasoning principles or reduce expressivity, respectively,
that make them unsuitable for programming.

\Citet{Eremondi:2019:ANG:3352468.3341692} propose \textit{approximate normalization},
where compile-time normalization of types and run-time evaluation of terms are given different semantics.
At compile-time, when missing type information means that termination cannot be guaranteed,
$\gqm$ is produced as an approximation. Run-time evaluation uses no approximations, so
expressivity is not lost.
We conjecture that approximate normalization could be introduced into \lang with little
difficulty. The main issue is finding a suitable termination argument,
since \citet{Eremondi:2019:ANG:3352468.3341692} provide a proof that does not apply to inductive types. The syntactic-model strategy
of \citet{bertrand:gcic} can likely be adapted. Also, a design decision
must be made about whether approximate or exact normalization should be used for run-time index calculations.
Witnesses are opaque to the programmer, so they could hide the cause of non-termination.

\subsubsection{Conjectures: EP Pairs, Composition, and Full Abstraction}
\Citet{bertrand:gcic} prove a stronger property than the gradual guarantees for GCIC.
They show that casts between precision-related types form an embedding-projection (EP) pair~\citep{10.1145/3236768}, so that increasing
then decreasing precision produces the same result modulo errors, and decreasing then increasing precision
produces an observationally-equivalent result.
While the gradual guarantees are helpful,
they are satisfied by trivial languages
where every cast produces $\gqm$.
Showing the EP pair property would prove that casts in \lang never lose run-time information,
giving more confidence in its ability to dynamically track constraints.
We conjecture that \lang fulfills the EP pair property, but suspect novel proof techniques are
needed to handle witness proofs.
Likewise, we conjecture that composition computes the greatest lower-bound
for semantic precision, so that each type forms a true semi-lattice.
This would establish that \lang never pre-maturely raises dynamic errors from witness composition,
since two witnesses would compose to $\err$ only when all other options are impossible.
Finally, we conjecture that there is a variant of CIC whose embedding into \lang is fully abstract, meeting the criteria
\citet{10.1145/3434288} set out for gradual languages.
Intuitively, we can form equalities between extensionally-equal functions,
and use those to cast between types indexed by them, so any property of a
function should apply to one that is extensionally-equal.
Full abstraction guarantees that all static equivalences
held in \lang, giving the programmer more tools with which to reason about their code.
Proving full abstraction for non-dependently typed
gradual languages is a recent development, so more investigation is needed to adapt them to dependent types.
The usual technique for full abstraction is to simulate the target language in the source, so every target
context can be translated into a source context that is unable to distinguish the terms.
As a consequence, the embedded variant of CIC must have capabilities for non-termination added.

\subsection{Related Work}
\label{subsec:related}

\boldpar{Flexible Dependent Types}
\lang builds on long line of work mixing dynamic and static enforcement of specifications,
in addition to GDTL and GCIC.
\Citet{dynamicDependent} support mixed static and dynamic checking of boolean-valued properties,
and \citet{lehmannTanter:popl2017} provide gradual typing for refinement types.
Similarly,
\citet{Tanter:2015:GCP:2816707.2816710} develop a system of casts for Coq, using an unsound axiom
to represent type errors. The casts supported subset types \ie a value paired with a proof
that some boolean-valued function returns true for that value, but not general inductive types.
\citet{Osera:2012:DI:2103776.2103779} present \textit{dependent interoperability}
for principled mixing of dependently typed and non-dependently typed programs.
Dependent interoperability was extended by \citet{partialTypeEquiv,dagandtabareautanter2018},
who provide a general mechanism for lifting higher-order programs to the dependently typed setting.
All of these approaches presuppose separate simple and dependent versions of types,
related by boolean-valued predicates. Our composition of witnesses provides similar checks,
but by keeping witnesses,
we do not need types to be reformulated in terms of subset types or boolean predicates.

\boldpar{Computational Equality}
\lang is not the only language to have a computational interpretation of equality.
Cubical Type Theory~\citep{ccha} defines an equality proof $\s{t_{1} ==_{T} t_{2}}$
to be a function $\s{f : \mathbb{I} -> T}$ over an abstract interval $\s{\mathbb{I}} = [\s{0},\s{1}]$
where $\s{f\ 0}$ is $\s{t_{1}}$ and $\s{f\ 1}$ is $\s{t_{2}}$.
Like in \lang, the $\s{\J}$ eliminator does not simply pattern match on an equality, but transforms
the input term to the output type by applying the function stored in the equality.
Our approach is not directly compatible with theirs, since our structural composition of types violates
univalence, but with modifications, the two notions of computing equality might be unified.

\bibliographystyle{ACM-Reference-Format}
\bibliography{myRefs}

\clearpage

\appendix

\binoppenalty=\maxdimen
\relpenalty=\maxdimen
\renewcommand{\ifapxCaption}[1]{\caption{Complete Rules - #1}}
\renewcommand{\ifapx}[1]{#1}
\renewcommand{\ifnotapx}[1]{}

\section{Full Rules}
\label{apx:rules}

\drawFigBcic
\drawFigBcicSem
\drawCastCIC
\drawFigElabGCIC
\drawClangTyping

\newcommand{\canon}{\ \mathsf{value}}
\newcommand{\neut}{\ \mathsf{neutral}}
\begin{figure}[H]
  \centering
  \begin{boxedarray}{@{}l@{}}
    \jform{$\gt \canon$ \qquad $\gt \neut$ }

    \begin{inferbox}
    \inferrule[]
    { \gt \neut }
    { \gt \canon }

    \inferrule[]
    { }
    {\gqmat{\gType{\ell}} \canon }

    \inferrule[]
    {}
    {\gqmat{C(\seq{t})} \canon }

    \inferrule[]
    { }
    {\gqmat{\gqmat{\gType{\ell}}} \canon }

    \inferrule[]
    { }
    {\gqmat{\errat{\gType{\ell}}} \canon }

    \inferrule[]
    { }
    {\errat{\gType{\ell}} \canon }

    \inferrule[]
    {}
    {\errat{C(\seq{t})} \canon }

    \inferrule[]
    { }
    {\errat{\gqmat{\gType{\ell}}} \canon }

    \inferrule[]
    { }
    {\errat{\errat{\gType{\ell}}} \canon }

    \inferrule[]
    { }
    {\gType{\ell} \canon}

    \inferrule[]
    { }
    {\g{(x : T_1) -> T_2} \canon}

    \inferrule[]
    { }
    {\g{C(\seq{\gt})} \canon}

    \inferrule[]
    { }
    {\g{t_1 ==_T t_2} \canon}

    \inferrule{}
    {\g{\lambda x \ldotp t} \canon }

    \inferrule{}
    {\g{D^C(\seq{t})} \canon}

    \inferrule[]
    { }
  {\grefl{t_w}{t_1}{t_2} \canon}

  \inferrule[]
  { }
  {\cast{\germ_i(\g{h})}{\gqmat{\gType{i}}}\gt \canon }
  \\\\\\
  \inferrule[]
  { }
  {\gx \neut}

  \inferrule[]
  {\gt \neut\\
  }
{\gt\ \g{t'} \neut}

\inferrule{ \g{t_{scrut}} \neut
}
{\g{\ind_C(t_{scrut}, \seq{\gt}) } \neut}

\inferrule[]
{\g{t_{eq}} \neut\\
}
{\g{\J(\seq{\gt}, t_{eq})} \neut}

\inferrule[]
{\g{T} \neut \\
}
{\g{t_1} \gcomp{\gT} \g{t_2} \neut}

\inferrule[]
{\g{t_1} \neut \\\\
\g{t_1} \text{ is not $\gqmat{T'}$ or $\errat{T'}$}}
{\g{t_1} \gcomp{\gT} \g{t_2} \neut}

\inferrule[]
{\g{t_1} \neut \\\\
\g{t_1} \text{ is not $\gqmat{T'}$ or $\errat{T'}$}}
{\g{t_2} \gcomp{\gT} \g{t_1} \neut}

\inferrule[]
{\g{t_1} \neut \\
  \gqmat{\gT} \neut}
{\g{t_1} \gcomp{\gT} \gqmat{\gT} \neut}

\inferrule[]
{\g{t_1} \neut \\
  \gqmat{\gT} \neut}
{\gqmat{\gT} \gcomp{\gT} \g{t_1} \neut}

\inferrule[]
{\g{t_1} \neut \\
  \g{T} \canon\\
  \errat{\gT} \neut}
{\g{t_1} \gcomp{\gT} \errat{\gT} \neut}

\inferrule[]
{\g{t_1} \neut \\
  \g{T} \canon\\
  \errat{\gT} \neut}
{\errat{\gT} \gcomp{\gT} \g{t_1} \neut}

\inferrule[]
{\g{T_1} \neut\\
}
{\cast{T_1}{T_2}\gt \neut}

\inferrule[]
{ \gT \neut}
{\cast{(x : T_1) -> T_2}{T}\gt \neut}

\inferrule[]
{ \gT \neut}
{\cast{C(\seq{\g{t}})}{T}\gt \neut}

\inferrule[]
{ \gT \neut}
{\cast{t_1 ==_{T'} t_2}{T}\gt \neut}

\inferrule[]
{ \gT \neut}
{\cast{\gType{\ell}}{T}\gt \neut}

\inferrule[]
{ \gT \neut}
{\cast{\gqmat{\gType{\ell}}}{T}\gt \neut}

\inferrule[]
{\gt \neut}
{\cast{(x : T_1) -> T_2}{(x : T'_1) -> T'_2}\gt \neut}

\inferrule[]
{\gt \neut}
{\cast{t_1 ==_{T} t_2}{t'_1 ==_{T'} t'_2}\gt \neut}

\inferrule[]
{\gt \neut}
{\cast{C(\seq{t})}{C(\seq{t'})}\gt \neut}
    \end{inferbox}
  \end{boxedarray}
  \caption{Values in \clang}
  \label{fig:values}
\end{figure}

\drawGCICPrec
\clearpage
\drawGeqPrec

\section{Proofs}
\label{apx:proofs}

\renewcommand{\genprec}{\defprec}
\renewcommand{\gensuprec}{\defsuprec}
\renewcommand{\gencst}{\defcst}

\printProofs

\end{document}
\endinput
